\documentclass[11pt,draftclsnofoot,peerreviewca,letterpaper,onecolumn]{IEEEtran}

\usepackage{cite}
\usepackage{graphicx,color,epsfig,rotating}
\usepackage{amsfonts,amsmath,amssymb,bbm}
\usepackage{algorithm,algorithmic}
\usepackage{subfigure}
\usepackage{amsmath}
\usepackage{cite}
\usepackage{mdwtab}
\usepackage{subfigure}
\usepackage{placeins}
\usepackage{psfrag, graphicx}
\usepackage[latin1]{inputenc}
\usepackage{amssymb}

\long\def\comment#1{}

\newfont{\bbb}{msbm10 scaled 700}

\newfont{\bb}{msbm10 scaled 1100}

\newcommand{\PP}{\mbox{\bb P}}

\newcommand{\EE}{\mbox{\bb E}}

\newcommand{\onev}{{\bf 1}}

\newcommand{\Ac}{{\cal A}}

\newcommand{\Ec}{{\cal E}}
\newcommand{\Fc}{{\cal F}}
\newcommand{\Gc}{{\cal G}}

\newcommand{\Lc}{{\cal L}}

\newcommand{\Tc}{{\cal T}}
\newcommand{\Uc}{{\cal U}}

\newcommand{\fsf}{{\sf f}}

\newcommand{\Asf}{{\sf A}}

\newcommand{\Gsf}{{\sf G}}

\newcommand{\Lsf}{{\sf L}}

\newcommand{\Nsf}{{\sf N}}

\newcommand{\Wsf}{{\sf W}}

\newcommand{\eqdef}{\stackrel{\Delta}{=}}

\newcommand{\be}{\begin{equation}}
\newcommand{\ee}{\end{equation}}
\newcommand{\bea}{\begin{eqnarray}}
\newcommand{\eea}{\end{eqnarray}}

\def\E{{\mathbb E}}

\def\fsf{ {\sf f}}

\def\Var{{\rm Var}}

\newtheorem{defn}{Definition}
\newtheorem{theorem}{Theorem}
\newtheorem{lemma}{Lemma}

\begin{document}

\title{The Throughput-Outage Tradeoff of Wireless One-Hop Caching Networks}

\author{Mingyue Ji,~\IEEEmembership{Student Member,~IEEE}, 
Giuseppe Caire,~\IEEEmembership{Fellow,~IEEE}, \\
and Andreas F. Molisch,~\IEEEmembership{Fellow,~IEEE}
\thanks{The authors are with the Department of Electrical Engineering,
University of Southern California, Los Angeles, CA 90089, USA. (e-mail: \{mingyuej, caire, molisch\}@usc.edu).
A short version of this work was presented at the 2013 IEEE International Symposium on Information Theory, ISIT 2013, Istanbul, July 7-11, 2013.}
\thanks{This work was supported by the Intel VAWN (Video Aware Wireless Networks) program and the 
National Science Foundation under grants CCF-1423140 and CIF-1161801.}
}

\maketitle

\newpage

\begin{abstract}
We consider a wireless device-to-device (D2D) network where the nodes 
have pre-cached information from a library of available files. 
Nodes request files at random. If the requested file is not in the on-board 
cache,  then it is downloaded from some neighboring node via {one-hop} 
``local'' communication.  
An outage event occurs when a requested file is not found in the neighborhood of the requesting node, 
or if the network {admission control} 
policy decides not to serve the request.   
{We characterize the optimal throughput-outage tradeoff 
in terms of tight scaling laws for various regimes of the system parameters, 
when both the number of nodes and the number of files in the library grow to infinity. 
Our analysis is based on Gupta and Kumar {\em protocol model}  for the underlying D2D wireless network, 
widely used in the literature on capacity scaling laws of wireless networks without caching.}
Our results show that the combination of D2D spectrum 
reuse and caching at the user nodes yields {a per-user throughput independent of the number of users, 
for any fixed outage probability in $(0,1)$. 
This implies that the D2D caching network is ``scalable'': even though the number of users increases, each user achieves 
constant throughput. This behavior is very different from the classical Gupta and Kumar result on ad-hoc wireless networks, 
for which the per-user throughput vanishes as the number of users increases. 
Furthermore, we show that the user throughput is directly proportional to the fraction of cached information 
over the whole file library size. Therefore, we can conclude that D2D caching networks can turn 
``memory'' into ``bandwidth'' (i.e., doubling the on-board cache memory on the user devices yields a 
100\% increase of the user throughout).}
\end{abstract}


\newpage

\section{Introduction}
\label{section: intro}

Data traffic generated by wireless and mobile devices is predicted to 
increase by something between one and two orders of magnitude \cite{cisco66} {in the next five years}, 
mainly due to wireless video  streaming.  Traditional methods for increasing the area spectral efficiency, such as using more spectrum and/or 
deploying more base stations, are either insufficient to provide the necessary wireless throughput increase, or are too expensive. 
Thus, exploring alternative strategies that leverage different and cheaper network resources is of great practical 
and theoretical interest. 

The bulk of wireless video traffic is due to asynchronous {\em video on demand}, where users request video files from some
library (e.g., iTunes, Netflix, Hulu or Amazon Prime) at arbitrary times. This type of traffic differs significantly from {\em live streaming}.
The latter is essentially a lossy multicasting problem, for which 
the broadcast nature of the wireless channel can be naturally exploited
(see for example \cite{ luo2008mobile, reimers1998digital, ladebusch2006terrestrial, 5755207, li2012three}). 
The theoretical foundation of schemes for live streaming relies on well-known information theoretic settings 
for one-to-many transmission of a  common message with possible refinement information, such as 
successive refinement \cite{equitz1991successive, bursalioglu2008lossy, chari2007flo} 
or multiple description coding \cite{goyal2001multiple, wang2005multiple, ahlswede1986multiple}. 

In contrast, the {\em asynchronous} nature of video on demand prevents from taking advantage 
of multicasting, despite the significant overlap of the requests (people wish to watch a few very popular files). Hence, 
even though users keep requesting the same few popular files, the asynchronism of their requests is large with respect to the 
duration of the video itself, such that the probability that a single transmission from the base station is 
useful for more than one  user is {\em essentially zero}. Due to this reason, 
current practical implementation of video on demand over wireless networks is handled at the application layer, requiring 
a dedicated data connection (typically TCP/IP) between each client (user)  and the server (base station), 
for each streaming user, as if users were requesting independent information. 

One of the most promising approaches to take advantage of the inherent {\em asynchronous content reuse}  
is {\em caching}, widely used in content distribution networks over the (wired) Internet \cite{nygren2010akamai}. 
In \cite{CommMag, DBLP:journals/corr/abs-1109-4179},  the idea of deploying dedicated ``helper nodes'' with large caches, that can be 
refreshed  via wireless at the cellular network off-peak  time, was proposed as a cost-effective alternative to providing large capacity wired backhaul 
to a network of densely deployed small cells. An even more radical view considers caching directly at the wireless users, 
exploiting the fact that modern devices have tens and even hundreds of GBytes of largely under-utilized storage space, 
which represents an enormous, cheap and yet untapped network resource. 

Recently, a {\em coded multicasting} scheme exploiting caching at the user nodes was 
proposed in \cite{maddah2012fundamental}.  In this scheme, the files in the library are divided in blocks (packets) and 
users cache carefully designed subsets {of} 
such packets. Then, for given set of user demands, 
the base station sends to all users (multicasting) a common codeword formed by a sequence of packets obtained 
as linear combinations of the original file packets.
{As noticed in} 
\cite{maddah2012fundamental}, 
coded multicasting can handle any form of asynchronism by suitable sub-packetization.  Hence, the scheme is able to create multicasting 
opportunities through coding, exploiting the overlap of demands while
eliminating the asynchronism problem. For the case of arbitrary (adversarial) demands, the coded multicasting scheme of \cite{maddah2012fundamental} is shown 
to perform within a small gap, independent of the number of users, of the cache size and of the library size,
from the cut-set bound of the underlying compound channel.\footnote{The compound nature of this model is due to the fact that
the scheme handles adversarial demands.}  However, the scheme has some significant drawbacks that makes it not easy to be implemented in practice: 
1) the construction of the caches is combinatorial and the sub-packetization explodes exponentially with the
library size and number of users;  2) changing even a single file in the library requires a significant 
reconfiguration of the user caches, making the cache update difficult.  
In \cite{maddah2013decentralized}, similar near-optimal performance of coded caching is shown to be achieved also through a  
{\em random caching scheme}, where each user caches a random selection of 
bits from each file in the library.  In this case, though, the combinatorial complexity of the coded caching scheme 
is transferred from the caching phase to the (coded) delivery phase, where the construction of the multicast codeword 
requires  solving multiple clique cover problems with fixed clique size (known to be NP-complete), for which a greedy algorithm 
is shown to be efficient. 

{\bf Our contributions:} In this paper, we focus on an alternative approach that involves random 
independent caching at the user nodes and device-to-device (D2D) communication. 
Instead of creating multicasting opportunities by coding, we exploit the spatial reuse provided by concurrent multiple short-range D2D transmissions.  
Inspired by the current standardization of a D2D mode for LTE (the 4-th generation of cellular systems) \cite{wu2010flashlinq}, 
we restrict to one-hop communication.  Under such assumption,  requiring that all users {must} 
be served for any request {configuration}
is too constraining. 
Therefore, we introduce the possibility of outages, i.e.,  that some requests are not served, {because of some network admission 
control policy (to be discussed in details later on).} 
For the system described in Section \ref{section: network model}, we define the throughput-outage region and obtain achievability
and converses that are sufficiently tight to characterize the throughput-outage scaling laws within a small gap of the {\em constants of the leading term}. 
Furthermore, our analysis shows very good agreement with with finite-dimensional simulation results.

In the relevant regime of small outage probability,  the throughput of the D2D one-hop caching network behaves in the same near-optimal 
way as the throughput of coded multicasting \cite{maddah2012fundamental,maddah2013decentralized}, while the system architecture is significantly 
more straightforward for a practical  implementation. 
In particular, for fixed cache size $M$,  as the number of users $n$ and the number of files $m$  become large with $nM \gg m$,  the throughput of the D2D one-hop caching  network grows linearly with $M$, and it is inversely proportional to $m$, but it is independent of $n$. 
Hence, D2D one-hop caching networks are very attractive to handle situations where a relatively small library of popular files (e.g., the 500 most popular 
movies and TV shows of the week) is requested by a large number of users (e.g., 10,000 users per km$^2$ in a typical urban environment). 
In this regime, the proposed system is able to efficiently {turn memory into bandwidth, in the sense that the per-user throughput increases proportionally to the cache capacity of the user devices.} 
We believe that this conclusion is important for the design of future wireless {systems}, 
since bandwidth is a  much more scarce and  expensive resource than storage capacity.  


{\bf Related literature:}
{The analysis of the capacity scaling laws\footnote{Scaling law order notation: 
given two functions $f$ and $g$, we say that: 1)  $f(n) = O\left(g(n)\right)$ if there exists a constant $c$ and integer $N$ such that  $f(n)\leq cg(n)$ for $n>N$. 2) $f(n)=o\left(g(n)\right)$ if $\lim_{n \rightarrow \infty}\frac{f(n)}{g(n)} = 0$. 
3) $f(n) = \Omega\left(g(n)\right)$ if $g(n) = O\left(f(n)\right)$. 4) 
$f(n) = \omega\left(g(n)\right)$ if $g(n) = o\left(f(n)\right)$. 
5) $f(n) = \Theta\left(g(n)\right)$ if $f(n) = O\left(g(n)\right)$ and $g(n) = O\left(f(n)\right)$.}
for large D2D (or ``ad-hoc'') wireless  networks has been the subject of a vast body of literature.
Gupta and Kumar  \cite{gupta2000capacity} proposed a network model where $n$ are randomly placed on some planar region and communicate through multiple hops. They characterized the transport capacity scaling as $n \rightarrow \infty$, under the same {\em protocol model} 
considered in our paper (see Section \ref{section: network model}). 
For random assignment of source-destination pairs,  \cite{gupta2000capacity} showed that the per-link capacity must vanish 
as $O(\frac{1}{\sqrt{n}})$. In addition, a multi-hop relaying scheme  was shown to achieve 
throughput scaling $\Omega(\frac{1}{\sqrt{n \log n}})$. The same results were confirmed, using a somehow simpler and more general analysis technique, 
in \cite{kulkarni2004deterministic}. 
The multicast capacity of large wireless networks has been investigated in \cite{shakkottai2010multicast, niesen2010balanced}. 
Finally, Franceschetti, Dousse, Tse and Thiran \cite{franceschetti2007closing} 
closed the $\sqrt{\log(n)}$ gap between upper bound and achievability
in \cite{gupta2000capacity,kulkarni2004deterministic} 
by creating an almost deterministically placed
grid sub-network with vertical and horizontal ``highways'' that relay messages with very short hops. 
The existence of such grid subnetwork is guaranteed with high probability by percolation theory.

Given the fact that randomly placed nodes yield the same scaling laws of 
nodes placed on a deterministic squared grid, in this work we assume a grid network from the start. 
This allows to focus on the essential aspect of caching at the nodes, and avoid the analytical complication of randomly placed nodes.} 
The same approach is taken in  \cite{gitzenis2012asymptotic},  where {\em multi-hop} D2D communication is considered under 
the protocol model for a network of nodes  placed deterministically on a squared grid.  If the aggregate distributed storage space  in the network is larger than the total size of the library, 
multi-hop guarantees that all user  requests can be served by the network.  Under the same assumption made here of the user requests distribution,  
\cite{gitzenis2012asymptotic} finds a deterministic replication  caching scheme and a multi-hop routing scheme that achieves 
order-optimal average throughput. Besides the consideration of multihop and single hop, there are several other key differences between 
our work and \cite{gitzenis2012asymptotic}. First, \cite{gitzenis2012asymptotic} considers a deterministic caching placement approach, which depends on  the files popularity distribution. This approach is not robust when users move between cells. In contrast, mobility is easily handled by our scheme 
which is based on independent random caching.  Next, in \cite{gitzenis2012asymptotic}, the transmission range is fixed, where each node can 
only reach its four neighbors.  Besides the deterministic caching placement, the key aspect of the problem is the design of the routing protocol and analyze the traffic through the {\em bottleneck link} of the network. Our work focuses on determining the transmission range within which nodes can access their neighbors 
caches in one hop. This, in turn, determines  the point of the throughput-outage tradeoff at which the system operates. 
Finally, \cite{gitzenis2012asymptotic} only gives the order of the throughput as the number of users $n$ goes to infinity,  but does not characterizes 
the multiplicative constant of the throughput leading term in the scaling law.
Therefore, it is difficult to understand in which regime of (large but finite $n$) the scaling laws become relevant. 
In passing, we notice that this is a common problem in several studies focused on scaling laws of large wireless networks.
In our case, we provide outer bounds and inner (achievable) bounds  to the throughput-outage tradeoff, which are tight enough to
characterize also the constants of the leading terms, within a bounded gap. In particular,  the analysis of our achievability 
scheme  matches well with finite-dimensional simulations.  

Preliminary work of the present paper is given in \cite{golrezaei2012wireless}, where only the sum throughput was 
considered irrespectively of user outage probability. The analysis in \cite{golrezaei2012wireless} considers 
a heuristic random caching policy, while here we find the optimal random caching distribution. 
More importantly, the total sum throughput is not a sufficient characterization 
of the performance of D2D one-hop caching networks:  in certain regimes of the number of users and file library size, 
it can be shown that  in order to achieve a large sum throughput only a small portion of the users should  be served, while leaving the majority 
of the users in outage.  In contrast, the throughput-outage tradeoff region considered here is able to capture the notion of {\em fairness}, 
since it focuses on the minimum per-user average throughput and on the fraction of users which are denied service. 

The paper is organized as follows. Section~\ref{section: network model} introduces the network model and the precise 
problem formulation of the throughput-outage tradeoff in D2D one-hop caching networks. 
The main results on the outer bound and achievability of the throughput-outage tradeoff are 
presented in Sections \ref{section: The Converse of Throughput-Outage Tradeoff} and
\ref{section: Achievable Throughput-Outage Tradeoff}, respectively. 
In Section~\ref{sec: Discussion and Conclusion} we presents some concluding remarks. 
All proofs are relegated in the Appendices, in order to maintain the flow of exposition.

\section{Network Model and Problem Formulation}
\label{section: network model}

\begin{figure}
\centering
\includegraphics[width=10cm]{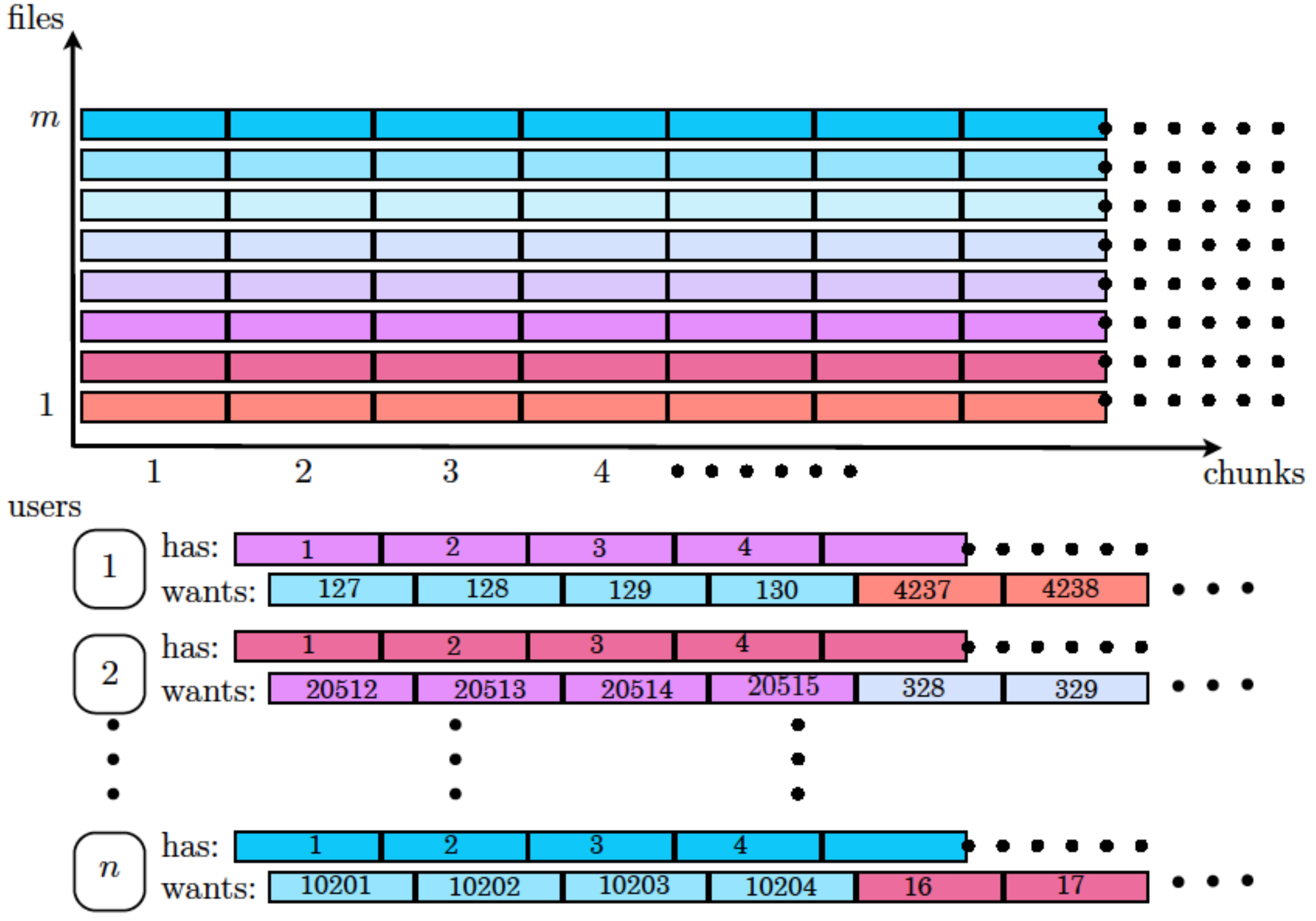} 
\caption{Qualitative representation of our system assumptions: each user caches an entire file, formed by a very large number of packets $L$.
Then, users place random requests of segments of $L'$ packets from files of the library, starting at random initial points.
In the figure, we have $L' = 4$.}
\label{library-and-asynchronous-demands}
\end{figure}

We consider a network deployed over a unit-area squared region 
and formed by $n$ nodes $\Uc = \{1, \ldots, n\}$ placed on a regular grid with minimum node distance  
$1/\sqrt{n}$ (see Fig.~\ref{fig: Grid_Network_D2D}).
Each user $u \in \Uc$ makes a request to a file $\fsf_u \in \Fc = \{1, \ldots, m\}$ in an i.i.d. manner, 
according to a given request probability mass function $\{ P_r(f) : f \in \Fc\}$.  
In order to model the {asynchronous content reuse} 
and forbid any form of ``for-free'' multicasting by ``overhearing'', we consider the following theoretical model. 
{We assume that each file is formed by a sequence of $L$ packets. Each user demand corresponds to 
a file index $f \in \Fc$ and a segment of $L' < L$ consecutive packets, starting at some initial index $\ell$, uniformly 
and independently distributed over $\{1, \ldots, L - L' + 1\}$. The packets of the requested segment are downloaded sequentially.} 
We measure the cache size in files, and in order to compute the system performance we consider first the limit for 
large file size ($L \rightarrow \infty$) with $L'$ finite, and then study the system scaling laws for $n,m \rightarrow \infty$.
Hence, the probability that users request overlapping segments vanishes for $L \rightarrow \infty$ for any finite $n,m,L'$, thus preventing
the trivial use of naive multicasting (i.e., overhearing common messages). 
In contrast,  the probability that two users request segments of the {\em same file} depends on the library 
size $m$ and on the request distribution $P_r$. 
We hasten to say that this model is just a way to {express in precise mathematical terms} the notion of 
{\em asynchronous content reuse}, 
such that the overlap of the demands and the overlap of concurrent transmissions are decoupled.\footnote{As a side note, we observe also that the segmentation of large files into smaller packets (or ``chunks'') to be sequentially downloaded is consistent with current video streaming protocols such as DASH \cite{sanchez2011improved, sanchez2011idash,pantos2012http,stockhammer2011dynamic}.}  
Fig.~\ref{library-and-asynchronous-demands} shows qualitatively our model assumptions. 

In our system, D2D communication obeys the following {\em protocol model} \cite{gupta2000capacity}:
a node $u$ can receive successfully a packet from node $v$ if $d(u,v) \leq R$ and if no other node $v'$ at distance
$d(u,v') < (1+\Delta) R$ is transmitting.  
{The transmission range $R$ is a} design parameter that can be set as a function of $m$ and $n$. 
We consider the following definitions:
 
\begin{figure}
\centering \includegraphics[width=10cm]{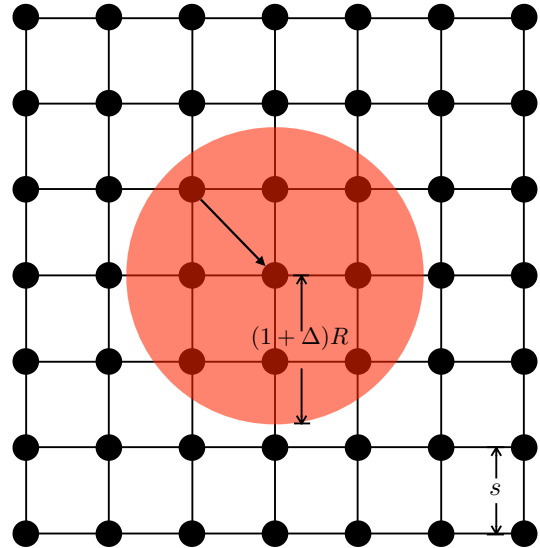}
\caption{Grid network with $n=49$ nodes (black circles) with minimum separation $s = \frac{1}{\sqrt{n}}$. 
The red area is the disk where the protocol model imposes no other concurrent transmission. 
$R$ is the case transmission range and $\Delta$ is the interference parameter, such that the forbidden disk around
the receiver has radius $(1+\Delta)R$.} 
\label{fig: Grid_Network_D2D}
\end{figure} 
 
\begin{defn} {\bf (Network)} A network if formed by a set of user nodes $\Uc$ 
and a set of files $\Fc = \{1, \ldots, m\}$. 
Nodes in $\Uc$ are placed in the two-dimensional unit-square region, 
and their transmissions obeys the protocol model. 
In general, all $n (n-1)$ directed links between  user nodes, 
subject to the protocol model, define an interference (conflict) graph. 
Only the links in an independent set in the interference graph can be  active simultaneously.
\hfill $\lozenge$
\end{defn}
 
\begin{defn}
{\bf (Cache placement)} The cache placement $\Pi_c$ is a rule to assign files from the library $\Fc$ 
to the user nodes $\Uc$ with ``replacement'' (i.e., with possible replication). 
Let $\Gsf = \{\Uc, \Fc, \Ec\}$ be a bipartite graph with 
``left'' nodes $\Uc$, ``right'' nodes $\Fc$ and edges $\Ec$ such that $(u,f) \in \Ec$ indicates that file 
$f$ is assigned to the cache of user $u$.
A bi-partite cache placement graph $\Gsf$ is feasible if the degree of each user node is 
not larger than the maximum user cache size equal to $M$ files. 
Let $\Gc$ denote the set of all feasible bi-partite graphs $\Gsf$. Then,  $\Pi_c$ is a probability mass function over $\Gc$,  i.e., a particular 
cache placement $\Gsf \in \Gc$ is assigned with probability $\Pi_c(\Gsf)$. 
\hfill $\lozenge$
\end{defn}
  
Notice that deterministic cache placements are special cases, corresponding to  a single probability mass equal to 1 on the desired assignment $\Gsf$. 
In contrast, we will be interested in ``decentralized'' random caching placements 
constructed as follows: each user node $u \in \Uc$ selects its cache content in an i.i.d. manner, 
by independently generating  at random $M$ indices $f_{u,1}, \ldots, f_{u,M}$ according to the same caching 
probability mass  function $\{P_c(f) : f \in \Fc \}$. 

\begin{defn} {\bf (Random requests)} At each request time (integer multiples of $L'$), 
each user $u \in \Uc$ makes a request to a segment of length $L'$ of chunks from file $\fsf_u \in \Fc$, 
selected independently with probability $P_r$. The vector of current requests $\fsf$ is 
a random vector taking on values in $\Fc^n$, with product joint probability mass function 
$\PP(\fsf = (f_{1}, \ldots, f_{n})) = \prod_{i=1}^n P_r(f_{i})$.
\hfill $\lozenge$
\end{defn} 

In this paper, we assume that $P_r$ is a Zipf distribution with 
parameter $0 < {\gamma} < 1$ \cite{breslau1999web}, i.e.,  $P_r(f) = \frac{f^{-{\gamma}}}{H({\gamma},1,m)}$ for $f = 1,\ldots, m$,  
and $H(\gamma,x,y) \eqdef \sum_{i=x}^y \frac{1}{i^\gamma}$.

\begin{defn} 
{\bf (Transmission policy)} The transmission policy $\Pi_t$ is a rule to activate the D2D links 
in the network. Let $\Lc$ denote the set of all directed links.  
Let $\Ac \subseteq 2^\Lc$ denote the set of all feasible subsets of links 
(this is a subset of the power set of $\Lc$, formed by all independent sets in the 
network interference graph).  Let $\Asf \subset \Ac$ denote a feasible set of simultaneously active links 
according to the protocol model. 
Then, $\Pi_t$ is a conditional probability mass function over $\Ac$ given $\fsf$ (requests) and 
$\Gsf$ (cache placement),  assigning probability $\Pi_t(\Asf |\fsf, \Gsf)$ to $\Asf \in \Ac$. \hfill $\lozenge$
\end{defn}

{We may think of $\Pi_t$ as a way of scheduling simultaneously compatible sets of links  (subject to the protocol model). 
Modeling the scheduling policy in a probabilistic manner allows the analytical convenience of 
defining the average per-user throughput (see below) as an ensemble average. As a matter of fact, 
deterministic link activation rules can be included by defining the average throughput as a time-average. For example, a bounded 
deterministic delay per user can be guaranteed by activating groups of compatible links (forming a maximal independent set in the network interference graph) 
in a deterministic round-robin sequence, such that each user is served with a deterministic delay.}

\begin{defn}
\label{definition: throughput} 
{\bf (Useful received bits per slot)} 
For given $P_r$, $\Pi_c$ and $\Pi_t$, and user $u  \in \Uc$, we define 
the random variable $T_u$ as the number of useful received information bits per slot 
unit time by user $u$ at a given scheduling time (irrelevant because of 
stationarity). This is given by 
\begin{equation}  \label{useful-throughput-i}
T_u = \sum_{v : (u,v) \in \Asf} c_{u,v} 1\{ \fsf_u \in  \Gsf(v) \}
\end{equation}
where $\fsf_u$ denotes the file requested by user node $u$,  $c_{u,v}$ denotes the rate of the link 
$(u,v)$,  and $\Gsf(v)$ denotes the content of the cache of node $v$, i.e., the neighborhood of node $v$
in  the cache placement graph $\Gsf$. 
\hfill $\lozenge$
\end{defn}

Consistently with the protocol model, $c_{u,v}$ depends only on the active link $(u,v) \in \Asf$ and not on the whole set of active 
links $\Asf$. Furthermore, we shall obtain our results under the simplifying assumption 
(usually made under the protocol model \cite{gupta2000capacity}) that $c_{u,v} = C$ for all $(u,v) \in \Asf$. 
The indicator function $1\{ \fsf_u \in \Gsf(v)\}$ expresses the fact that only the bits relative to the file 
$\fsf_u$ requested by user $u$ are ``useful'' and count towards the throughput. 
Obviously, scheduling links $(u,v)$ for which $\fsf_u \notin \Gsf(v)$  is useless for the sake of the throughput defined above. 
Hence, we can restrict our transmission policies to those activating only links $(u,v)$ for which $\fsf_u \in \Gsf(v)$. 
These links are referred to as ``potential links'', i.e., links potentially carrying useful data. 
Potential links included in $\Asf$  are ``active links'',  at the given scheduling slot. 

The average throughput for user $u \in \Uc$ is given by $\overline{T}_u = \EE[T_u]$, where expectation  is with respect 
to the random triple  $(\fsf, \Gsf, \Asf) \sim \prod_{i=1}^n P_r(f_i) \Pi_c(\Gsf) \Pi_t(\Asf|\fsf,\Gsf)$.
We say that user $u$ is in outage if $\EE[T_u | \fsf, \Gsf] = 0$. 
This condition captures the event that no link $(u,v)$ with $\fsf_u \in \Gsf(v)$ is scheduled with positive probability, 
for given requests vector $\fsf$ and cache placement $\Gsf$. In other words, a user $u$ for which $\EE[T_u | \fsf, \Gsf] = 0$ 
experiences  a ``long'' lack of service (zero rate), as far as the cache placement is $\Gsf$ and the request vector is $\fsf$. 

\begin{defn}
{\bf (Number of nodes in outage)}  The number of  nodes in outage is given by 
\begin{equation} \label{outage-nodes}
\Nsf_o = \sum_{u \in \Uc} 1\{ \EE[T_u |\fsf, \Gsf]  = 0\}. 
\end{equation}
Notice that $\Nsf_o$ is a random variable, function of $\fsf$ and $\Gsf$. 
\hfill $\lozenge$
\end{defn}

\begin{defn}
\label{definition: outage}
{\bf (Average outage probability)}  The average (across the users) outage probability is given by 
\begin{equation} \label{outage-probability}
p_o = \frac{1}{n} \EE[ \Nsf_o ] = \frac{1}{n} \sum_{u \in \Uc} \PP\left ( \EE[T_u |\fsf, \Gsf]  = 0 \right ). 
\end{equation}
\hfill $\lozenge$
\end{defn}

In this work we focus on max-min fairness, i.e., we express the throughput-outage tradeoff 
in terms of the minimum  average user throughput, defined as
\begin{equation}
\overline{T}_{\rm min} = \min_{u \in \Uc} \; \left \{ \overline{T}_u  \right \}. 
\end{equation}
Notice that that the max-min fairness criterion in our setting is {\em essential} to make 
the outage probability $p_o$ defined in (\ref{outage-probability}) meaningful. In fact, for $0 \leq p_o' < p_o \leq 1$, consider a system that 
achieves outage probability $p_o$ by serving only a fraction $1 - \lambda$ of users with outage probability 
$p_o' = \frac{p_o - \lambda}{1 - \lambda}$, while leaving the remaining fraction $\lambda$ of users permanently idle.  
In this case, we have $\overline{T}_{\min} = 0$ since there are $\lambda n > 0$ users with identically zero throughput. 
Hence, a system that permanently excludes some users in favor of others is certainly not optimal in terms of the 
throughput-outage tradeoff as defined below:

\begin{defn} {\bf (Throughput-Outage Tradeoff)}  
For a given network and request probability distribution $P_r$,  an throughput-outage pair $(T,p)$ is {\em achievable} 
if there exists a cache placement $\Pi_c$ and a transmission policy $\Pi_t$ with outage probability
$p_o \leq p$ and minimum per-user average throughput $\overline{T}_{\min} \geq T$. The throughput-outage achievable region
$\Tc$ is the closure of all achievable throughput-outage pairs $(T,p)$. 
In particular, we let $T^*(p) = \sup \{ T : (T,p) \in \Tc \}$. 
\hfill $\lozenge$
\end{defn}

Notice that $T^*(p)$ is the result of the following optimization problem:
\begin{eqnarray} \label{sucaminchia}
\mbox{maximize} & & \overline{T}_{\min} \nonumber \\
\mbox{subject to} & & p_o \leq p, 
\end{eqnarray}
where the maximization is with respect to the cache placement and transmission policies 
$\Pi_c, \Pi_t$.  Hence, it is immediate to see that $T^*(p)$ is non-decreasing in $p$. 
The range of feasible outage probability, in general, is an interval $[p_{o, \min}, 1]$ for some $p_{o,\min} \geq 0$. 
We say that an achievable point $(T,p)$ dominates an achievable point $(T',p')$  if $p \leq p'$ and $T \geq T'$. 
The Pareto boundary of $\Tc$ consists of all achievable points that  are not dominated by other achievable points, i.e., it is given by $\{(T^*(p),p)  : p \in [p_{o, \min}, 1]\}$. 

It is also immediate to see that the throughput-outage tradeoff region is convex. In fact, consider two achievable points
$(\overline{T}^{(1)}_{\min}, p_o^{(1)})$ and $(\overline{T}^{(2)}_{\min}, p_o^{(2)})$ corresponding to 
caching placements $\Gsf_1$ and $\Gsf_2$, with probability assignments $\Pi_c^{(1)}$ and $\Pi_c^{(2)}$. 
For $\lambda \in [0,1]$, the caching placement $\Gsf$ with mixture probability assignment 
$\Pi_c = \lambda \Pi_c^{(1)} + (1 - \lambda) \Pi_c^{(2)}$ achieves $p_o = \lambda p_o^{(1)} + (1 - \lambda) p_o^{(2)}$. 
For this value of outage probability, the best possible strategy achieves
\[ T^*(p_o) \geq \min_u \left \{ \lambda \overline{T}_u^{(1)} + (1 - \lambda) \overline{T}_u^{(2)} \right \} \geq 
\lambda \min_u \overline{T}_u^{(1)} + (1 - \lambda) \min_u \overline{T}_u^{(2)},  \]
where, by definition, $\overline{T}_{\min}^{(1)} = \min_u \overline{T}_u^{(1)}$ and 
$\overline{T}_{\min}^{(2)} = \min_u \overline{T}_u^{(2)}$. Hence, the segment joining any two achievable 
throughput-outage points $(\overline{T}^{(1)}_{\min}, p_o^{(1)})$ and $(\overline{T}^{(2)}_{\min}, p_o^{(2)})$ is contained into
the achievable throughput-outage region.

\begin{figure}
\centering
\subfigure[]{
\centering \includegraphics[width=5cm]{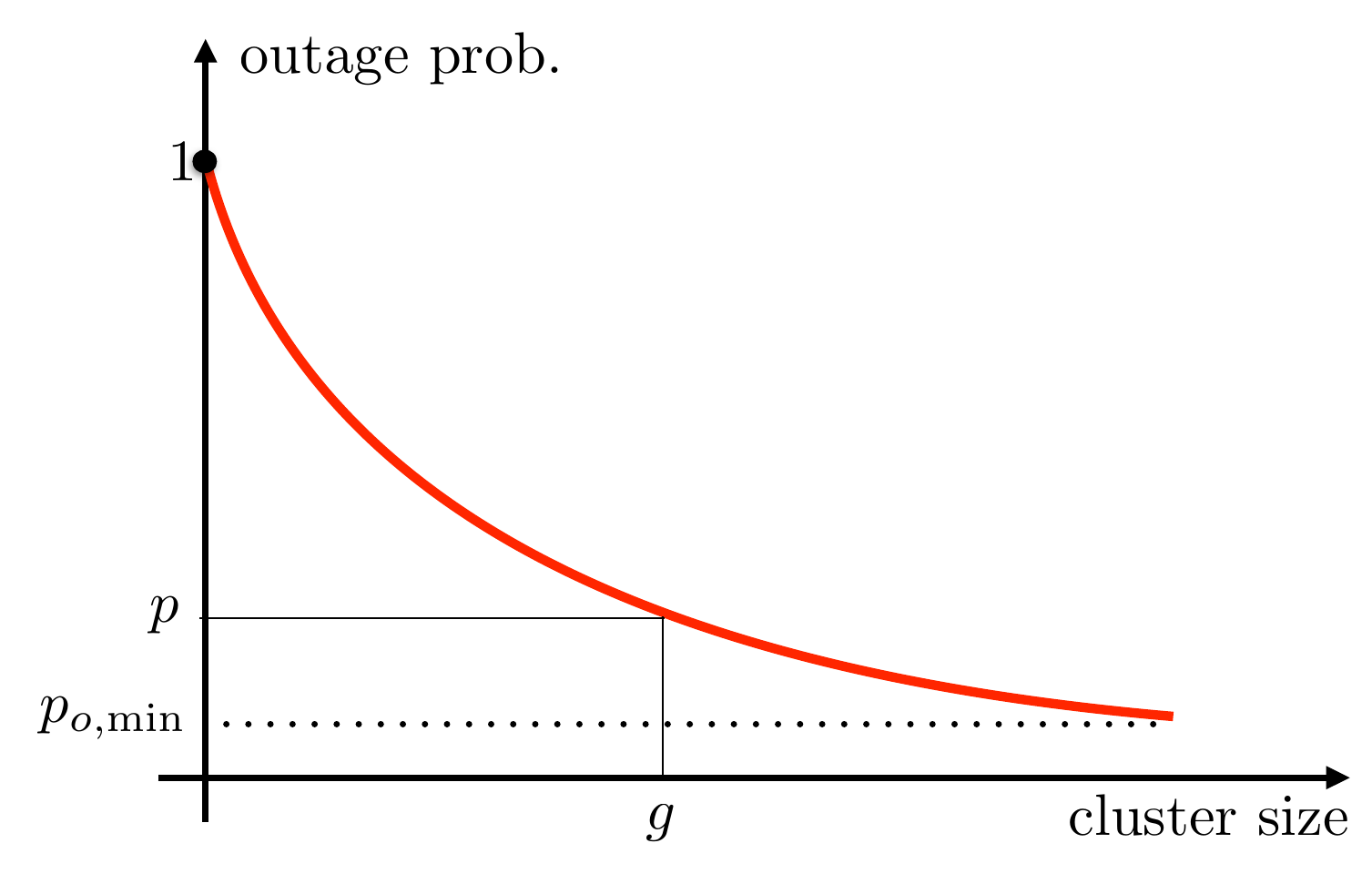}
\label{intuitive-1}
}
\subfigure[]{
\centering \includegraphics[width=5cm]{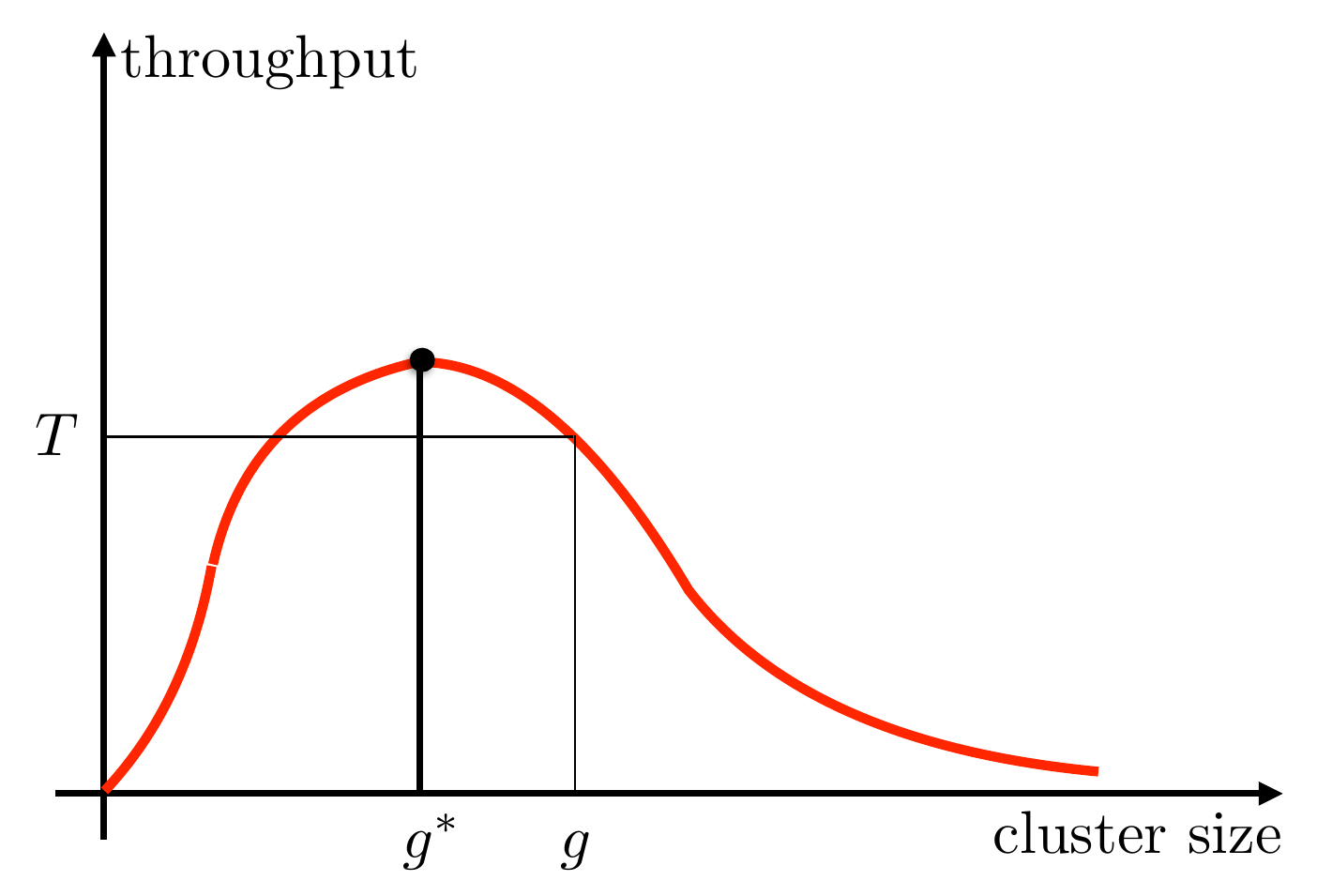}
\label{intuitive-2}
}
\subfigure[]{
\centering \includegraphics[width=5cm]{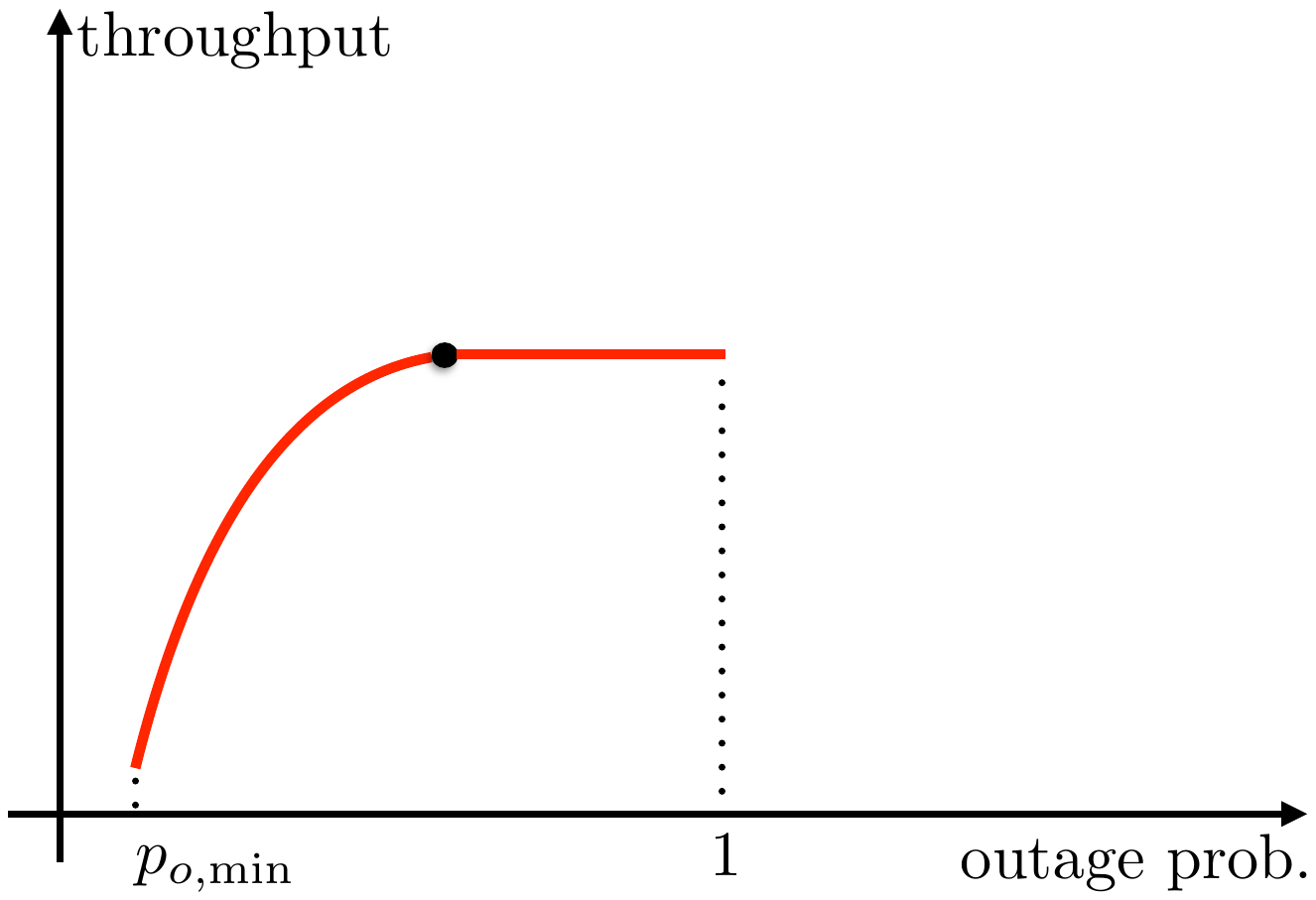}
\label{intuitive-3}
}
\caption{Qualitative behavior of the tradeoff between throughput and outage probability, by ways of the tradeoff parameter $g$, which
represents the size of the cluster of nodes over which any node can look for its desired requested file and download it by D2D on-hop communication.}
\end{figure}

{We conclude this section by providing the intuition behind the tension between outage and throughput, and explaining through 
an intuitive argument  why $T^*(p)$ is non-decreasing for $p \in [p_{o,\min}, 1]$. 
The key tradeoff quantity here is the cooperation cluster size $g$, that is, the size of the set of nearest neighbor nodes among which any node $u$ can look for
its desired file $f_u$. On one hand, we would like to have $g$ large, in order to take advantage of the content reuse, i.e., the larger $g$, the larger the probability that any user can find and retrieve its desired file. On the other hand, we would like to have $g$ small, in order to take advantage of the spatial reuse, i.e., 
the smaller $g$, the larger the number of simultaneously active links that the network can support. 
Therefore, $g$ describes the tradeoff between content reuse and spatial reuse. 

As $g$ increases the probability that user $u$ does not find its desired file decreases.  
Hence, $p_o$ is a decreasing function of $g$ (see Fig.\ref{intuitive-1}). 

Since nodes can retrieve their desired files within a cluster of size $g$, 
then the communication range of the D2D links must be enough to communicate across such clusters. The average number of active links that can be activated
without violating the protocol model is of the order of the number of disjoint clusters in the network, i.e., $\frac{n}{g}$. 
With a probability $1 - p_o$ that any user cannot find its requested file, the average per-user throughput is roughly given by 
$T \propto \frac{1 - p_o}{n} \times \frac{n}{g} = \frac{1 - p_o}{g}$. Since for small $g$ we have $p_o \uparrow 1$, and for large $g$ we have 
$p_o \downarrow p_{o,\min}$, where the latter is the probability that a node $u$ does not find its requested file in the whole network, 
we clearly see that $T$ must be increasing for small $g$ and decreasing for large $g$ (see Fig.\ref{intuitive-2}). 

Now, consider the constraint on the outage probability $p_o \leq p$. If $p$ is small, then the constraint must be satisfied with
equality, yielding the corresponding value of $g$ (Fig.\ref{intuitive-1}) which in turns yields a corresponding value of $T$ 
(Fig.\ref{intuitive-2}). As $p$ increases, we obtain the concave increasing part of the throughput vs outage Pareto boundary 
curve  qualitatively depicted in Fig. \ref{intuitive-3}. However, when $p$ becomes larger than some threshold, the optimal
throughput is obtained by letting $g = g^*$, which is the size that achieves the maximum unconstrained throughput (see Fig. \ref{intuitive-2}). 
This means that for values of $p$ beyond this threshold value, the throughput curve reaches a bound (horizontal line), equal to the unconstrained 
maximum throughput, as shown in Fig. \ref{intuitive-3}. 

It follows from the upper bounds developed in Section \ref{section: The Converse of Throughput-Outage Tradeoff} that 
this intuitive argument, even though it is developed for a cluster-based achievability strategy, holds true also
for the upper bounds, despite the fact that the latter do not assume any a priori transmission strategy other than the 
one-hop constraint and the protocol model.}

\section{Outer Bounds}
\label{section: The Converse of Throughput-Outage Tradeoff}

Under the one-hop restriction, network topology and protocol model given in Section~\ref{section: network model}, 
we can provide an outer bound $T^{\rm ub}(p)$ on the throughput-outage tradeoff, such that the ensemble of points 
$\{(T^{\rm ub}(p), p) :  p \in [0,1]\}$ dominates the tradeoff region $\Tc$.  In what follows, the quantity
\begin{equation} \label{alpha} 
\alpha \eqdef \frac{1- {\gamma}}{2 - {\gamma}}
\end{equation}
plays an important role. Notice that $0 \leq \alpha < 1/2$ under the assumption made here that $0 < {\gamma} < 1$.  
The following results are proved in Appendices \ref{sec: the proof of theorem 1 and 2} and \ref{sec: the proof of theorem 3}:

\begin{theorem}
\label{theorem: 1}
When $\lim_{n \rightarrow \infty} \frac{m^\alpha}{n} = 0$, the throughput-outage region is dominated by the set of points 
$(T^{\rm ub}(p),p)$ given by :
\begin{align}
&T^{\rm ub}(p) = \notag\\
&\left\{\begin{array}{ll} 
\frac{16CM}{\Delta^2 m (1-p)^{\frac{1}{1-{\gamma}}}} + o\left(\frac{1}{m(1-p)^{\frac{1}{1-{\gamma}}}} \right), & p = 1 - \left(\frac{M g_R(m)}{m}\right)^{1-{\gamma}} \\
\min\left\{ \frac{16CM}{\Delta^2 m (1-p)^{\frac{1}{1-{\gamma}}}},  f_{\rm ub}(\rho') m^{-\alpha} \right\} + o\left(m^{-\alpha} \right), & 1 - {(M\rho')}^{1-{\gamma}} m^{-\alpha} \leq 
p < 1 - {(M \rho^*)}^{1-{\gamma}} m^{-\alpha} \\
f_{\rm ub}(\rho^*) m^{-\alpha} + o\left(m^{-\alpha} \right), & 1 - {(M\rho^*)}^{1-{\gamma}} m^{-\alpha} \leq p \leq 1, \\
\end{array} \right. \label{ub-th1}
\end{align}
where $\rho' \geq \rho^*$ and $\rho^*$ is the solution (with respect to $\rho$) of the equation 
\begin{equation} \label{eq: theorem 1}
\zeta(\rho) = \log(1 + (2 - {\gamma}) \zeta(\rho)) 
\end{equation}
with 
\begin{equation} \label{upsilon}
\zeta(\rho) \eqdef \left(\left(1+\frac{3\Delta}{2}\right)^\frac{2}{2-{\gamma}} \rho \right)^{2-{\gamma}}M^{1-{\gamma}}, 
\end{equation} 
$g_R(m)$ is any function such that $g_R(m) = \omega\left(m^\alpha \right)$  and $g_R(m) \leq \min\{\frac{m}{M}, n\}$, 
and where
\begin{equation} \label{f1rho} 
f_{\rm ub}(\rho) \eqdef \frac{16C}{\Delta^2\rho}\left(1- e^{-\zeta(\rho)} \right ). 
\end{equation}
\hfill  $\square$
\end{theorem}

\begin{theorem}
\label{theorem: 2}
When there exists a positive constant $\xi$ such that $\xi \leq \lim_{n \rightarrow \infty}  \frac{m^\alpha}{n} \leq \frac{16}{\Delta^2\rho^*}$, 
the throughput-outage region is dominated by the set of points  $(T^{\rm ub}(p),p)$ given by:
\begin{align}
&T^{\rm ub}(p) = \notag\\
&\left\{\begin{array}{ll}\min\left\{ \frac{16CM}{\Delta^2m(1-p)^{\frac{1}{1-{\gamma}}}}, f_{\rm ub}(\rho')m^{-\alpha}\right\} + o(m^{-\alpha}), & 
1 - {(M\rho')}^{1-{\gamma}}m^{-\alpha} \leq p < 1 - {(M \rho^*)}^{1-{\gamma}}m^{-\alpha} \\
f_{\rm ub}(\rho^*)m^{-\alpha} + o(m^{-\alpha}), & 1 - {(M\rho^*)}^{1-{\gamma}}m^{-\alpha} \leq p \leq 1, 
\end{array} \right.
\end{align}
where $\rho^*$ is the solution of (\ref{eq: theorem 1}),  and $\rho'  \in [\rho^*, \frac{16}{\Delta^2}\frac{n}{m^{\alpha}}]$. 
\hfill  $\square$
\end{theorem}

\begin{theorem}
\label{theorem: 3}
When $\lim_{n \rightarrow \infty} \frac{m^\alpha}{n} >  \frac{16}{\Delta^2\rho^*}$ ($\rho^*$ being the solution of (\ref{eq: theorem 1})), 
the throughput-outage region is dominated by the set of points  $(T^{\rm ub}(p),p)$ given by :
\begin{equation}
T^{\rm ub}(p) = C \left ( \frac{M n}{m} \right )^{1-{\gamma}} + o\left (\left ( \frac{n}{m} \right )^{1-{\gamma}} \right ), \;\;\;  1 - \left ( \frac{M n}{m} \right )^{1-{\gamma}}  \leq p \leq 1.
\end{equation}
\hfill  $\square$
\end{theorem}

{Notice that the range of $p$ in Theorems \ref{theorem: 2} and \ref{theorem: 3} is limited to $[p_{o,\min},1]$ with
$p_{o,\min} = 1 - {(M\rho')}^{1-{\gamma}}m^{-\alpha}$ (for Theorem \ref{theorem: 2}) and 
$p_{o,\min} = 1 - \left ( \frac{M n}{m} \right )^{1-{\gamma}}$ (for Theorem  \ref{theorem: 3}), showing that in these regimes 
the outage probability cannot be small. As a matter of fact,} of all the regimes identified by Theorems \ref{theorem: 1}, \ref{theorem: 2} and \ref{theorem: 3}, 
the only {\em practically interesting} one is the first regime of Theorem \ref{theorem: 1}. 
In particular, Theorem \ref{theorem: 2} and \ref{theorem: 3} show that, when 
$\lim_{n \rightarrow \infty} \frac{m^{\alpha}}{n}$ is bounded away from zero, 
any scheme for the one-hop D2D caching network yields outage probability that goes to 1, which is clearly not an acceptable.  In contrast, 
$\lim_{n \rightarrow \infty} \frac{m}{n} = \kappa < \infty$, there might exist schemes  achieving some fixed target outage probability value
$p \in [0,1)$, as $n,m \rightarrow \infty$. 
{Intuitively, the function $g_R(m)$ plays the role of the size of the cooperation cluster of neighboring nodes within which
each user can find its requested file.} For example, choosing $g_R(m) = \beta m$ for some constant  
$\beta \leq \min \left \{\frac{1}{M}, \frac{1}{\kappa} \right \}$, both conditions $g_R(m) = \omega(m^\alpha)$
and $g_R(m) \leq \min \left \{\frac{m}{M},n \right \}$ are satisfied, for all sufficiently large $n$. 
Notice that for $\kappa \leq M$, the choice $\beta = 1/M$ yields that the outer bound contains points 
of the type $(O(M/m), 0)$ (zero outage probability). We shall see in the next section that throughput-outage points 
with throughput $\Omega(M/m)$ and fixed $p$ bounded away from 1 are achievable. 

For a conventional unicast system where users are served by a single omniscient node (e.g., a base station) that can store the whole library, 
the throughput scaling is $O(1/n)$.\footnote{This is obviously achieved by TDMA, serving users on different time slots in a round-robin fashion. Notice that even if a more refined physical layer including a Gaussian broadcast channel is considered, 
the throughput scaling  remains the same.} 
Hence, in the case of $nM \gg m$, the combination of caching and D2D spatial reuse yields a very large throughput 
relative gain with respect to a conventional system.  It is also interesting to notice that despite the first regime 
of Theorem  \ref{theorem: 1}  requires that $m$ grows more slowly than $n^{1/\alpha}$,  the only practically interesting 
sub-regime is $m = o(n)$, otherwise conventional unicast from the base station yields better throughput 
scaling and zero outage probability (all users are served). 

All other regimes in Theorems \ref{theorem: 1} -- \ref{theorem: 3} are included for completeness, 
in order to prove mathematically a somehow intuitively expected ``negative'' result: 
unless the library  size $m$ is small with respect to the aggregate caching memory $nM$, caching cannot 
achieve significant throughput gains with respect  to conventional unicast from a single base station.  
This result is expected since, in this case,  the {\em asynchronous content reuse} that the D2D caching network tries 
to exploit is essentially non-existent. 

It is also important to notice that here we are considering  the (realistic) case of a ``heavy tail'' Zipf request distribution with ${\gamma} \in (0,1)$. 
If ${\gamma} > 1$, then a finite number of files collects essentially all the request probability mass,  
and this case is similar to the case of $m = O(1)$, which is a special  case of $m = o(n)$.   
As a matter of fact, Zipf-distributed  requests with ${\gamma} \in (0,1)$ have been observed experimentally \cite{Zink,zink2009,cachingrefs2}. 

In the next section, we show that the upper bounds obtained here are tight in the scaling laws, 
and that the constants of the leading terms can be determined within constant gaps. 
This is obtained by exhibiting and analyzing a specific achievability strategy.

\section{Achievable Throughput-Outage Tradeoff}
\label{section: Achievable Throughput-Outage Tradeoff}

Consistently with the outer bounds in Theorems \ref{theorem: 1} -- \ref{theorem: 3} and the concluding {remarks} 
in 
Section \ref{section: The Converse of Throughput-Outage Tradeoff}, we consider achievability 
only in the ``small library'' regime $\lim_{n \rightarrow \infty} \frac{m^{\alpha}}{n} = 0$, for which there is hope to 
achieve some target outage probability strictly less than 1. We obtain an achievable inner bound on the achievable throughput-outage 
tradeoff region by considering a transmission policy based on {\em clustering}  and a caching policy based on {\em independent random caching}. 

\begin{figure}
\centering \includegraphics[width=10cm]{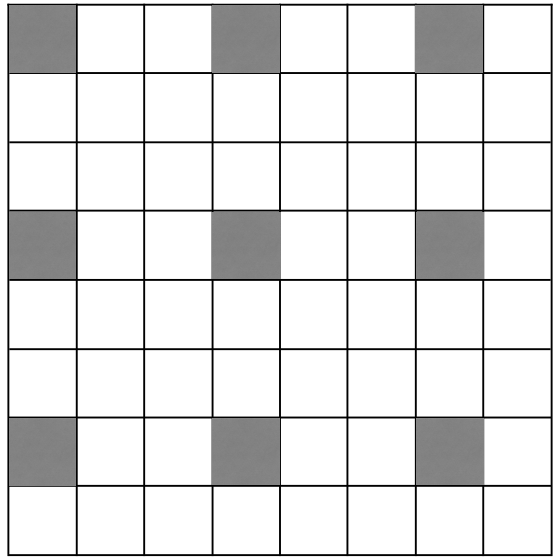}
\caption{
Example of TDMA reuse scheme: each square represents a cooperation cluster.  Gray squares represent the concurrently active 
clusters in a given time slot. In other times slots, other patterns of concurrently active clusters obtained by shifting the 
pattern in the figure are activated. In this particular example, the TDMA reuse parameter is $K=9$.}
\label{fig: Grid_TDMA}
\end{figure}

{\bf Clustering:}  the network is divided into clusters of equal size, denoted by $g_c(m)$, and 
independent of the users' demands and  cache placement realizations.
A user can only look for the requested file inside its own cluster.  For each user whose demand is found inside its cluster, 
we say that a \emph{potential link} exists in the cluster.   If a cluster contains at least one potential link, we say that this cluster is \emph{good}. 
We use an {\em interference avoidance} transmission policy $\Pi^*_t$ for which at most one concurrent transmission 
is allowed in each cluster, over any time-frequency slot (transmission resource).  
Furthermore, potential links inside the same cluster are scheduled with equal probability 
(or, equivalently, in round robin), such that all users have the same throughput $\overline{T}_u = \overline{T}_{\min}$. 
To avoid interference between clusters, we use a {conventional TDMA with spatial 
reuse scheme \cite[Ch. 17]{molisch2011wireless}, very similar to the spatial reuse scheme of a cellular network, where 
each cluster acts as a cell.  In short, a ``coloring'' scheme with $K$ colors is applied to the clusters such that 
clusters with the same color can be concurrently active on the same time slot, without violating the protocol model. The resulting groups of
clusters are assigned to $K$ orthogonal time slots, and are activated in a round-robin fashion. 
In particular, we use $K = \left(\left\lceil\sqrt{2}(1+\Delta)\right\rceil+1\right)^2$ in order to guarantee that concurrently active clusters 
do not interference with each other. Fig.~\ref{fig: Grid_TDMA} shows an example for $K = 9$.}


{\bf Random Caching:} we consider a caching policy $\Pi^*_c$ where each node independently caches $M$ files according 
to a common probability distribution $P^*_c$, given by the following result (proved in Appendix \ref{sec: the proof of theorem 4}):

\begin{theorem}
\label{theorem: optimal caching distribution}
Under the system model assumptions and the clustering scheme described above, the 
caching distribution $P_c^*$  that maximizes the probability that any user $u \in \Uc$ finds its requested file inside its corresponding  
cluster is given by
\begin{equation}
P_c^*(f) = \left[1 - \frac{\nu}{z_{f}}\right]^+,  \;\;\; f = 1,\ldots, m,
\end{equation}
where $\nu = \frac{m^*-1}{\sum_{j=1}^{m^*} \frac{1}{z_{j}}}$, 
$z_j = P_r(j)^{\frac{1}{M(g_c(m) - 1)-1}}$,  and $m^* = \Theta \left (\min \{\frac{M}{{\gamma}}g_c(m), m\}\right )$.
\hfill  $\square$
\end{theorem}

The following theorem (proved in Appendices \ref{sec: the proof of theorem 5 and 6}) yields an inner bound on the 
achievable outage-throughput tradeoff region:

\begin{theorem} \label{theorem: 4}
Assume $\lim_{n \rightarrow \infty} \frac{m^{\alpha}}{n} = 0$. Then,  the throughput-outage tradeoff achievable 
by random caching and  clustering behaves as:
\begin{align}
\label{eq: theorem 4}
&T(p) = \left\{\begin{array}{ll}
\frac{C}{K}\frac{M}{\rho_1 m} +  o(1/m), &  p = (1-{\gamma}) e^{{\gamma} - \rho_1} \\
\frac{CA}{K} \frac{M}{m (1-p)^{\frac{1}{1-{\gamma}}}} + o\left(\frac{1}{m(1-p)^{\frac{1}{1-{\gamma}}}} \right),  & p = 1 - a \left(\frac{g_c(m)}{m}\right)^{1-{\gamma}} \\
\frac{CB}{K } m^{-\alpha} + o\left(m^{-\alpha}\right),  & 1 -  a \rho_2^{1-{\gamma}} m^{-\alpha} \leq  p \leq 1 - a b^{1 - {\gamma}} m^{-\alpha}  \\
\frac{CD}{K} m^{-\alpha} + o\left(m^{-\alpha}\right),  & 1 - a b^{1 - {\gamma}}  m^{-\alpha} \leq p \leq 1,
\end{array}\right.
\end{align}
where we define $a = {\gamma}^{{\gamma}}M^{1-{\gamma}}$, $b = \left ( \frac{1 - {\gamma}}{a} \right )^{\frac{1}{2 - {\gamma}}}$, 
$A \eqdef {{\gamma}}^{\frac{{\gamma}}{1-{\gamma}}}$,
$B \eqdef  \frac{a \rho_2^{1-{\gamma}}}{1+ a \rho_2^{2-{\gamma}}}$, 
$D \eqdef  \frac{ab^{1-{\gamma}}}{1+a b^{2-{\gamma}}}$ and where $\rho_1$ and $\rho_2$ 
are positive parameters satisfying $\rho_1 \geq {\gamma}$ and $\rho_2 \geq  b$.  The cluster size $g_c(m)$ is any function of $m$ satisfying 
$g_c(m) = \omega\left(m^{\alpha} \right)$ and $g_c(m) \leq {\gamma} m/M$. 
\hfill $\square$
\end{theorem}
In all cases, the achievable throughput scaling law both for $p$ bounded away from 1 and $p \rightarrow 1$
coincide with the outer bounds of Theorem \ref{theorem: 1}.  Therefore, these throughput 
scaling laws are tight up to some gap in the constants of the leading terms.

{In the rest of this section we compare the achievable throughput scaling law of Theorem \ref{theorem: 4}
with the outer bounds of Section \ref{section: The Converse of Throughput-Outage Tradeoff} and with the 
performance achievable by other schemes. In particular,  we focus on the interesting regime of small library 
(Theorems \ref{theorem: 1} and \ref{theorem: 4}). 
Since $\alpha < 1/2$, even in this regime the library size $m$ can grow faster than $n^2$.
However, we restrict to the practically relevant regime of $m = O(n)$ (linear or sub-linear in $n$). 
Choosing $g_c(m) = \beta m$ for some  $\beta > 0$, it is apparent from the first and second line of (\ref{eq: theorem 4}) that
$p$ strictly bounded away from 1. By fixing a small but positive target outage probability, the per-user average throughput of the D2D one-hop 
caching network with random (decentralized) caching  scales as  $T^*(p) = \Theta\left (\max \left \{\frac{1}{n}, \frac{M}{m} \right \} \right )$, 
where the scaling $\Theta\left(\frac{1}{n}\right)$ can be trivially achieved by letting the whole network to be a single cluster
(e.g., transmission radius $R = \sqrt{2}$) and serving one demand per unit time. This scaling is equivalent to conventional 
unicast from a single omniscient node which can be regarded as the state of the art of today's (single cell) 
systems, with a base station or access  point serving individual requests without exploiting the asynchronous content reuse. 
We notice that, when $nM \gg m$, the throughput of the D2D caching network achieves per-user throughput that increases 
linearly with $M$.  In this regime, caching in the user nodes and exploiting the dense spatial reuse of the D2D network 
is a very attractive approach, since storage space is much ``cheaper''   than scarce resources such as bandwidth 
or dense base station deployment (the reader will forgive this vague statement in this context).  

It is interesting to notice that our analysis is able to characterize also the constant of the leading term within 
a bounded gap. This is a fortunate fact that does not happen often for the scaling analysis of wireless network capacity (e.g., see \cite{gupta2000capacity,kulkarni2004deterministic,shakkottai2010multicast,niesen2010balanced,franceschetti2007closing}).  
For example, upper and lower bounds (Theorem \ref{theorem: 1} and \ref{theorem: 4}, respectively) 
and finite-dimensional simulation results are compared in Fig.~\ref{fig: result_1}, which shows both theoretical (solid lines) and simulated 
(dashed lines) curves of the throughput ($y$-axis) vs. outage ($x$-axis) tradeoff for different values of $\gamma$. In this simulation, the throughput is normalized by $C$, so that it is independent of the link rate. In particular, the theoretical curves show the dominant term in (\ref{eq: theorem 4}) divided by $C$. 
In these examples we used $m = 1000$, $n = 10000$, $M = 1$ and the spatial reuse factor $K = 4$. 
The Zipf parameter $\gamma$ varies from $0.1$ to $0.6$, corresponding to the curves from the left (blue) to the right (cyan).

\begin{figure}[ht]
\centerline{\includegraphics[width=10cm]{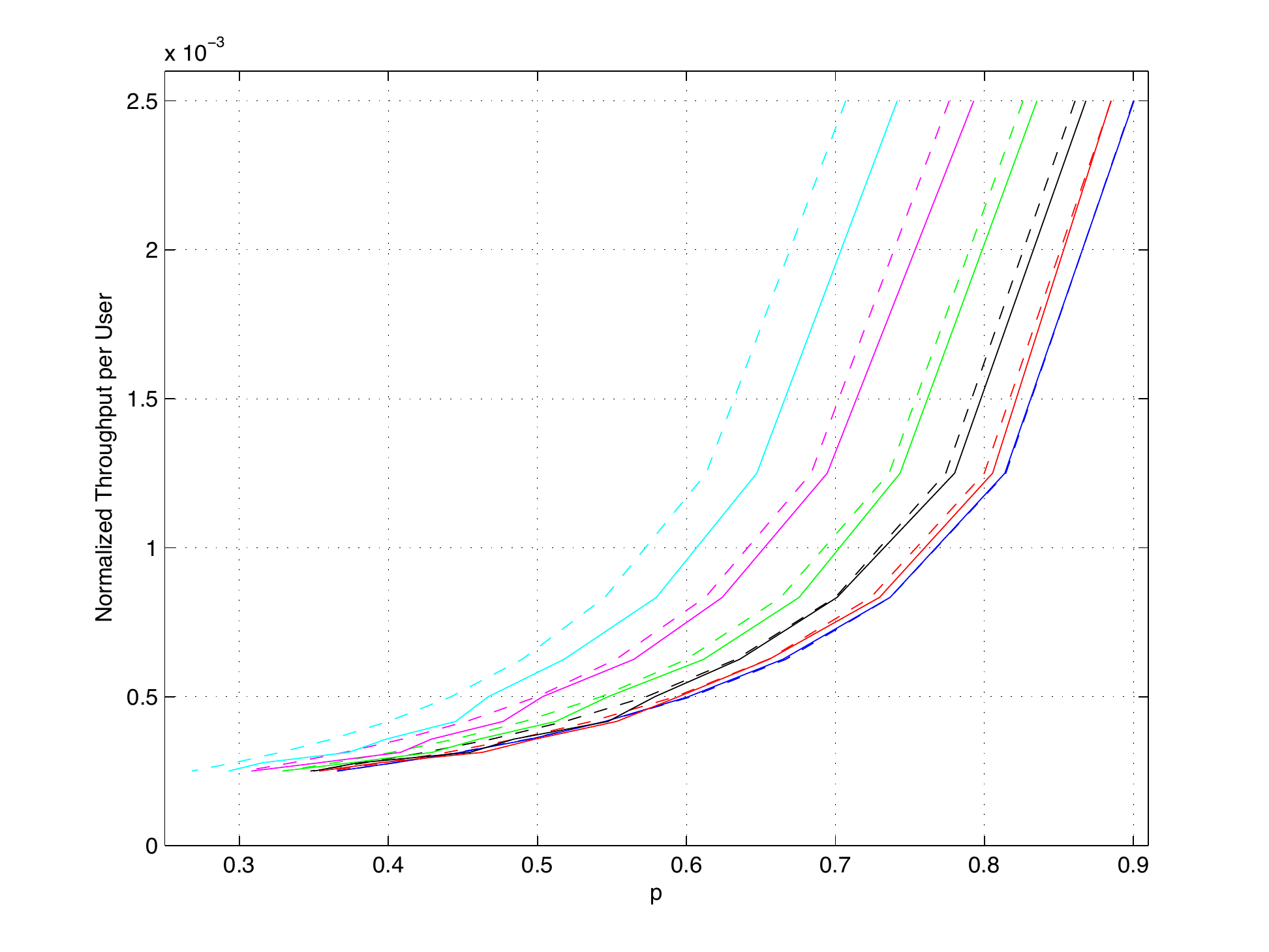}}
\caption{{Comparison between the normalized theoretical result (solid lines) and normalized simulated result (dashed lines) in terms of the minimum throughput per user vs. outage probability, where $m = 1000$, $n = 10000$, $M = 1$ and spatial reuse factor $K = 4$. The Zipf parameter $\gamma$ varies from $0.1$ to $0.6$ (from the left (blue) to the right (cyan)).}}
\label{fig: result_1}
\end{figure}

As anticipated in Section \ref{section: intro}, in order to understand the potential of the combination of D2D one-hop communication and caching
in the user devices, it is instructive to compare the scaling laws achieved by the D2D caching network with 
those achievable by other possible approaches. We have already discussed conventional unicast, achieving $\Theta(\frac{1}{n})$ throughput. 

When the number of files $m$ is less than the number of users $n$, an alternative consists of broadcasting all files on orthogonal channels. 
In order to guarantee that any requesting user can start its playback within a delay much shorter than the whole file duration, also in the presence of asynchronous requests,  a well-known approach consists of Harmonic Broadcasting \cite{juhn1997harmonic, paris1998efficient, paris1998low,engebretsen2006harmonic}.
This scheme broadcasts continuously to all users a common message formed by the $m' \leq m$ most probable files, 
each of which is encoded in the way proposed in  \cite{juhn1997harmonic}, refined in \cite{paris1998efficient, paris1998low} and whose optimality
in an information theoretic sense was established in \cite{engebretsen2006harmonic}. 
Without entering the details, each file of length $L$ packets is divided into blocks of length $L/N$ packets 
and encoded with bandwidth expansion factor $H(1,1,N) = \sum_{i=1}^N \frac{1}{i}$, such that 
the storage space at each user node is $0 < M < 1$ (less than one entire file is stored at each given time), 
and the time that any user must wait between the instant at which the streaming session is requested and the instant at which 
the playback starts is not larger than $L/N$ packets. 
In order to allow the users to start playback within a finite delay while $L \rightarrow \infty$, the ratio $L/N$ must be finite, i.e.,  
it must be $N = \Theta(L)$. Furthermore, for $m \rightarrow \infty$ and $0 < {\gamma} < 1$, in order to have outage probability 
bounded away from 1, we need $m' = \Theta(m)$.  Therefore, the bandwidth expansion factor of harmonic broadcasting in this regime is 
$m' \log N = \Theta(m \log L)$.  It follows that the throughput of Harmonic Broadcasting 
scales as  $\Theta(\frac{1}{m \log L})$.  From a strictly technical viewpoint, since in our system assumptions we study the system performance 
by first letting $L \rightarrow \infty$, and then considering $n,m$ that simultaneously grow in some relation, 
the throughput of harmonic broadcasting under our assumption is identically zero. 
In practice, for large but finite $L$, the gain of D2D caching over Harmonic Broadcasting can be appreciated by 
by comparing the term $\frac{M}{m}$ with the term $\frac{1}{m \log L}$ in the per-user throughput. 
It is clear that Harmonic Broadcasting does not take advantage of 
the user nodes storage memory, and in addition suffers an arbitrarily large  multiplicative penalty 
as the length of the files $L$ increases.
 
Finally, we examine the coded multicasting scheme of  \cite{maddah2012fundamental}, already briefly described in Section \ref{section: intro}, which
represent another example of one-hop network with caching in the user nodes, able to make efficient (and in fact, near-optimal in an information theoretic sense)
use of caching.  The rate analysis provided in \cite{maddah2012fundamental} shows that the number of equivalent file multicast transmissions 
from the base station in order to satisfy any set of users' requests is given by 
\[ N(n,m,M) =  n \left ( 1 - \frac{M}{m} \right ) \frac{1}{1 + \frac{Mn}{m}}, \]
such that the minimum average throughput per user is given by 
\[ T = \frac{C_0  \left (1 + \frac{Mn}{m} \right )}{ n \left ( 1 - \frac{M}{m} \right )},  \]
where $C_0$ is the common downlink rate at which the base station can send the multicast coded message to all users. 
For large $m,n$ and finite $M$, the scaling of the per-user throughput given again by  $\Theta\left (\max \left \{\frac{1}{n}, \frac{M}{m} \right \} \right )$, 
where the two terms  inside the max are realized depending on  whether $nM \gg m$, or $nM \ll m$. 
Interestingly and somehow surprisingly,  this is the same scaling behavior of the
D2D caching network studied in this paper. \footnote{For a performance comparison between the D2D caching network 
of the present work, coded multicasting in \cite{maddah2012fundamental}, Harmonic Broadcasting in \cite{juhn1997harmonic} 
and conventional unicasting under realistic assumptions on the underlying D2D and cellular physical channels, please see \cite{JiCaireMolisch2013}.}

}

\section{Conclusions}
\label{sec: Discussion and Conclusion}

{In this paper we have considered a wireless device-to-device (D2D) network where the nodes 
have pre-cached information from a fixed library of possible files, users request files at random and, if 
the requested file is not in the on-board  cache,  then it is downloaded from some neighboring 
node via one-hop ``local'' communication.  To model the wireless network, we have followed the simple
{\em protocol model}, widely considered in the analysis of the transport capacity scaling laws of wireless ad-hoc networks. 

We have proposed a model that captures mathematically the {\em asynchronous content reuse} typical of 
on-demand video streaming, where the users' requests have strong overlap and concentrate on a small set of popular movies, 
but the demands are completely asynchronous, such that ``naive multicasting'' is not effective.  

In our model, a user is in outage when its requested file is not found within the allowed transmission range. 
We have defined the optimal tradeoff between minimum per-user average throughput and the average fraction of 
users in outage, that we refer to as outage probability. Then, we have characterized such optimal tradeoff in terms of
tight scaling laws in all the scaling regimes of the system parameters, when both the number of nodes and the number of files 
in the library grow to infinity. 

The main result of this work is that, in the relevant regime ``small library'', i.e., when $m = O(n)$ and the aggregate cache capacity of the network, 
$nM$ is much larger that the library size $m$,  
the throughput of the D2D one-hop caching network is proportional to $M/m$ and independent of $n$. 
Hence, D2D one-hop caching networks are very attractive to handle situations where a relatively small library of 
popular files (e.g., the 500 most popular 
movies and TV shows of the week) is requested by a large number of users (e.g., 10,000 users per km$^2$ in a typical 
urban environment).  In this regime, the proposed system is able to efficiently turn memory into bandwidth, in the sense that the per-user throughput increases proportionally to the cache capacity $M$ of the user devices. Since the latter follows the doubling rate of Moore's law, caching in the user devices 
can achieve orders of magnitude throughput gains without requiring more bandwidth.

Interestingly, the same throughput scaling law is achieved by coded multicasting \cite{maddah2012fundamental,maddah2013decentralized} for a different
one-hop network topology with caching at the user nodes, where a single central node (e.g., a base station) multicast network-coded codewords
formed by EXORing data packets.  It is worthwhile to point out that, although these schemes yield the same throughput scaling law, 
they achieve their (order-optimal) ``caching gain'' according to two completely different principles.
The D2D caching network exploits the dense spatial reuse provided by caching, i.e., by replicating the same file many times in the network, 
any user with high probability can find its requested file at short distance, such that many simultaneously active links can be supported 
on the same time slot. In contrast, coded multicasting achieves its gain by using network coding in order to create a single
message which is simultaneously useful for many users. While in the D2D network transmissions should be ``as local as possible''
in order to exploit spatial reuse, in the coded multicast network transmissions should be as global as possible, in order to benefit 
the largest number of users. In a recent follow-up paper \cite{ji2014fundamental}, 
we have 
investigated a decentralized version of network-coded scheme of \cite{maddah2012fundamental} for the same
D2D network of the present work, without any omniscient node that has the whole file library. 
It turns out that coded multicasting gain and spatial reuse gain do not cumulate. Thus, it seems that
the throughput scaling law obtained here is somehow an inherent limitation of one-hop networks with caching in the user nodes.}

\appendices

\section{Proof of Theorems~\ref{theorem: 1} and \ref{theorem: 2}}
\label{sec: the proof of theorem 1 and 2}

We first provide an outline of the proof and then dig into the details.
\begin{enumerate}
\item We define $T_{\rm sum} = \sum_{u=1}^n \overline{T}_u$ and let ($T_{\rm sum}^*(p), p$) be the solution of 
\begin{eqnarray}  \label{max-Tsum}
\mbox{maximize} & & T_{\rm sum} \nonumber \\
\mbox{subject to} & & p_o \leq p, 
\end{eqnarray}
where the maximization is with respect to the cache placement and transmission policies $\Pi_c, \Pi_t$. 
As for  $T^*(p)$, also $T_{\rm sum}^*(p)$ is non-decreasing in $p$. Furthermore, the inequality
$T^*(p) \leq \frac{1}{n} T_{\rm sum}^*(p)$ follows immediately from the definition of $T_{\rm sum}^*(p)$ and $T^*(p)$. 
\item We parameterize problem (\ref{max-Tsum}) with respect to the number of nodes in a disk of radius $R$, referred to (for brevity) as ``disk size'' and 
indicated by $g_R(m)$, where $R$ denotes the one-hop transmission range of the protocol model.  
For any value $g_R(m) = g$, let $\Tc^*_{\rm sum}(g)$ denote 
the largest  achievable sum throughput with disk size $g$, and let $p^*_o(g)$ denote the corresponding outage probability. 
While obtaining exact expressions for $\Tc^*_{\rm sum}(g)$ and for $p^*_o(g)$ is difficult, we shall obtain an upper bound $\Tc^{\rm ub}_{\rm sum}(g) \geq \Tc^*_{\rm sum}(g)$ 
and a lower bound $p^{\rm lb}(g) \leq p^*_o(g)$. By the monotonicity property said above, it follows that 
$(\Tc^{\rm ub}_{\rm sum}(g), p^{\rm lb}(g))$ dominates $(\Tc^*_{\rm sum}(g), p^*_o(g))$ and, as a consequence, 
$(\frac{1}{n} \Tc^{\rm ub}_{\rm sum}(g), p^{\rm lb}(g))$ dominates $(T^*(p), p)$ for $p = p^*_o(g)$. 
Also, we have that the set of outage probability values $p^*_o(g)$ obtained by varying $g$ includes the feasibility domain $[p_{o,\min}, 1]$ 
of the original problem (\ref{sucaminchia}). This implies that the set of points  $(\frac{1}{n} \Tc^{\rm ub}_{\rm sum}(g), p^{\rm lb}(g))$, obtained by varying $g$, 
dominates the Pareto boundary of the throughput-outage region $\Tc$. 
\item Finally, we shall consider separately the different regimes of the outer bound, by ``eliminating'' the parameter $g_R(m)$. 
Conceptually, this can be obtained by letting $p = p^{\rm lb}(g)$, solving for $g$ as a function of $p$ and replacing 
the result into $\Tc^{\rm ub}_{\rm sum}(g)$. The resulting outer bound shall be denoted simply by $(T^{\rm ub}(p), p)$, 
given by Theorems \ref{theorem: 1} and \ref{theorem: 2}. 
\end{enumerate}

We focus first on Theorem~\ref{theorem: 1}, where $\lim_{n \rightarrow \infty} \frac{m^\alpha}{n} = 0$, and  
consider in details step 2) of the above outline. 
In the following, we shall implicitly ignore the non-integer effects when they are irrelevant for the scaling laws.
For example, recalling that the network has node density $n$ (we have $n$ nodes in the unit square), the 
disk size  is given (up to integer rounding) by 
\be  \label{suca2}
g_R(m) = \pi R^2 n \eqdef g.
\ee
For given disk size $g$, a lower bound on $p_o$ can be obtained by observing that $1 - p_o$ is upper bounded by the maximum 
over the users $u=1, \cdots, n$,  of the probability that  user $u$ can be served by the D2D network. 
A necessary condition for this to happen is that the message $\fsf_u$ is found in the cache of some node inside a disk
of size $g$ centered at node $u$. We denote such event by  $\Fc_g^u$.\footnote{Notice: events are defined in the probability space
of the triple $(\fsf, \Gsf, \Asf) \sim \prod_{i=1}^n P_r(f_i) \Pi_c({\sf G}) \Pi_t({\sf A}| {\sf f}, {\sf G})$, 
of requests, cache placements and transmission scheduling decisions.} 
If $g \geq m/M$, then the outage probability lower bound is zero, since we can arrange the files in the caches such that
at least one node $u$ finds all files in the library within a radius $R$. 
Hence, assuming $g < m/M$, we have
\begin{eqnarray} \label{eq: 1-po}
1-p_o &\leq& \max_{u} \PP\left(\Fc_g^u\right) \notag\\
& \buildrel (a) \over \leq & \sum_{f=1}^{Mg} P_r(f) = \sum_{f=1}^{Mg} \frac{f^{-{\gamma}}}{H({\gamma},1,m)} \notag\\
& = & \frac{H({\gamma},1,Mg)}{H({\gamma},1,m)}, 
\end{eqnarray}
where $(a)$ follows by caching all most popular $M g$ files within a disk of radius $R$ form a given user.
In order to estimate the value of $H(\cdot,\cdot,\cdot)$, we have the following lemma:
\begin{lemma}
\label{lemma: H}
For $\gamma \neq 1$, then 
\begin{align}
\frac{1}{1-\gamma}(y+1)^{1-\gamma} - \frac{1}{1-\gamma}x^{1-\gamma} \leq H(\gamma,x,y) \leq 
\frac{1}{1-\gamma}y^{1-\gamma} - \frac{1}{1-\gamma}x^{1-\gamma} + \frac{1}{x^{\gamma}}.
\end{align}
For $\gamma = 1$, then  
\begin{align}
\log(y+1) - \log(x) \leq H(\gamma,x,y) \leq \log(y) - \log(x)+\frac{1}{x}.
\end{align}
\end{lemma}

\begin{IEEEproof} 
See Appendix~\ref{proof: lemma H}.    
\end{IEEEproof}

From (\ref{eq: 1-po}) and Lemma~\ref{lemma: H}, we have the lower bound
\be
\label{eq: pL 1}
p^*_o(g) \geq  p^{\rm lb}(g) \eqdef \left \{ 
\begin{array}{ll} 0 & \mbox{for} \;\;  g\geq \frac{m}{M} \\
1 - \frac{\frac{1}{1-{\gamma}}(Mg)^{1-{\gamma}} - \frac{1}{1-{\gamma}} + 1}{\frac{1}{1-{\gamma}}m^{1-{\gamma}}-\frac{1}{1-{\gamma}}} & \mbox{for} \;\; g < \frac{m}{M}
\end{array} \right .
\ee
Next, we seek an upper bound on $\Tc^*_{\rm sum}(g)$ as a function of the disk size $g$.  
According to the protocol model (see Section \ref{section: network model}),  the throughput $T_{\rm sum}$ is given by 
\be  \label{TsumL}
T_{\rm sum} = C \cdot \EE \left [ \Lsf \right], 
\ee
where $\Lsf$ is the number of active links over any strategy with transmission radius $R$. 
Letting $(i, j)$ and $(k,l)$ denote  two distinct transmitter-receiver pairs, using the triangle inequality and the protocol model constraints, we have
\begin{eqnarray}
\label{eq: d 1}
d(j,l) & \geq & d(k,j) - d(k,l) \notag\\
& \geq & (1+\Delta) R - d(k,l) \notag\\
& \geq & (1+\Delta) R - R = \Delta R.
\end{eqnarray}
Hence, any two receivers must be separated by distance not smaller than $\Delta R$. Equivalently, disks of radius
$\frac{\Delta}{2} R$ around any receiver must be disjoint.
Since there is at least a fraction $1/4$ of the area of such disks inside the unit square containing our network, 
the number of such disjoint disks in the unit square is upper bounded by  $\left \lceil \frac{16}{\pi \Delta^2 R^2} \right \rceil$. 

\begin{figure}[ht]
\centerline{\includegraphics[width=8cm]{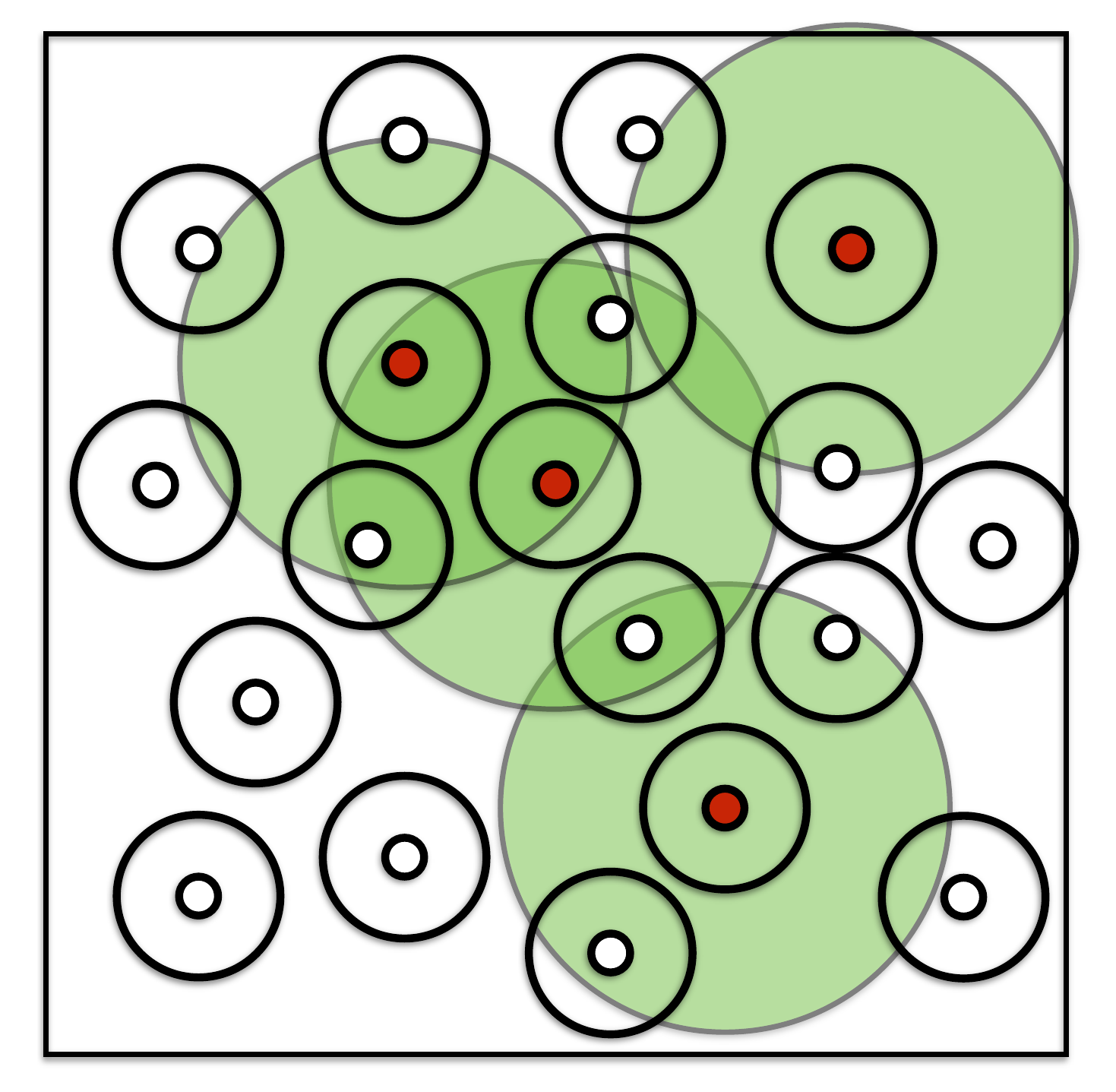}}
\caption{Illustration of the fact that the number of small disks intersecting the union of the big disks centered at the active receivers is necessarily an upper bound
to the number of active receivers.}
\label{disks}
\end{figure}

We wish to upper bound the number of simultaneously active receivers $\Lsf$. In order to do so, 
consider the situation in Fig.~\ref{disks}, where the potentially active receivers (those that can receive according to the protocol model)
are at the centers of mutually exclusive disks of radius $\frac{\Delta}{2} R$. Now, any of these receivers $u$ can effectively receive only if 
$\Fc_g^u$ occurs. From (\ref{eq: pL 1}), we have $\PP(\Fc_g^u) \leq 1 - p^{\rm lb}(g)$. Now, consider a disk of radius $(1 + \Delta)R$ around 
each active receiver (shown as filled dots in Fig.~\ref{disks}), and let $U(R,\Delta,\Lsf)$ denote the union of all such disks. 
It is clear that the number of active receivers $\Lsf$ is less than or equal to the number of 
small disks of radius $\frac{\Delta}{2}R$ with non-empty intersection with $U(R,\Delta,\Lsf)$. 
Since, as argued before, there are at most $\left \lceil \frac{16}{\pi \Delta^2 R^2} \right \rceil$ such disks,  we can write 
\begin{equation} \label{suca1}
\Lsf \leq \sum_{i=1}^{\left \lceil \frac{16}{\pi \Delta^2 R^2} \right \rceil} 1 \left \{\mbox{disk $i$} \cap U(R,\Delta, \Lsf) \right \}.
\end{equation}
Taking expectation of both sides of (\ref{suca1}), and denoting the disks of radius $\frac{\Delta}{2}R$ centered around the receivers simply as
``disk'', we can write 
\begin{eqnarray}
\EE[\Lsf] & \leq & \sum_{i=1}^{\frac{16n}{\Delta^2 g}} \PP \left (\mbox{disk $i$} \cap U(R,\Delta, \Lsf)  \right ) \notag\\ 
& \leq & \frac{16n}{\Delta^2 g} \cdot \PP\left( \mbox{Any disk} \cap U(R,\Delta, \Lsf)  \right). 
\label{suca3}
\end{eqnarray}
Then we introduce the following lemma.

\begin{lemma}
\label{lemma: grid}
\begin{align}
&\PP\left(  \mbox{Any disk} \cap U(R,\Delta, \Lsf)  \right) \leq  \PP\left ( \exists \; \mbox{an active receiver in a disk of radius $\left(1+\frac{3\Delta}{2}\right)R$} \right).
\end{align}
\end{lemma}
\begin{IEEEproof}
See Appendix \ref{sec: lemma grid}.
\end{IEEEproof}
Using Lemma \ref{lemma: grid} in (\ref{suca3}), we obtain
\begin{eqnarray}
\EE[\Lsf] & \leq & \frac{16}{\Delta^2}  \left ( \frac{n}{g} \right ) \cdot 
\PP\left( \exists \; \mbox{an active receiver in a disk of radius $\left(1+\frac{3\Delta}{2}\right)R$} \right) \notag\\
& \buildrel (a) \over \leq & \frac{16}{\Delta^2} \left ( \frac{n}{g} \right ) \cdot \left (1 - \left (  p^{\rm lb}(g) \right )^{\left(1+\frac{3\Delta}{2}\right)^2 g} \right ), 
\label{suca4}
\end{eqnarray}
where (a) follows from the fact that, recalling (\ref{suca2}), the number of users in a disk of radius $\left(1+\frac{3\Delta}{2}\right)R$ is given by 
\be
n \pi \left(1+\frac{3\Delta}{2}\right)^2 R^2 = g \left(1+\frac{3\Delta}{2}\right)^2, 
\ee
and the probability that no users in such disk find their requested content within the transmission range $R$ 
can be lower bounded as 
\[ \PP\left ( \bigcap_{i=1}^{g \left(1+\frac{3\Delta}{2}\right)^2}  (\Fc_g^i)^c \right ) \geq \PP \left ( \bigcap_{i=1}^{g \left(1+\frac{3\Delta}{2}\right)^2} \{ \fsf_i  > gM\} \right ) 
\geq \left (  p^{\rm lb}(g) \right )^{\left(1+\frac{3\Delta}{2}\right)^2 g}. \]


Using (\ref{suca4}) in (\ref{TsumL}), we obtain the sought upper bound $\Tc^{\rm ub}_{\rm sum}(g)$ on $\Tc^*_{\rm sum}(g)$ as
\begin{eqnarray} \label{eq: upper bound 1}
\Tc^*_{\rm sum}(g)  & \leq & \Tc^{\rm ub}_{\rm sum}(g)  \eqdef \frac{16 C}{\Delta^2} \cdot \left [ \left(1-\left (  p^{\rm lb}(g) \right )^{\left(1+\frac{3\Delta}{2}\right)^2 g} \right)  \frac{n}{ g} \right ]. \nonumber \\
& &
\end{eqnarray}
In order to discuss the different regimes of the outer bound, we start by 
considering the maximum throughput regime and the corresponding outage lower bound. 
This is obtained by maximizing $\Tc^{\rm ub}_{\rm sum}(g)$ in (\ref{eq: upper bound 1}) with respect to $g$, and is given by the following result:
\begin{lemma}
\label{lemma: upper bound}
As $m \rightarrow \infty$, the maximum of the quantity 
$\left [ \left(1 - \left ( p^{\rm lb}(g) \right )^{\left(1+\frac{3\Delta}{2}\right)^2 g} \right) \frac{n}{g}\right ]$ is given by 
$\frac{1}{\rho^*} \left(1-e^{-\zeta(\rho^*)} \right)\frac{n}{m^{\alpha}}$, 
where $\rho^*$ is the solution of (\ref{eq: theorem 1}) with $\zeta(\rho)$ given by (\ref{upsilon}), 
and where the optimal $g$ takes on the form $g^* =  \rho^* m^\alpha$,  with $\alpha$ given in (\ref{alpha}). 
\end{lemma}

\begin{IEEEproof} 
See Appendix~\ref{Proof of lemma: upper bound}.
\end{IEEEproof}

Using Lemma~\ref{lemma: upper bound}, the resulting maximum (with respect to $g$) of the sum throughput upper bound is given by:
\begin{eqnarray}
\Tc^{\rm ub}_{\rm sum} (g^*) & = & 
f_{\rm ub}(\rho^*)\frac{n}{m^{\alpha}},  \label{Tub-max}
\end{eqnarray}
where $f_{\rm ub}(\rho)$ is defined in (\ref{f1rho}). 

By replacing $g = g^*$ into (\ref{eq: pL 1}), the corresponding value of the outage probability lower bound is given by 
\begin{eqnarray}
p^{\rm lb} (g^*)
& = & 1 - \frac{\frac{1}{1-{\gamma}}\left(M\rho^*m^{\alpha}\right)^{1-{\gamma}} - \frac{1}{1-{\gamma}} + 1}{\frac{1}{1-{\gamma}}m^{1-{\gamma}}-\frac{1}{1-{\gamma}}}.
\end{eqnarray}
For large $m$,  we have
\be \label{zak}
p^{\rm lb} (g^*) = 1 - {(M \rho^*)}^{1-{\gamma}} m^{-\alpha} + o(m^{-\alpha}), 
\ee
where we used the identity $(1 - {\gamma})(1 - \alpha) = \alpha$.

At this point, we have essentially captured the throughput-outage tradeoff outer bound in the third line in expression (\ref{ub-th1}) of Theorem \ref{theorem: 1}. 
There is one small technical point that needs to be settled in order to obtain the desired result from (\ref{Tub-max}) and (\ref{zak}), namely, 
we have to show that by introducing a perturbation of size $o(m^{-\alpha})$ in the outage probability lower bound $p^{\rm lb}(g^*)$, 
the corresponding perturbation of the throughput upper bound  is $n o(m^{-\alpha})$. 
This fact follows from the continuity of $\Tc^{\rm ub}_{\rm sum}(g)$ and $p^{\rm lb}(g)$ in $g$,  and it is proved in 
Appendix \ref{continuity-perturbation}.  After this perturbation argument, the throughput-outage point corresponding to the maximization 
of $\Tc^{\rm ub}_{\rm sum}(g)$ with respect to $g$ shall be denoted by  $((T^{\rm ub})^*, (p^{\rm lb})^*)$, 
with coordinates  $(p^{\rm lb})^* = 1 - {(M\rho^*)}^{1-{\gamma}}m^{-\alpha}$ 
 and  $(T^{\rm ub})^* = f_{\rm ub}(\rho^*)m^{-\alpha} + o(m^{-\alpha})$.   
The point $((T^{\rm ub})^*, (p^{\rm lb})^*)$ dominates the achievable throughput-outage 
tradeoff boundary  $(T^*(p), p)$, for all $p \geq (p^{\rm lb})^*$, yielding the third line in expression (\ref{ub-th1}) of Theorem \ref{theorem: 1}.

Next, we characterize the other regimes of the outer bound on the throughput-outage tradeoff region by using (\ref{eq: pL 1}) and 
(\ref{eq: upper bound 1}), for different regimes of the disk size $g_R(m)$. 
It is clear from (\ref{eq: pL 1}) that by increasing $g_R(m)$ beyond $g_R^*(m) = g^*$ given in Lemma \ref{lemma: upper bound}, the
outage probability lower bound decreases. We consider two cases: 1) $g_R(m) = \Theta\left(m^{\alpha}\right)$ with $g_R(m) > \rho^* m^{\alpha}$; 
2) $\omega\left(m^{\alpha}\right) = g_R(m) \leq \min\{\frac{m}{M},n\}$. 

{\em Case 1)} In this case, we let $g = g_R(m) = \rho' m^\alpha$ with $\rho' > \rho^*$. 
Letting $m \rightarrow \infty$ in (\ref{eq: pL 1}) and in (\ref{eq: upper bound 1}), we obtain 
\[ 1 - {(M\rho')}^{1-{\gamma}} m^{-\alpha} + o(m^{-\alpha}) \leq p^{\rm lb}(g) <  1 - {(M \rho^*)}^{1-{\gamma}} m^{-\alpha} + o(m^{-\alpha}), \]
and
\begin{equation}
\Tc^{\rm ub}_{\rm sum}(g) = f_{\rm ub}(\rho') \frac{n}{m^{\alpha}}. 
\end{equation}
With a derivation similar to what done in Appendix \ref{continuity-perturbation}, and not included in the paper for the sake of brevity, 
this yields part of second line in expression (\ref{ub-th1}) of Theorem \ref{theorem: 1} (one of the two terms of the minimum).

{\em Case 2)} 
When $\omega\left (m^\alpha \right ) = g_R(m) \leq \min\{\frac{m}{M},n\}$, we use (\ref{eq: pL 1}) and the probability bound as in (\ref{suca4}) and write 
\begin{align}
& \PP\left( \exists \; \mbox{an active receiver in a disk of radius $\left(1+\frac{3\Delta}{2}\right)R$}  \right) \notag\\
& \leq 1- \left ( p^{\rm lb}(g) \right )^{\left(1+\frac{3\Delta}{2}\right)^2 g} \notag\\
&  =  1- \left ( 1 - \frac{\frac{1}{1-{\gamma}}(Mg)^{1-{\gamma}} - \frac{1}{1-{\gamma}} + 1}{\frac{1}{1-{\gamma}}m^{1-{\gamma}}-\frac{1}{1-{\gamma}}} \right )^{\left(1+\frac{3\Delta}{2}\right)^2 g} \label{jumbolo} \\
& \leq  1 - o(1), \label{eq: p good 1}
\end{align}
where the last line follows from the fact that, writing the second term in (\ref{jumbolo}) as
\begin{equation} 
\left [ \left ( 1 -  M^{1 - {\gamma}} \left ( \frac{g}{m} \right )^{1 - {\gamma}} (1 + o(1))  \right )^g \right ]^{\left(1+\frac{3\Delta}{2}\right)^2}, \label{quantitity} 
\end{equation}
we see that the condition for (\ref{quantitity}) to be non-vanishing in the limit for $g,m \rightarrow \infty$ is that
\[ \left ( \frac{g}{m} \right )^{1 - {\gamma}} = \Theta(g^{-1}), \]
or, equivalently, that
\[ g = \Theta(m^\alpha), \]
where $\alpha = \frac{1 - {\gamma}}{2 - {\gamma}}$ is the familiar quantity defined in (\ref{alpha}). Hence, in the case
$g = \omega(m^\alpha)$, the disk size $g$ grows rapidly and the limit of (\ref{quantitity}) vanishes. 

By using (\ref{eq: p good 1}) into (\ref{suca3}) with $g = g_R(m)$ we eventually obtain
\begin{equation} \label{ziobababa} 
\Tc^{\rm ub}_{\rm sum} = \frac{16C}{\Delta^2} \left ( \frac{n}{g_R(m)}\right ) + o\left (\frac{n}{g_R(m)} \right ). 
\end{equation}
Moreover, from (\ref{eq: pL 1}) we have
\begin{eqnarray}
p^{\rm lb}(g) & = & 1 - \frac{\frac{1}{1-{\gamma}}(M g_R(m))^{1-{\gamma}} - \frac{1}{1-{\gamma}} + 1}{\frac{1}{1-{\gamma}}m^{1-{\gamma}}-\frac{1}{1-{\gamma}}} \notag\\
& = & 1 - \left(\frac{Mg_R(m)}{m}\right)^{1-{\gamma}} + o\left(\left(\frac{g_R(m)}{m}\right)^{1-{\gamma}} \right).
\end{eqnarray}
Expressing $g_R(m)$ as a function of $p = p^{\rm lb}$, we find 
\[ g_R(m) = \frac{m}{M} (1 - p)^{\frac{1}{1 - {\gamma}}}. \]
Using this into (\ref{ziobababa}) and following a perturbation argument similar to Appendix \ref{continuity-perturbation},  we find the desired form
\begin{equation}
\Tc^{\rm ub}_{\rm sum}(g) = T^{\rm ub}_{\rm sum}(p) = n \left ( \frac{16 C M}{\Delta^2 m(1-p)^{\frac{1}{1-{\gamma}}}} + o \left (  \frac{1}{m (1 - p)^{\frac{1}{1 - {\gamma}}}} \right ) \right ),
\end{equation}
which yields the first line and the second term in the minimum of the second line 
in expression (\ref{ub-th1}) of Theorem \ref{theorem: 1}. 

By following into the same footsteps, Theorem~\ref{theorem: 2} can be proved along the same lines with the only difference that, 
when there exists a positive constant $\xi$ such that $\xi \leq \lim_{n \rightarrow \infty}\frac{m^\alpha}{n} \leq \frac{16}{\Delta^2\rho^*}$, 
the case $g_R(m) = \omega\left(m^{\alpha}\right)$ does not exist.

\section{Proof of Theorem~\ref{theorem: 3}}
\label{sec: the proof of theorem 3}

In the case  $\lim_{n \rightarrow \infty}\frac{m^\alpha}{n} > \frac{16+\Delta^2}{\Delta^2\rho_2}$, 
an obvious upper bound of the sum throughput $\Tc_{\rm sum}^*(p)$ is provided by
\begin{eqnarray}
T_{\rm sum} & = & C \cdot \EE[\Lsf] \notag \\
&\buildrel (a) \over   \leq & C \sum_{u=1}^n \sum_{f = 1}^{Mn} P_r(f) = C n \frac{H({\gamma},1,Mn)}{H({\gamma},1,m)} \notag\\
&\buildrel (b) \over \leq & Cn \frac{\frac{1}{1-{\gamma}}(Mn)^{1-{\gamma}} - \frac{1}{1-{\gamma}}+1}{\frac{1}{1-{\gamma}}(m+1)^{1-{\gamma}} - \frac{1}{1-{\gamma}}} \notag\\
& \leq & n \left ( CM^{1-{\gamma}}\frac{n^{1-{\gamma}}}{m^{1-{\gamma}}} + o\left(\frac{n^{1-{\gamma}}}{m^{1-{\gamma}}}\right) \right ),
\end{eqnarray}
where 
(a) is because we use a deterministic caching scheme (see Appendix~\ref{Proof of lemma: upper bound}) 
which makes the network store the most $n$ popular messages, and (b) follows 
from Lemma~\ref{lemma: H}.  Dividing by $n$, we obtain the upper bound
\be  \label{ziobadola}
T^*(p) \leq T^{\rm ub}(p) \eqdef CM^{1-{\gamma}} \frac{n^{1-{\gamma}}}{m^{1-{\gamma}}} + o\left(\frac{n^{1-{\gamma}}}{m^{1-{\gamma}}}\right).
\ee
Moreover, as $n$ goes to infinity, the outage probability in this case can be computed as
\be
p_o 
\geq 1 - \frac{H({\gamma},1,Mn)}{H({\gamma},1,m)} \geq 1 - M^{1-{\gamma}}\frac{n^{1-{\gamma}}}{m^{1-{\gamma}}} + o\left(\frac{n^{1-{\gamma}}}{m^{1-{\gamma}}}\right).
\ee
Again, following a perturbation argument similar to Appendix \ref{continuity-perturbation}), 
for $p \geq 1 - M^{1-{\gamma}}\frac{n^{1-{\gamma}}}{m^{1-{\gamma}}}$, we have  $T^*(p) \leq T^{\rm ub}(p)$ 
in (\ref{ziobadola}).  Otherwise, the problem is infeasible.

\section{Proof of Theorem~\ref{theorem: optimal caching distribution}}
\label{sec: the proof of theorem 4}

As mentioned in Section \ref{section: Achievable Throughput-Outage Tradeoff}, we divide the network into clusters, each of which contains $g_c(m)$ nodes. 
In this case, let $\Fc_{g_c(m)}^u$ denote the event that user $u$ can find the requested message inside 
its cluster of size $g_c(m)$. Letting $\mathbf{1}_u = 1\{\Fc_{g_c(m)}^u\}$, we define
\begin{equation}
p_u^c = \EE[\mathbf{1}_u] = \PP(\Fc_{g_c(m)}^u).
\end{equation}
Our goal here is to find the caching distribution $P^*_c(f)$ that maximizes $p_u^c$.  With independent random caching, 
the probability that a user $u$ finds its request $\fsf_u = f$ in its cluster is given by 
$\PP(\Fc_{g_c(m)}^u | \fsf_u = f) = 1 - (1 - P_c(f))^{M(g_c(m)-1)}$ (notice that we do not consider requests to files in the user own cache, 
since these do not generate any traffic).  By the law of total probability, we can write
\begin{align}
p_u^c = \sum_{f=1}^m P_r(f) \left((1-(1-P_c(f))^{Mg_c(m) - M}\right).
\end{align}
Letting $g_c(m) = g$ for simplicity of notation, and assuming $g > 2$, we have the convex optimization problem
\begin{eqnarray} 
\mbox{minimize} & &  \sum_{f=1}^m P_r(f) (1 - P_c(f))^{Mg-M} \nonumber \\
\text{subject to} & & \sum_{f=1}^m P_c(f) = 1, \;\; P_c(f) \geq 0 \;\;\; \forall \; f 
\end{eqnarray}
The Lagrangian function for the problem is
\begin{equation}
\Lc(P_c, \xi) = \sum_{f=1}^m P_r(f) (1 - P_c(f))^{Mg-M}  + \xi' \left ( \sum_{f=1}^m P_c(f) - 1 \right )
\end{equation}
Taking the partial derivative with respect to $P_c(f)$ and using the KKT conditions \cite{boyd2004convex} we obtain
\begin{equation} 
P_c(f) = \left [ 1 - \left ( \frac{\xi'}{P_r(f) M(g - 1)} \right )^{1/(M(g-1)-1)} \right ]^+ 
\end{equation}
It is immediate to see that the minimum is obtained when the sum probability constraint holds with equality. 
In order to solve for the Lagrangian multiplier that imposes the constraint with equality, it is convenient to re-parameterize the problem
by defining $(\frac{\xi'}{M(g-1)})^{\frac{1}{M(g-1)-1}} = \nu$ and $z_f = P_r(f)^{\frac{1}{M(g-1)-1}}$ where the coefficients $z_f$ are 
non-increasing since $P_r(f)$ is non-increasing by 
assumption. Hence, we wish to solve
\[ \sum_{f=1}^m \left [ 1 - \frac{\nu}{z_f} \right ]^+ = 1 \]
The unique solution must be found among the following conditions:
\begin{eqnarray}
1 - \frac{\nu}{z_1} = 1 & \mbox{with} & \frac{\nu}{z_2} \geq 1 \nonumber \\
2 - \frac{\nu}{z_1} - \frac{\nu}{z_2} = 1 & \mbox{with} & \frac{\nu}{z_3} \geq 1 \nonumber \\
3 - \frac{\nu}{z_1} - \frac{\nu}{z_2} - \frac{\nu}{z_3} = 1 & \mbox{with} & \frac{\nu}{z_4} \geq 1 \nonumber \\
& \vdots & \nonumber \\
m - \sum_{f=1}^m \frac{\nu}{z_f} = 1 & &  
\end{eqnarray}
which can be rewritten compactly as finding the unique index $m^*$ for which the equation
\begin{eqnarray} 
\nu \left ( \sum_{f=1}^{m^*} \frac{1}{z_f} \right )  = m^* - 1
\end{eqnarray}
has a solution in the interval for $\nu \geq z_{m^* + 1}$ and $\nu \leq z_{m^*}$. Since we are guaranteed that such $m^*$ exists, we can write
\begin{equation}
\nu(m^*) = \frac{m^* - 1}{\sum_{f=1}^{m^*} \frac{1}{z_f}} 
\end{equation}
From the conditions $\nu(m^*) \geq z_{m^*+1}$ and $\nu(m^*) \leq z_{m^*}$, we find that $m^*$ is explicitly given as the unique integer in $\{1,2,\cdots,m\}$ 
such that
\begin{equation} \label{eqmstar}
m^* \geq 1 + z_{m^*+1} \sum_{f=1}^{m^*} \frac{1}{z_f}, 
\end{equation}
and
\begin{equation} \label{eqmstar2}
m^* \leq 1 + z_{m^*} \sum_{f=1}^{m^*} \frac{1}{z_f}. 
\end{equation}
Next, we wish to determine $m^*$ as a function of $g = g_c(m)$ in the assumption that $g_c(m) \rightarrow \infty$
as $m \rightarrow \infty$. In order to do so, we shall evaluate the terms in the right-hand side of (\ref{eqmstar}) and (\ref{eqmstar2}).
Recalling the expression of $z_f$ in terms of $P_r(f) = \frac{\kappa}{f^{{\gamma}}}$ (recall that we assume a Zipf distribution for the demands, 
with exponent ${\gamma} \in (0,1)$), we have
\begin{equation} 
z_{m^*+1} \sum_{f=1}^{m^*} \frac{1}{z_f} = \sum_{f=1}^{m^*} \left ( \frac{f}{m^*+1} \right )^{a'},
\end{equation}
and
\begin{equation} 
z_{m^*} \sum_{f=1}^{m^*} \frac{1}{z_f} = \sum_{f=1}^{m^*} \left ( \frac{f}{m^*} \right )^{a'},
\end{equation}
where we let ${a'} = \frac{{\gamma}}{M(g-1)-1}$ for brevity. 
We use the following integral lower and upper bounds
\begin{equation} \label{lbub}
\frac{1}{(m^*+1)^{a'}}  + \frac{1}{(m^*+1)^{a'}} \int_1^{m^*} x^{a'} dx \leq  \sum_{f=1}^{m^*} \left ( \frac{f}{m^*+1} \right )^{a'} \leq 
\frac{1}{(m^*+1)^{a'}} \int_1^{m^*+1} x^{a'} dx,
\end{equation}
and
\begin{equation} \label{lbub2}
\frac{1}{(m^*)^{a'}}  + \frac{1}{(m^*)^{a'}} \int_1^{m^*} x^{a'} dx \leq  \sum_{f=1}^{m^*} \left ( \frac{f}{m^*} \right )^{a'} \leq 
\frac{1}{(m^*)^{a'}} \int_1^{m^*+1} x^{a'} dx.
\end{equation}
Solving the integrals, we obtain the lower bound (LB 1) and the upper bound (UB 1) in (\ref{lbub}) in the form
\begin{eqnarray*}
\mbox{LB 1} & = & \frac{a'}{a'+1} \frac{1}{(m^*+1)^{a'}} + \frac{m^*}{a'+1} \left (\frac{m^*}{m^*+1} \right )^{a'} \\
\mbox{UB 1} & = & \frac{m^*+1}{a'+1} - \frac{1}{a'+1}\frac{1}{(m^*+1)^{a'}},
\end{eqnarray*}
and we obtain the lower bound (LB 2) and the upper bound (UB 2) in (\ref{lbub2}) in the form
\begin{eqnarray*}
\mbox{LB 2} & = & \frac{a'}{a'+1} \frac{1}{(m^*)^{a'}} + \frac{m^*}{a'+1} \\
\mbox{UB 2} & = & \frac{m^*+1}{a'+1}\left(\frac{m^*+1}{m^*}\right)^{a'} - \frac{1}{(m^*)^{a'}}\frac{1}{a'+1} .
\end{eqnarray*}
We let 
$m^* = c/{a'}$ for some constant $c$, and notice that ${a'} \downarrow 0$ as $g_c(m) \rightarrow \infty$ and that
$\lim_{{a'} \downarrow 0} (1 + c/{a'})^{a'} = 1$, $\lim_{{a'} \downarrow 0} (1 + {a'}/c)^{a'} = 1$ and $\lim_{{a'} \downarrow 0} (c/{a'})^{a'} = 1$. Hence, in the limit of ${a'} \downarrow 0$ we can write
\begin{eqnarray*}
\mbox{LB 1} & = & \frac{c/{a'}}{{a'}+1} (1 - \delta_1({a'})) + \delta_2({a'}) \\
\mbox{UB 1} & = & \frac{c/{a'} + 1}{{a'}+1} - 1 + \delta_3({a'}),
\end{eqnarray*}
and
\begin{eqnarray*}
\mbox{LB 2} & = & \frac{c/{a'}}{{a'}+1}  + \delta_4({a'}) \\
\mbox{UB 2} & = & \frac{c/{a'} + 1}{{a'}+1} ( 1 + \delta_5({a'})) - 1 + \delta_6({a'}),
\end{eqnarray*}
where $\delta_i$, $i = 1, \cdots, 6$ tend to zero from above as $a \downarrow 0$. 
It follows that
\begin{equation*}
\frac{c/{a'}}{{a'}+1}  ( 1 - \delta_1({a'})) + \delta_2({a'}) \leq z_{m^*+1} \sum_{f=1}^{m^*} \frac{1}{z_f} \leq \frac{c/{a'} + 1}{{a'}+1} - 1 + \delta_3({a'}),
\end{equation*}
and
\begin{equation*}
\frac{c/{a'}}{{a'}+1}  + \delta_4({a'}) \leq z_{m^*} \sum_{f=1}^{m^*} \frac{1}{z_f} \leq \frac{c/{a'} + 1}{{a'}+1} ( 1 + \delta_5({a'})) - 1 + \delta_6({a'}),
\end{equation*}
as $m^* = c/{a'}$ and $a \downarrow 0$.  Replacing the common leading term in the LB 1, UB 1, LB 2 and UB 2 above into (\ref{eqmstar}) and (\ref{eqmstar2}), we obtain 
\[ \frac{c}{{a'}} \gtrsim 1 + \frac{c/{a'}}{{a'}+1},  \]
and
\[ \frac{c}{{a'}} \lesssim1 + \frac{c/{a'}}{{a'}+1},  \]
which yields
\[ \frac{c}{{a'}+1} \cong 1 \]
Therefore, we obtain $c = 1$, which yields 
\[ m^* = \frac{1}{{a'}} = \frac{M(g_c(m)-1)-1}{{\gamma}} + O(1) \]
i.e., $m^* = \frac{M}{{\gamma}} g_c(m)$ to the leading order. Clearly, if $\frac{M}{{\gamma}} g_c(m) > m$, then $m^*=m$.

\section{Proof of Theorem~\ref{theorem: 4}}
\label{sec: the proof of theorem 5 and 6}

Recall that Theorem~\ref{theorem: 4} deals with the small library 
regime $\lim_{n \rightarrow \infty}\frac{m^\alpha}{n} = 0$. 
We define the probability
\begin{equation}
p_{uu'}^c = \EE[\mathbf{1}_u \mathbf{1}_{u'}] = \PP(\Fc_{g_c(m)}^u \cap \Fc_{g_c(m)}^{u'}),
\end{equation}
i.e., the probability that both user $u$ and user $u'$ can find the requested files in the corresponding 
cluster. We let 
\begin{equation}
\Wsf = \sum_{u=1}^{g_c(m)} \mathbf{1}_u,
\end{equation} 
denote the number of potential links in a cluster.

Given the random and independent caching placement $\Pi_c^*$ and the random (or round robin) transmission policy $\Pi_t^*$ as given at the beginning of 
Section \ref{section: Achievable Throughput-Outage Tradeoff}, we let $T(p)$ denote achievable values of
$\overline{T}_{\rm min}$ subject to the outage constraint $p_o \leq p$.
Also, we define $\overline{T}_{\rm sum} = \sum_{u=1}^n \overline{T}_{u}$.

 We provide first an outline of the proof and then dig into the details. 
 \begin{enumerate}
 \item Under policies $P_c^*(f)$ and  $\Pi_t^*$, we notice that both $\overline{T}_{\min}$ and $p_o$ are uniquely determined by the cluster size $g_c(m)$. Hence, 
 the maximum throughput is obtained by solving: 
\begin{eqnarray}
\label{optimization problem 2}
\mbox{maximize} & & \overline{T}_{\min} \notag\\
\mbox{subject to} & & 0 < g_c(m) \leq n.
\end{eqnarray}
Since the exact solution of (\ref{optimization problem 2}) is difficult to obtain, 
we instead compute a lower bound and, for the maximizing $g_c(m) = g_c^*(m)$,  the corresponding value of the outage probability $p^*_o$. 
\item It follows immediately form the definition that $T(p) = T(p_o^*)$ for all $p \geq p_o^*$ is achievable, by keeping $g_c(m) = g^*_c(m)$ and using the same
caching and scheduling policies. This yields a lower bound on the achievable throughput-outage tradeoff when $p \geq p_o^*$.
\item In order to obtain a tradeoff for $p < p_o^*$, we increase $g_c(m)$ above $g^*_c(m)$.  By letting $g_c(m)$ grow and calculating 
the corresponding value of $p_o$ and (a lower bound on) $\overline{T}_{\min}$,  
we obtain a lower bound $T(p)$ for $p = p_o$ on the achievable throughput-outage tradeoff. 
\end{enumerate}

\subsection{Achievable $T(p)$ when $p \geq p_o^*$}
\label{section: A Lower Bound of T}

We first compute a lower bound on $\overline{T}_{\rm sum}$ and the corresponding outage probability $p_o$ for the caching and transmission policies 
$\Pi_c^*$ with $\Pi_t^*$, with cluster size $g_c(m)$. Since the resulting system is symmetric with respect to any user, 
it follows that each user has exactly the same average throughput, such that $\overline{T}_{\min} = \frac{1}{n} \overline{T}_{\rm sum}$. 
Then, we shall maximize the resulting (lower bound on) $\overline{T}_{\min}$ with respect to $g_c(m)$ in order to find $g^*_c(m)$, $p^*_o$ and
$T(p)$ for $p \geq p_o^*$.  For simplicity of notations, in the following we ignore some of the smaller order terms as $m$ and $n$ goes to infinity.

The main tool to obtain a lower bound on $\overline{T}_{\rm sum}$ is the Paley-Zygmund Inequality (see \cite{ozgur2010operating} and references therein). 
Letting again $\Lsf$ denote the number of active links, we have
\begin{eqnarray}
\label{eq: T 2}
\overline{T}_{\rm sum} & = & C \cdot \EE[\Lsf] \notag\\
&\buildrel (a) \over = & C \cdot \EE[\text{number of active clusters}],
\end{eqnarray}
where (a) is because that in $\Pi_t^*$, only one transmission is allowed in each cluster. Moreover, 
\begin{align}
\label{eq: T 1}
&\EE[\text{number of active clusters}] \notag\\
&\geq \frac{1}{K} \EE[\text{number of good clusters}] \notag\\
&= \frac{1}{K} \left(\text{total number of clusters in the network} \cdot \PP(\Wsf > 0) \right),
\end{align}
where $K$ is the TDMA reuse factor and we use the fact that a cluster is good if $\Wsf > 0$. 

From (\ref{eq: T 1}), we have that  a lower bound of $\overline{T}_{\rm sum}$ can b obtained by lower bounding $\PP(\Wsf > 0)$. 
The distribution of $\Wsf$ is not obvious since the random variables $\mathbf{1}_u$ and $\mathbf{1}_{u'}$ are dependent when $u$ and $u'$ are in the same 
cluster and $u \neq u'$. Nevertheless, it is possible to compute the first and second moments of $\Wsf$. Then, with the help of the Paley-Zygmund Inequality, 
we can obtain a lower bound on $\PP(\Wsf > 0)$ which is good enough for our purposes. 
For completeness, the Paley-Zygmund Inequality is provided in the following lemma:

\begin{lemma}
\label{lemma: Paley-Zygmund Inequality}
Let $X$ be a non-negative random variable such that $\EE[X^2] < \infty$. Then for any $t \geq 0$ such that $t < \EE[X]$,  we have
\be
\PP(X > t) \geq \frac{(\EE[X] - t)^2}{\EE[X^2]}.
\ee
\hfill  $\square$
\end{lemma}

By using Lemma~\ref{lemma: Paley-Zygmund Inequality} with $t = 0$ and $X = \Wsf$, we get
\be  \label{paleypaley}
\PP(\Wsf > 0) \geq \frac{\EE[\Wsf]^2}{\EE[\Wsf^2]}.
\ee
Therefore, our goal is to find a lower bound for $\EE[\Wsf]$ and an upper bound $\EE[\Wsf^2]$ under the optimal caching distribution $P_c^*$, 
given by Theorem~\ref{theorem: optimal caching distribution}. First, we focus on $\EE[\Wsf]$. 
Using the expression $\EE[\Wsf] = \sum_{u = 1}^{g_c(m)}p_u^c = g_c(m) p_u^c$, we shall focus on the computation of $p_u^c$ as follows:
\begin{align}
\label{eq: puc 111}
p_u^c &= \sum_{f=1}^{m^*} P_r(f)\left(1-\left(\frac{\nu}{z_f}\right)^{M(g_c(m)-1)}\right) \notag\\
&\buildrel (a) \over \leq \sum_{f=1}^{m^*} P_r(f)\left(1-\left(\frac{z_{m^*+1}}{z_f}\right)^{M(g_c(m)-1)}\right) \notag\\
&= \sum_{f=1}^{m^*} P_r(f) \left(1-\left(\frac{P_r(m^*+1)}{P_r(f)}\right)^\frac{M(g_c(m)-1)}{M(g_c(m)-1)-1}\right) \notag\\
&= \sum_{f=1}^{m^*} P_r(f) - \sum_{f=1}^{m^*} P_r(f) \left(\frac{P_r(m^*+1)}{P_r(f)}\right)^\frac{M(g_c(m)-1)}{M(g_c(m)-1)-1} \notag\\
&= \sum_{f=1}^{m^*} P_r(f) - \sum_{f=1}^{m^*} P_r(f) \left(\frac{P_r(m^*+1)}{P_r(f)}\right)\left(\frac{P_r(m^*+1)}{P_r(f)}\right)^\frac{1}{M(g_c(m)-1)-1} \notag\\
&= \sum_{f=1}^{m^*} P_r(f) - P_r(m^*+1) \sum_{f=1}^{m^*}\left(\frac{f}{m^*+1}\right)^\frac{{\gamma}}{M(g_c(m)-1)-1} \notag\\
&= \frac{H({\gamma},1,m^*)}{H({\gamma},1,m)} - \frac{{(m^*+1)}^{(-{\gamma})}}{H({\gamma},1,m)} \sum_{f=1}^{m^*}\left(\frac{f}{m^*+1}\right)^\frac{{\gamma}}{M(g_c(m)-1)-1},
\end{align}
where (a) is because $\nu \geq z_{m^*+1}$ (see Theorem~\ref{theorem: optimal caching distribution} and its proof in Section \ref{sec: the proof of theorem 4}).  
Similarly, we have
\begin{align}
\label{eq: puc 112}
p_u^c &= \sum_{i=1}^{m^*} P_r(f)\left(1-\left(\frac{\nu}{z_f}\right)^{M(g_c(m)-1)}\right) \notag\\
&\buildrel (a) \over \geq \sum_{i=1}^{m^*} P_r(f)\left(1-\left(\frac{z_{m^*}}{z_f}\right)^{M(g_c(m)-1)}\right) \notag\\
&= \frac{H({\gamma},1,m^*)}{H({\gamma},1,m)} - \frac{{(m^*)}^{(-{\gamma})}}{H({\gamma},1,m)} \sum_{f=1}^{m^*}\left(\frac{f}{m^*}\right)^\frac{{\gamma}}{M(g_c(m)-1)-1},
\end{align}
where (a) is because $\nu \leq z_{m^*}$ (again, see Theorem~\ref{theorem: optimal caching distribution} and its proof in Section \ref{sec: the proof of theorem 4}).

By (\ref{eq: puc 111}), (\ref{eq: puc 112}) and Lemma~\ref{lemma: H}, we have 
\begin{eqnarray}
\label{eq: puc 1}
p_u^c &\leq& \frac{1}{\frac{1}{1-{\gamma}}(m+1)^{1-{\gamma}} - \frac{1}{1-{\gamma}}}\left(\left(\frac{1}{1-{\gamma}}{m^*}^{1-{\gamma}} - \frac{1}{1-{\gamma}} + 1\right) - {m^*}^{(-{\gamma})}\sum_{f=1}^{m^*}\left(\frac{f}{m^*+1}\right)^\frac{{\gamma}}{M(g_c(m)-1)}\right) \notag\\
&\leq& \frac{1-{\gamma}}{(m+1)^{1-{\gamma}}-1} \cdot \left(\frac{1}{1-{\gamma}}{m^*}^{1-{\gamma}} - {m^*}^{(-{\gamma})}\sum_{f=1}^{m^*}\left(\frac{f}{m^*+1}\right)^\frac{{\gamma}}{M(g_c(m)-1)} - \frac{{\gamma}}{1-{\gamma}}\right) \notag\\
&\buildrel (a) \over =& \frac{1-{\gamma}}{(m+1)^{1-{\gamma}}} \cdot \left(\frac{1}{1-{\gamma}}{\left(\frac{M}{{\gamma}}g_c(m)\right)}^{1-{\gamma}} - \left({\frac{M}{{\gamma}}g_c(m)}\right)^{(-{\gamma})}  \right. \notag\\
& & \left. \cdot \sum_{f=1}^{\frac{M}{{\gamma}}g_c(m)}\left(\frac{f}{\frac{M}{{\gamma}}g_c(m)+1}\right)^\frac{{\gamma}}{M(g_c(m)-1)-1} - \frac{{\gamma}}{1-{\gamma}}\right) \notag\\
& \leq & {\gamma}^{{\gamma}-1} \left(\frac{Mg_c(m)}{m+1}\right)^{1-{\gamma}} - (1-{\gamma}){\gamma}^{{\gamma}}\frac{(Mg_c(m))^{-{\gamma}}}{(m+1)^{1-{\gamma}}}\left(\frac{1}{\frac{M}{{\gamma}}g_c(m)+1}\right)^{\frac{{\gamma}}{M(g_c(m)-1)-1}} \notag\\
& & \cdot \left(1 +  \int_{1}^{\frac{M}{{\gamma}}g_c(m)}x^\frac{{\gamma}}{M(g_c(m)-1)-1}dx \right) -\frac{{\gamma}}{(m+1)^{1-{\gamma}}} \notag\\
& = & {\gamma}^{{\gamma}-1} \left(\frac{Mg_c(m)}{m+1}\right)^{1-{\gamma}} - (1-{\gamma}){\gamma}^{{\gamma}}\frac{(Mg_c(m))^{-{\gamma}}}{(m+1)^{1-{\gamma}}}\left(\frac{1}{\frac{Mg_c(m)}{{\gamma}}+1}\right)^{\frac{{\gamma}}{M(g_c(m)-1)-1}} \notag\\
& & \cdot \left(\frac{\frac{{\gamma}}{M(g_c(m)-1)-1}}{\frac{{\gamma}}{M(g_c(m)-1)-1}+1} +  \frac{\left(\frac{Mg_c(m)}{{\gamma}}\right)^{\frac{{\gamma}}{M(g_c(m)-1)-1}}}{\frac{{\gamma}}{M(g_c(m)-1)-1}+1}\frac{Mg_c(m)}{{\gamma}} \right) -\frac{{\gamma}}{(m+1)^{1-{\gamma}}} \notag\\
& = & {\gamma}^{{\gamma}}\left(\frac{Mg_c(m)}{m}\right)^{1-{\gamma}} + o\left(\left(\frac{Mg_c(m)}{m}\right)^{1-{\gamma}}\right),
\end{eqnarray}
where (a) is because $m^*=\frac{Mg_c(m)}{{\gamma}}$. 
and
\begin{eqnarray}
\label{eq: puc 2}
p_u^c &\geq& \frac{1}{\frac{1}{1-{\gamma}}m^{1-{\gamma}} - \frac{1}{1-{\gamma}}+1}\left(\left(\frac{1}{1-{\gamma}}{(m^*+1)}^{1-{\gamma}} - \frac{1}{1-{\gamma}}\right) - {m^*}^{(-{\gamma})}\sum_{f=1}^{m^*}\left(\frac{f}{m^*}\right)^\frac{{\gamma}}{M(g_c(m)-1)}\right) \notag\\
&\geq& \frac{1-{\gamma}}{m^{1-{\gamma}}-{\gamma}} \cdot \left(\frac{1}{1-{\gamma}}{m^*}^{1-{\gamma}} - {m^*}^{(-{\gamma})}\sum_{f=1}^{m^*}\left(\frac{f}{m^*}\right)^\frac{{\gamma}}{M(g_c(m)-1)} - \frac{1}{1-{\gamma}}\right) \notag\\
&\buildrel (a) \over =& \frac{1-{\gamma}}{m^{1-{\gamma}}-{\gamma}} \left(\frac{1}{1-{\gamma}}{\left(\frac{Mg_c(m)}{{\gamma}}\right)}^{1-{\gamma}} - {\left(\frac{Mg_c(m)}{{\gamma}}\right)}^{(-{\gamma})} \right. \notag\\
& & \left. \cdot
\sum_{f=1}^{\frac{Mg_c(m)}{{\gamma}}}\left(\frac{f}{\frac{Mg_c(m)}{{\gamma}}}\right)^\frac{{\gamma}}{M(g_c(m)-1)-1} - \frac{1}{1-{\gamma}}\right) \notag\\
& \geq & {\gamma}^{{\gamma}-1} \left(\frac{Mg_c(m)}{m}\right)^{1-{\gamma}}\left(1 + \frac{{\gamma}}{m^{1-{\gamma}}-{\gamma}}\right) - (1-{\gamma}){{\gamma}}^{{\gamma}}\left(\frac{Mg_c(m)}{m}\right)^{-{\gamma}} \notag\\
& & \cdot \left(\frac{m^{-{\gamma}}}{m^{1-{\gamma}}-{\gamma}}\right) \left(\frac{{\gamma}}{Mg_c(m)}\right)^{\frac{{\gamma}}{M(g_c(m)-1)-1}}\int_1^{\frac{Mg_c(m)}{{\gamma}}+1}x^{\frac{{\gamma}}{M(g_c(m)-1)-1}}dx - \frac{1}{m^{1-{\gamma}}-{\gamma}} \notag\\
& = & {\gamma}^{{\gamma}-1} \left(\frac{Mg_c(m)}{m}\right)^{1-{\gamma}}\left(1 + \frac{{\gamma}}{m^{1-{\gamma}}-{\gamma}}\right) - (1-{\gamma}){{\gamma}}^{{\gamma}}\left(\frac{Mg_c(m)}{m}\right)^{-{\gamma}} \notag\\
& & \cdot \left(\frac{m^{-{\gamma}}}{m^{1-{\gamma}}-{\gamma}}\right) \left(\frac{{\gamma}}{Mg_c(m)}\right)^{\frac{{\gamma}}{M(g_c(m)-1)-1}}\frac{1}{\frac{{\gamma}}{M(g_c(m)-1)-1}+1} \notag\\
& & \cdot \left(\left(\frac{Mg_c(m)}{{\gamma}}+1\right)^{\frac{{\gamma}}{M(g_c(m)-1)-1}}\frac{Mg_c(m)}{{\gamma}} + \left(\frac{Mg_c(m)}{{\gamma}}+1\right)^{\frac{{\gamma}}{M(g_c(m)-1)-1}}-1\right) \notag\\
& & - \frac{1}{m^{1-{\gamma}}-{\gamma}} \notag\\
& = & {\gamma}^{{\gamma}}\left(\frac{Mg_c(m)}{m}\right)^{1-{\gamma}} + o\left(\left(\frac{Mg_c(m)}{m}\right)^{1-{\gamma}}\right).
\end{eqnarray}
where (a) is because $m^*=\frac{Mg_c(m)}{{\gamma}}$.

Therefore, by using (\ref{eq: puc 1}) and (\ref{eq: puc 2}), we obtain
\be
\label{eq: puc 1-1}
p_u^c = {\gamma}^{{\gamma}}\left(\frac{Mg_c(m)}{m}\right)^{1-{\gamma}} + o\left(\left(\frac{Mg_c(m)}{m}\right)^{1-{\gamma}}\right).
\ee

By using (\ref{eq: puc 1-1}), we have 
\be
\label{eq: EW 4}
\EE[\Wsf] = {\gamma}^{{\gamma}}g_c(m)\left(\frac{Mg_c(m)}{m}\right)^{1-{\gamma}} + o\left(g_c(m)\left(\frac{Mg_c(m)}{m}\right)^{1-{\gamma}}\right).
\ee

Now, since here we deal with an achievability strategy and we can choose the clustering strategy at wish,  
we choose $g_c(m) = c_2 m^{\alpha}$. By Theorem~\ref{theorem: optimal caching distribution}, it follows that $m^* = c_1 m^{\alpha}$ with
$\frac{c_1}{c_2}=\frac{M}{{\gamma}}$.  
Clearly, this requires that $n \geq g_c(m) = c_2 m^{\alpha}$ for all sufficiently large $n$. 
Then, by using (\ref{eq: EW 4}), 
as $m \rightarrow \infty$, we have
\begin{eqnarray}
\label{eq: EW 31}
\EE[\Wsf] &=&  {\gamma}^{{\gamma}}g_c(m)\left(\frac{Mg_c(m)}{m}\right)^{1-{\gamma}} + o\left(g_c(m)\left(\frac{Mg_c(m)}{m}\right)^{1-{\gamma}}\right) \notag\\
& = & {\gamma}^{{\gamma}}c_2 m^{\alpha}\left(\frac{Mc_2 m^{\alpha}}{m}\right)^{1-{\gamma}} + o\left(c_2 m^{\alpha}\left(\frac{Mc_2 m^{\alpha}}{m}\right)^{1-{\gamma}}\right) \notag\\
& = & {\gamma}c_1^{1-{\gamma}}c_2 - o(1).
\end{eqnarray}
Next, we compute $\EE[\Wsf^2]$. Since
\begin{eqnarray}
\label{eq: EW^2 1}
\EE[\Wsf^2] & = & \EE\left[\left(\sum_{u=1}^{g_c(m)}\mathbf{1}_u\right)^2\right]\notag\\
& = & \EE\left[\sum_{u=1}^{g_c(m)}\mathbf{1}_u\right] + \sum_{u=1}^{g_c(m)}\sum_{u'=1,u \neq u'}^{g_c(m)} \EE[\mathbf{1}_u\mathbf{1}_{u'}] \notag\\
& = & g_c(m)p_u^c + g_c(m)(g_c(m)-1)p_{uu'}^c,
\end{eqnarray}
then under the optimal caching distribution $P_c^*$, we need to compute $p_{uu'}^c$. 

Let $B_u^f$ be the event that user $u$ requests file $f$ and can find message $f$ in its cluster, 
such that $\Fc^u_{g_c(m)} = \bigcup_{f=1}^m B_u^f$. 
Then, we can write
\begin{eqnarray}
p_{uu'}^c & = & \PP(\{\onev_u = 1\} \cap \{\onev_{u'}=1\}) \notag\\
& = & \PP\left(\left(\cup_{i=1}^m B_u^i\right) \cap \left(\cup_{j=1}^m B_{u'}^j\right)\right) \notag\\
& = & \PP\left(\cup_{i=1}^m\cup_{j=1}^m \left(B_u^i \cap B_{u'}^j\right)\right) \notag\\
& \buildrel (a) \over = & \sum_{i=1}^m\sum_{j=1}^m\PP\left(B_u^i \cap B_{u'}^j\right) \notag\\
& = & \sum_{i=1}^m\sum_{j=1}^m\PP\left(B_u^i\right)\PP\left(B_{u'}^j|B_u^i\right) \notag\\
& = & \sum_{i=1}^m\sum_{j=1, j \neq i}^m \PP\left(B_u^i\right)\PP\left(B_{u'}^j|B_u^i\right) + \sum_{i=1}^m\PP\left(B_u^i\right)\PP\left(B_{u'}^i|B_u^i\right) \notag\\
& \leq & \sum_{i=1}^m\sum_{j=1, j \neq i}^m \left(P_r(i)(1-(1-P_c(i))^{M(g_c(m)-1)})\right)\left(P_r(j)(1-(1-P_c(j))^{M(g_c(m)-1)-1})\right) 
\notag\\
& &+ \sum_{i=1}^m \left(P_r(i)(1-(1-P_c(i))^{M(g_c(m)-1)})\right) P_r(i) \notag
\end{eqnarray}
\begin{eqnarray}
\label{eq: puu 1}
& = & \sum_{i=1}^m\left(P_r(i)(1-(1-P_c(i))^{M(g_c(m)-1)})\right) \sum_{j=1, j \neq i}^m \left(P_r(j)(1-(1-P_c(j))^{M(g_c(m)-1)-1})\right) \notag\\
& & + \sum_{i=1}^m \left(P_r(i)(1-(1-P_c(i))^{M(g_c(m)-1)})\right)P_r(i),
\end{eqnarray}
where (a) is because that $B_u^i \cap B_{u'}^j$ are disjoint for different pairs of $(i,j)$. Replacing $P_c(f) = P_c^*(f)$ in (\ref{eq: puu 1}), we can continue as
\begin{align}
\label{eq: puu 2}
p_{uu'}^c &\leq \sum_{i=1}^{m^*}\left(P_r(i)(1-(1-P_c^*(i))^{M(g_c(m)-1)})\right) \sum_{j=1, j \neq i}^{m^*} \left(P_r(j)(1-(1-P_c^*(j))^{M(g_c(m)-1)-1})\right) \notag\\
&+ \sum_{i=1}^{m^*} \left(P_r(i)(1-(1-P_c^*(i))^{M(g_c(m)-1)})\right)P_r(i) \notag\\
&\leq \left(\sum_{i=1}^{m^*}\left(P_r(i)(1-(1-P_c^*(i))^{M(g_c(m)-1)})\right)\right)^2 + \sum_{i=1}^{m^*} \left(P_r(i)^2(1-(1-P_c^*(i))^{M(g_c(m)-1)})\right) \notag\\
&\buildrel (a) \over = {p_u^c}^2 + \sum_{i=1}^{m^*} \left(P_r(i)^2(1-(1-P_c^*(i))^{M(g_c(m)-1)})\right),
\end{align}
where (a) is because that $p_u^c = \sum_{i=1}^{m^*}\left(P_r(i)(1-(1-P_c^*(i))^{M(g_c(m)-1)})\right)$.

The second term in (\ref{eq: puu 2}) can be upper  bounded by the following lemma. 
\begin{lemma}
\label{lemma: second term}
$\sum_{i=1}^{m^*} \left(P_r(i)^2(1-(1-P_c^*(i))^{M(g_c(m)-1)})\right)$ upper bounded by $o\left({p_u^c}^2\right)$. 
\end{lemma}
\begin{IEEEproof}
See Appendix \ref{sec: Proof of Lemma second term}.
\end{IEEEproof}
At this point we are ready to obtain a lower bound on $\PP(\Wsf > 0)$ via Lemma \ref{lemma: Paley-Zygmund Inequality} and (\ref{paleypaley}). 
From (\ref{eq: EW^2 1}) we can write
\begin{eqnarray}
\label{eq: EW21}
\EE[\Wsf^2] & \leq & g_c(m)p_u^c + g_c(m)^2p_{uu'}^c  \notag\\
&\leq & g_c(m)p_u^c + g_c(m)^2\left({p_u^c}^2 + o\left({p_u^c}^2\right) \right).
\end{eqnarray}
Then, Lemma~\ref{lemma: Paley-Zygmund Inequality}, (\ref{eq: EW 31}) and (\ref{eq: EW21}) yield
\begin{eqnarray} 
\label{eq: PW 1}
\PP(\Wsf > 0) & \geq & \frac{\EE[\Wsf]^2}{\EE[\Wsf^2]} \notag\\
& \geq & \frac{({\gamma}c_1^{1-{\gamma}}c_2)^2}{g_c(m)p_u^c + g_c(m)^2\left({p_u^c}^2 + o\left({p_u^c}^2\right)\right)} \notag\\
& \geq & \frac{({\gamma}c_1^{1-{\gamma}}c_2)^2}{{\gamma}c_1^{1-{\gamma}}c_2 +({\gamma}c_1^{1-{\gamma}}c_2)^2 + o(1) } \notag\\
& \buildrel (a) \over = & \frac{{{\gamma}}^{{\gamma}}M^{1-{\gamma}}c_2^{2-{\gamma}}}{1+{{\gamma}}^{{\gamma}}M^{1-{\gamma}}c_2^{2-{\gamma}}} + o(1),
\end{eqnarray}
where (a) is because that we pick $\frac{c_1}{c_2}=\frac{M}{{\gamma}}$.

By using (\ref{eq: T 1}), we obtain
\begin{eqnarray}
\EE[\text{number of good clusters}] & = & \frac{n}{g_c(m)} \cdot \PP(\Wsf > 0) \notag\\
& \geq & \frac{n}{c_2m^{\alpha}} \cdot \frac{{{\gamma}}^{{\gamma}}M^{1-{\gamma}}c_2^{2-{\gamma}}}{1+{{\gamma}}^{{\gamma}}M^{1-{\gamma}}c_2^{2-{\gamma}}} + o\left(\frac{n}{m^\alpha}\right) \notag\\
& = & \frac{n}{m^{\alpha}}\frac{{{\gamma}}^{{\gamma}}M^{1-{\gamma}}c_2^{1-{\gamma}}}{1+{{\gamma}}^{{\gamma}}M^{1-{\gamma}}c_2^{2-{\gamma}}} + o\left(\frac{n}{m^\alpha}\right) \notag\\
& = & \frac{n}{m^{\alpha}}\frac{ac_2^{1-{\gamma}}}{1+ac_2^{2-{\gamma}}}  + o\left(\frac{n}{m^\alpha}\right),
\end{eqnarray}
where $a = {{\gamma}}^{{\gamma}}M^{1-{\gamma}}$.
Since $m^* = \frac{M}{{\gamma}}g_c(m) = \frac{c_2M}{{\gamma}}m^{\alpha}$ by Theorem~\ref{theorem: optimal caching distribution}, 
then as $m \rightarrow \infty$, by using (\ref{eq: puc 1}) and (\ref{eq: puc 2}), the corresponding average outage probability 
is given by
\begin{eqnarray}
\label{eq: p 1}
p_o & = & 1 - p_u^c \notag\\
& = & 1 - {\gamma}^{{\gamma}}M^{1-{\gamma}}c_2^{1-{\gamma}}m^{-\alpha} + o\left(m^{-\alpha}\right) \notag\\
& = & 1 - ac_2^{1-{\gamma}}m^{-\alpha} + o\left(m^{-\alpha}\right). 
\end{eqnarray}
Therefore, we have
\begin{equation}
\overline{T}_{\rm sum} \geq \frac{C}{K}\frac{ac_2^{1-{\gamma}}}{1+ac_2^{2-{\gamma}}}\frac{n}{m^{\alpha}} + o\left(\frac{n}{m^\alpha}\right).
\end{equation}
By the symmetry of the system and of the caching and transmission policies  $\Pi_c^*$ and  $\Pi_t^*$, the achievable throughput is lower bounded by 
\be
\label{eq: T(p_o)}
\overline{T}_{\min} = \frac{1}{n}\overline{T}_{\rm sum} \geq \frac{C}{K}\frac{ac_2^{1-{\gamma}}}{1+ac_2^{2-{\gamma}}}\frac{1}{m^{\alpha}} + o\left(\frac{1}{m^\alpha}\right).
\ee
Next, we wish to find $c_2$ that maximizes the coefficient $\frac{ac_2^{1-{\gamma}}}{1+ac_2^{2-{\gamma}}}$ in the throughput lower bound (\ref{eq: T(p_o)}). 
Setting the derivative to zero and looking for a maximum point, we find the unique solution $c_2 = b = \left(\frac{1-{\gamma}}{a}\right)^{\frac{1}{2-{\gamma}}}$.   
Let $D = \frac{ab^{1-{\gamma}}}{1+ab^{2-{\gamma}}}$, 
by using (\ref{eq: T(p_o)}), we have
\be
\overline{T}_{\min} \geq \frac{CD}{K}\frac{1}{m^{\alpha}} + o\left(\frac{1}{m^\alpha}\right),
\ee 
with outage probability
\begin{eqnarray}
\label{eq: p 2}
p_o & = & 1 - p_u^c \notag\\
& = & 1 - a b^{1-{\gamma}}m^{-\alpha} + o\left(m^{-\alpha}\right).
\end{eqnarray}
Letting $p_o^* = 1 - a b^{1-{\gamma}}m^{-1/\alpha}$, following a perturbation argument similar to Appendix \ref{continuity-perturbation}, we have that for 
all $p \geq p_o^*$,
\be
T(p) = \frac{CD}{K}\frac{1}{m^{\alpha}} + o\left(m^{-\alpha}\right),
\ee
is achievable.  Thus, we have proved the last regime in (\ref{eq: theorem 4}) in Theorem~\ref{theorem: 4}. 

\subsection{Achievable $T(p)$ for $p < p_o^*$}

By choosing a throughput-suboptimal value 
$c_2 = \rho_2 > b$ in (\ref{eq: p 1}) and (\ref{eq: T(p_o)}), we have that for $p_o = 1 - a \rho_2^{1-{\gamma}}m^{-\alpha} \leq p \leq p_o^*$, then
\begin{eqnarray}
T(p)  = \frac{CB}{K}\frac{1}{m^{\alpha}} + o\left(m^{-\alpha}\right),
\end{eqnarray}
with $B = \frac{a\rho_2^{1-{\gamma}}}{1+a\rho_2^{2-{\gamma}}}$, is achievable.  
This yields the third regime in Theorem~\ref{theorem: 4}.

Next, we turn our attention to the case of $p_o = 1 - \omega\left(m^{-\alpha}\right)$. This is obtained by increasing the cluster size in order to
decrease the outage probability and correspondingly decrease the throughput. 
As before, we find expressions for $p_o$ and lower bounds on $\overline{T}_{\min}$ as a function of $g_c(m)$. 
We consider two cases for the value of $g_c(m)$. One is when $g_c(m) = \omega\left(\frac{n}{m^{\alpha}}\right)$ 
and $g_c(m) \leq {\gamma} m/M$. The other is when $g_c(m)=\rho_1 m/M$, where $\rho_1 \geq {\gamma}$.

\subsubsection{Case $g_c(m) = \omega\left(\frac{n}{m^{\alpha}}\right)$ and $g_c(m) \leq {\gamma} m/M$}
In this case, the cluster size is so large that $\PP(\Wsf > 0) \rightarrow 1$ as $m \rightarrow \infty$. 
In order to show this, we shall show that for arbitrary $\varepsilon_1 > 0$, with high probability 
$\Wsf \in [(1-\varepsilon_1)\E[\Wsf], (1+\varepsilon_1)\E[\Wsf]]$ with $\EE[\Wsf] \rightarrow \infty$ as $m \rightarrow \infty$. 
This will be proved using Chebyshev's Inequality, which requires the computation of $\EE[\Wsf]$ and $\Var[\Wsf]$. By using (\ref{eq: EW 4}), we obtain $\EE[\Wsf]$. 
Since $g_c(m) = \omega\left(m^{\alpha}\right)$, then 
$\lim_{m \rightarrow \infty}\EE[\Wsf] = \infty$.


Next, we need to compute 
\begin{eqnarray}
\Var[\Wsf] & = & \EE[\Wsf^2] - \EE[\Wsf]^2 \notag\\
& = & \E\left[\sum_{u=1}^{g_c(m)}\onev_u\right] + \sum_{u=1}^{g_c(m)}\sum_{u'=1,u \neq u'}^{g_c(m)} \EE[\onev_u\onev_{u'}] - \left(\sum_{u=1}^{g_c(m)}\EE[\onev_u]\right)^2\notag\\
& = & g_c(m)p_u^c + g_c(m)(g_c(m)-1)p_{uu'}^c - g_c(m)^2{p_u^c}^2 \notag\\
& = & g_c(m)(p_u^c-p_{uu'}^c) + g_c(m)^2(p_{uu'}^c - {p_u^c}^2).
\end{eqnarray}
We focus now on the term $p_{uu'}^c$, which is given by the following lemma.
\begin{lemma}
\label{lemma: puuc upper bound}
$p_{uu'}^c$ is given by:
\be
p_{uu'}^c \leq  {\gamma}^{2{\gamma}} \left(\frac{Mg_c(m)}{m}\right)^{2(1-{\gamma})} + o\left(\left(\frac{Mg_c(m)}{m}\right)^{2(1-{\gamma})}\right).
\ee
\end{lemma}
\begin{IEEEproof}
See Appendix \ref{sec: Proof of Lemma lemma: puuc upper bound}.
\end{IEEEproof}

Therefore, as $m \rightarrow \infty$, by using Lemma \ref{lemma: puuc upper bound}, we have
\begin{eqnarray} \label{varW-gnot1/2}
\Var[\Wsf] & = & g_c(m)(p_u^c-p_{uu'}^c) + g_c(m)^2(p_{uu'}^c - {p_u^c}^2) \notag\\
& \leq & g_c(m)\left({\gamma}^{{\gamma}} \left(\frac{Mg_c(m)}{m}\right)^{1-{\gamma}} - {\gamma}^{2{\gamma}} \left(\frac{Mg_c(m)}{m}\right)^{2(1-{\gamma})}\right) \notag\\
& & + g_c(m)^2 \left({\gamma}^{2{\gamma}} \left(\frac{Mg_c(m)}{m}\right)^{2(1-{\gamma})} - {\gamma}^{2{\gamma}}\left(\frac{Mg_c(m)}{m}\right)^{2(1-{\gamma})}\right) \notag\\
& & + o\left(g_c(m)\left(\frac{Mg_c(m)}{m}\right)^{1-{\gamma}}\right) \notag\\
& = & {\gamma}^{{\gamma}}g_c(m) \left(\frac{Mg_c(m)}{m}\right)^{1-{\gamma}} + o\left(g_c(m)\left(\frac{Mg_c(m)}{m}\right)^{1-{\gamma}}\right).
\end{eqnarray}

Thus, by using (\ref{eq: EW 4}) 
and (\ref{varW-gnot1/2}) 
into Chebyshev's Inequality,  
it is not difficult to show that, for any $\varepsilon_1> 0$,
\begin{align}
\PP\left(|\Wsf - \EE[\Wsf]| \leq \varepsilon_1\EE[\Wsf]\right) & \geq 
 1 - o\left(1\right).
\end{align}
as $m \rightarrow \infty$.


Since, as observed before, $\lim_{m \rightarrow \infty} \EE[\Wsf] = \infty$, we conclude that for any $0 < {\gamma} < 1$, 
as $m \rightarrow \infty$, $\PP(\Wsf > 0) = 1 - o(1)$. It follows that all clusters are good, such that
\be
\label{eq: TA 2}
\overline{T}_{\rm sum} = \frac{C}{K}\frac{n}{g_c(m)} + o\left(\frac{n}{g_c(m)}\right).
\ee
As $m \rightarrow \infty$, by using (\ref{eq: puc 1}) and (\ref{eq: puc 2}), the corresponding outage probability is given by
\begin{eqnarray}
p_o & = & 1 - p_u^c \notag\\
& = & 1 -  {\gamma}^{{\gamma}} \left(\frac{Mg_c(m)}{m}\right)^{1-{\gamma}} + o\left(\left(\frac{Mg_c(m)}{m}\right)^{1-{\gamma}}\right).
\end{eqnarray}
By the usual symmetry argument, we have
\be
\label{eq: TA 3}
\overline{T}_{\min} = \frac{1}{n} \overline{T}_{\rm sum} = \frac{C}{K}\frac{1}{g_c(m)}  + o\left(\frac{1}{g_c(m)}\right).
\ee
Finally, letting $p = p_o$, we can solve for $g_c(m) = \frac{1}{{\gamma}^{\frac{{\gamma}}{1-{\gamma}}}}\frac{m}{M}(1-p)^{\frac{1}{1-{\gamma}}} + o\left(\frac{m}{M}(1-p)^{\frac{1}{1-{\gamma}}}\right)$. By using (\ref{eq: TA 3}) and letting $A = {\gamma}^{\frac{{\gamma}}{1-{\gamma}}}$, with the similar perturbation argument shown in Appendix \ref{continuity-perturbation}, we have that 
when $p = 1 -  {\gamma}^{{\gamma}} \left(\frac{Mg_c(m)}{m}\right)^{1-{\gamma}}$, then
\be
T(p) = \frac{CA}{K} \frac{M}{m(1-p)^{\frac{1}{1-{\gamma}}}} + o\left(\frac{M}{m(1-p)^{\frac{1}{1-{\gamma}}}}\right)
\ee
is achievable.  This settles the second regime of Theorem~\ref{theorem: 1}.

\subsubsection{Case $g_c(m)=\rho_1m/M$, where $\rho_1 \geq {\gamma}$}

In this case, by using Theorem~\ref{theorem: optimal caching distribution}, we have $m^* = m$. We can obtain that $\PP(\Wsf > 0) = 1 - o(1)$ as $m \rightarrow \infty$. Thus, we have 
\be
\overline{T}_{\min} = \frac{1}{n} \frac{C}{K}\frac{n}{g_c(m)} = \frac{C}{K}\frac{M}{\rho_1m} + o\left(\frac{M}{m}\right).
\ee
The corresponding outage probability is computed next. Here, we need to find a different bounding technique other than the one we used before. 
In this case, we directly plug $\nu = \frac{m^*-1}{\sum_{j=1}^{m^*}\frac{1}{z_{f_j}}}$ into $p_u^c$ and use the integral approximations of summations 
to obtain the lower bound of $p_u^c$, instead of using that fact that $\nu \leq z_{m^*}$ and $\nu \geq z_{m^*+1}$ as we used before. The reason is that in this case, $m^* = m$ as shown by Theorem \ref{theorem: optimal caching distribution}, which means that $m^* \neq \frac{Mg_c(m)}{{\gamma}}$ when $\rho_1 > {\gamma}$, this makes $z_{m^*}$ and $z_{m^*+1}$ not good approximations anymore.

Operating along these lines, $p_u^c$ can be computed as
\begin{eqnarray}
\label{eq: puc c1}
p_u^c &=& \sum_{f=1}^{m^*} P_r(f)\left(1-\left(\frac{\nu}{z_f}\right)^{M(g_c(m)-1)}\right) \notag\\
& = & \sum_{f=1}^{m} P_r(f)\left(1-\left(\frac{\nu}{z_f}\right)^{M(g_c(m)-1)}\right) \notag\\
& = & 1 - \nu^{M(g_c(m)-1)}\sum_{f=1}^{m} \frac{P_r(f)}{z_f^{M(g_c(m)-1)}} \notag\\
& = & 1 - \nu^{M(g_c(m)-1)}\sum_{f=1}^{m}\frac{P_r(f)}{P_r(f)^{\frac{M(g_c(m)-1)}{M(g_c(m)-1)-1}}} \notag\\
& = & 1 - \left(\frac{m-1}{\sum_{i=1}^{m}\frac{1}{z_i}}\right)^{M(g_c(m)-1)}\sum_{f=1}^{m} P_r(f)^{-\frac{1}{M(g_c(m)-1)-1}} \notag\\
& = & 1 - \left(\frac{m-1}{ \sum_{i=1}^{m}P_r(i)^{-\frac{1}{M(g_c(m)-1)-1}}}\right)^{M(g_c(m)-1)}\sum_{f=1}^{m} P_r(f)^{-\frac{1}{M(g_c(m)-1)-1}} \notag\\
& = & 1 - (m-1)^{M(g_c(m)-1)}\left(\sum_{f=1}^{m} P_r(f)^{-\frac{1}{g_c(m)-2}}\right)^{-(g_c(m)-2)} \notag \\
& = & 1 - (m-1)^{M(g_c(m)-1)}\left(\sum_{f=1}^{m} \left(\frac{f^{-{\gamma}}}{H({\gamma},1,m)}\right)^{-\frac{1}{M(g_c(m)-1)-1}}\right)^{-(M(g_c(m)-1)-1)} \notag\\
& = & 1 - \frac{(m-1)^{M(g_c(m)-1)}}{H({\gamma},1,m)} \frac{1}{\left(\sum_{f=1}^{m} f^{\frac{{\gamma}}{M(g_c(m)-1)-1}}\right)^{M(g_c(m)-1)-1}}.
\end{eqnarray}
The lower bound of $p_u^c$ is given by
\begin{eqnarray}
p_u^c & \geq & 1 - \frac{(m-1)^{M(g_c(m)-1)}}{\frac{1}{1-{\gamma}}(m+1)^{1-{\gamma}} - \frac{1}{1-{\gamma}}} \cdot \frac{1}{\left(1+\int_1^mx^{\frac{{\gamma}}{M(g_c(m)-1)-1}}dx\right)^{M(g_c(m)-1)-1}} \notag\\
& = & 1 - \frac{(m-1)^{M(g_c(m)-1)}}{\frac{1}{1-{\gamma}}(m+1)^{1-{\gamma}} - \frac{1}{1-{\gamma}}} \cdot \frac{1}{\left(1 + \frac{1}{1+\frac{{\gamma}}{M(g_c(m)-1)-1}}(m^{\frac{{\gamma}}{M(g_c(m)-1)-1}+1}-1)\right)^{M(g_c(m)-1)-1}} \notag\\
& = & 1 - (1-{\gamma})\frac{(m-1)^{M(g_c(m)-1)}}{(m+1)^{1-{\gamma}}-1}\frac{1}{\left(\frac{1}{1+\frac{{\gamma}}{M(g_c(m)-1)-1}}m^{\frac{{\gamma}}{M(g_c(m)-1)-1}+1} + 1 - \frac{1}{\frac{{\gamma}}{M(g_c(m)-1)-1}+1}\right)^{M(g_c(m)-1)-1}} \notag\\
& = & 1 - (1-{\gamma})\frac{(m-1)^{M(g_c(m)-1)}}{m^{1-{\gamma}}} \frac{m^{1-{\gamma}}}{(m+1)^{1-{\gamma}}-1} \frac{1}{m^{M(g_c(m)-1)-1+{\gamma}}} \notag\\
& & \cdot \frac{m^{M(g_c(m)-1)-1+{\gamma}}}{\left(\frac{1}{1+\frac{{\gamma}}{M(g_c(m)-1)-1}}m^{\frac{{\gamma}}{M(g_c(m)-1)-1}+1} + 1 - \frac{1}{\frac{{\gamma}}{M(g_c(m)-1)-1}+1}\right)^{M(g_c(m)-1)-1}} \notag\\
& = & 1 - (1-{\gamma})\frac{(m-1)^{M(g_c(m)-1)}}{m^{1-{\gamma}}} \frac{m^{1-{\gamma}}}{(m+1)^{1-{\gamma}}-1} \frac{1}{m^{M(g_c(m)-1)-1+{\gamma}}} \frac{1}{\left(\frac{1}{1+\frac{{\gamma}}{M(g_c(m)-1)-1}}\right)^{M(g_c(m)-1)-1}} \notag\\
& & \cdot \frac{m^{M(g_c(m)-1)-1+{\gamma}}}{\left(m^{\frac{{\gamma}}{M(g_c(m)-1)-1}+1} + \frac{{\gamma}}{M(g_c(m)-1)-1}\right)^{M(g_c(m)-1)-1}} \notag\\
& = & 1 - (1-{\gamma})\left(1-\frac{1}{m}\right)^{M(g_c(m)-1)} \frac{1}{\left(1-\frac{{\gamma}}{M(g_c(m)-1)-1+{\gamma}}\right)^{\frac{M(g_c(m)-1)-1+{\gamma}}{{\gamma}}\frac{M(g_c(m)-1)-1}{M(g_c(m)-1)-1+{\gamma}}{\gamma}}} \notag\\
& & \cdot \frac{m^{1-{\gamma}}}{(m+1)^{1-{\gamma}}-1} \frac{m^{M(g_c(m)-1)-1+{\gamma}}}{\left(m^{\frac{{\gamma}}{M(g_c(m)-1)-1}+1} + \frac{{\gamma}}{M(g_c(m)-1)-1}\right)^{M(g_c(m)-1)-1}} \notag\\
& = & 1 - (1-{\gamma})\left(1-\frac{1}{m}\right)^{M(\rho_1m/M-1)} \frac{1}{\left(1-\frac{{\gamma}}{M(g_c(m)-1)-1+{\gamma}}\right)^{\frac{M(g_c(m)-1)-1+{\gamma}}{{\gamma}}\frac{M(g_c(m)-1)-1}{M(g_c(m)-1)-1+{\gamma}}{\gamma}}} \left(1 + o(1)\right) \notag\\
& = & 1 - (1-{\gamma})\left(\frac{1}{e}\right)^{\rho_1} \frac{1}{\left(\frac{1}{e}\right)^{{\gamma}}} \left(1 + o(1)\right) \notag\\
& = & 1 - (1-{\gamma})\left(\frac{1}{e}\right)^{\rho_1-{\gamma}}\left(1 + o(1)\right).
\end{eqnarray}
Thus, we have 
\begin{eqnarray}
p_o & = & 1 - p_u^c \notag \\
& \leq  & 1 - \left(1 - (1-{\gamma})\left(\frac{1}{e}\right)^{\rho_1-{\gamma}}\left(1 + o(1)\right)\right) \notag\\
& = & (1-{\gamma})\left(\frac{1}{e}\right)^{\rho_1-{\gamma}}\left(1 + o(1)\right).
\end{eqnarray}
Therefore, letting $p = (1-{\gamma})e^{{\gamma} - \rho_1}$, and following a perturbation argument similar to  Appendix \ref{continuity-perturbation}, we have that the throughput
\be
T(p) = \frac{C}{K}\frac{M}{\rho_1m} + o\left(\frac{M}{m}\right)
\ee
is achievable. This settles the first regime of Theorem \ref{theorem: 4}.

\section{Proof of Lemma~\ref{lemma: H}}
\label{proof: lemma H}

When $\gamma \neq 1$, then, since $\frac{1}{x^\gamma}$ is an decreasing function, we have
\begin{align}
H(\gamma,x,y) &= \sum_{i=x}^y \frac{1}{i^\gamma} \geq \int_a^{b+1} \frac{1}{{x'}^\gamma}dx' \notag\\
&= \frac{1}{1-\gamma}(y+1)^{1-\gamma} - \frac{1}{1-\gamma}x^{1-\gamma},
\end{align}
and
\begin{align}
H(\gamma,x,y) &= \sum_{i=x}^y \frac{1}{i^\gamma} = \frac{1}{x^{\gamma}} + \sum_{i=x-1}^y \frac{1}{i^\gamma} \notag\\
&\leq \int_{x-1+1}^y \frac{1}{{x'}^\gamma}dx' + \frac{1}{x^{\gamma}} \notag\\
&= \frac{1}{1-\gamma}y^{1-\gamma} - \frac{1}{1-\gamma}x^{1-\gamma} + \frac{1}{x^{\gamma}}.
\end{align}

When $\gamma = 1$, similarly, since $\frac{1}{x}$ is an decreasing function, we have
\begin{align}
H(\gamma,x,y) &= \sum_{i=x}^y \frac{1}{i} \geq \int_x^{y+1} \frac{1}{x'}dx' \notag\\
&= \log(y+1) - \log(x),
\end{align}
and
\begin{align}
H(\gamma,x,y) &= \sum_{i=x}^y \frac{1}{i} = \frac{1}{x} + \sum_{i=x-1}^y \frac{1}{i} \notag\\
&\leq \int_{x-1+1}^y \frac{1}{x'}dx' + \frac{1}{x} \notag\\
&= \log(y) - \log(x)+\frac{1}{x}.
\end{align}

\section{Proof of Lemma \ref{lemma: grid}}
\label{sec: lemma grid}


Recall that we denote the disks of radius $\frac{\Delta}{2}R$ centered around the receivers as
``disk'', our goal is to show that
\begin{align}
&\PP\left(  \mbox{Any disk} \cap U(R,\Delta, \Lsf)  \right) \leq  \PP\left ( \exists \; \mbox{an active receiver in a disk of radius $\left(1+\frac{3\Delta}{2}\right)R$} \right).
\end{align}
which is equivalent to show that
\begin{align}
\label{eq: grid 1}
&\{ \mbox{Any disk} \cap U(R,\Delta, \Lsf) \} \notag\\
&\subseteq \{\exists \; \mbox{an active receiver in a disk of radius $\left(1+\frac{3\Delta}{2}\right)R$}\}.
\end{align}

\begin{figure}[ht]
\centerline{\includegraphics[width=10cm]{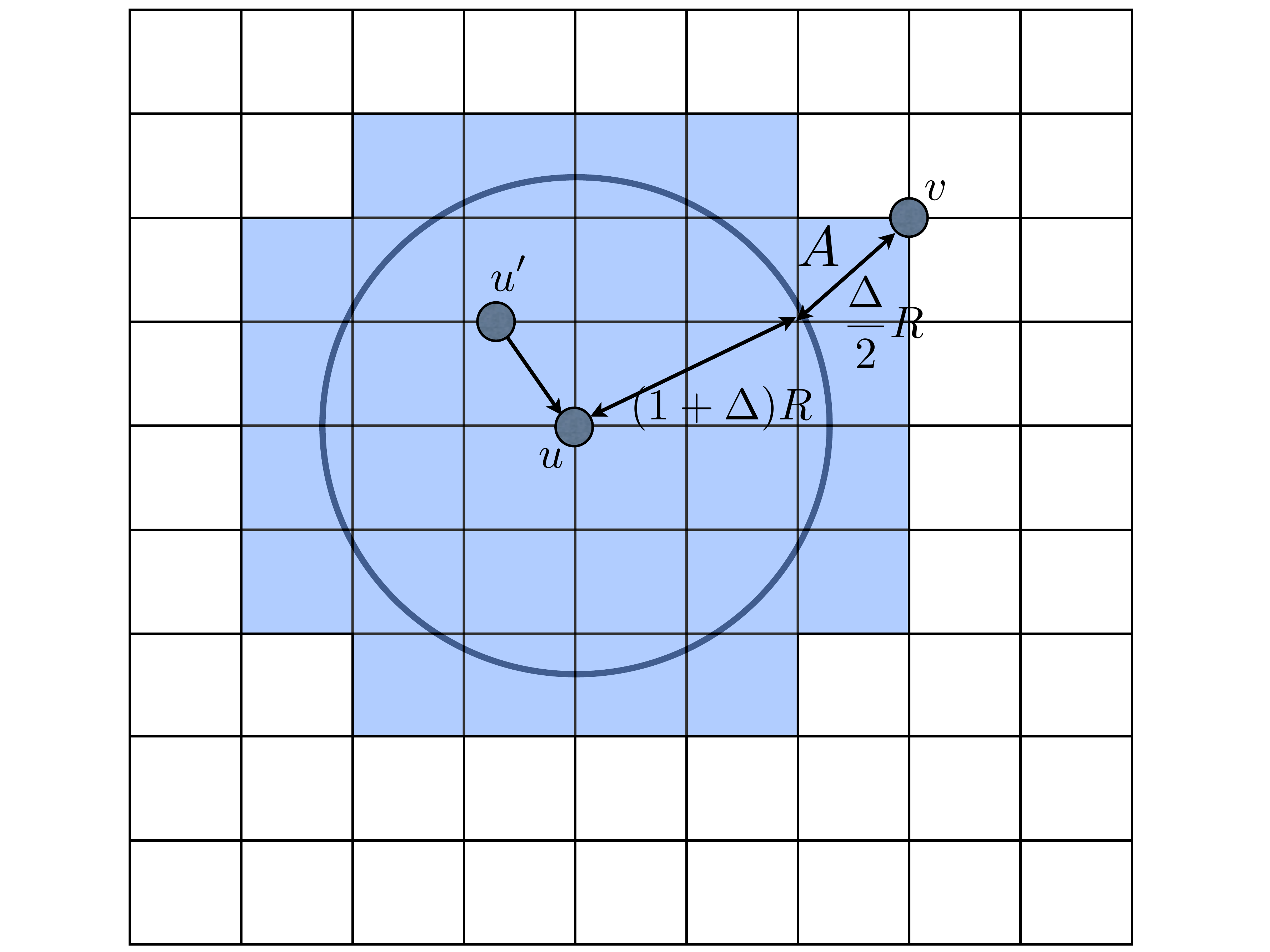}}
\caption{In this figure, $u'$ is a transmitter and $u$ is a receiver. $v$ is another receiver corresponding to another transmitter. The diagonal of each squarelet is $\frac{\Delta}{2}R$. The maximum area that are consumed by receiver $u$ is the disk centered at $u$, with radius $(1+\Delta)R$. The blue squarelets are the maximum activated squarelets that are caused by the active receiver $u$. $A$ indicates the squarelet containing receiver $v$.} 
\label{fig: grid_1}
\end{figure}

To see (\ref{eq: grid 1}), we first consider a simple illustration which is easier to explain, but it is not accurate. As shown in Fig. \ref{fig: grid_1}, the network is divided into squarelets whose diagonal are $\frac{\Delta}{2}R$. These squarelets are the analogue to the the sectors with radius of $\frac{\Delta}{2}R$ (a quarter of disk with radius $\frac{\Delta}{2}R$). Now we want to see which events can cause a squarelet to intersect with $U(R, \Delta, \Lsf)$, which means that the area of this squarelet is consumed due to communicating links according to the protocol model. From Fig. \ref{fig: grid_1}, we can see that if the link from user $u'$ to $u$ is activated, then the upper bound of the maximum area this link can consume is the area of all the blue squarelets. If we consider squarelet $A$ and let user $v$ be a receiver, we can see that $A$ cannot intersect with $U(R, \Delta, \Lsf)$ if there is no any active receiver in a disk centered at $v$, with radius $\left(1+\frac{3\Delta}{2}\right)R$. Therefore, if there is at least one active receiver in a disk centered at $v$, with radius $\left(1+\frac{3\Delta}{2}\right)R$, then it is possible that $A$ can intersect with $U(R, \Delta, \Lsf)$. 

\begin{figure}[ht]
\centerline{\includegraphics[width=10cm]{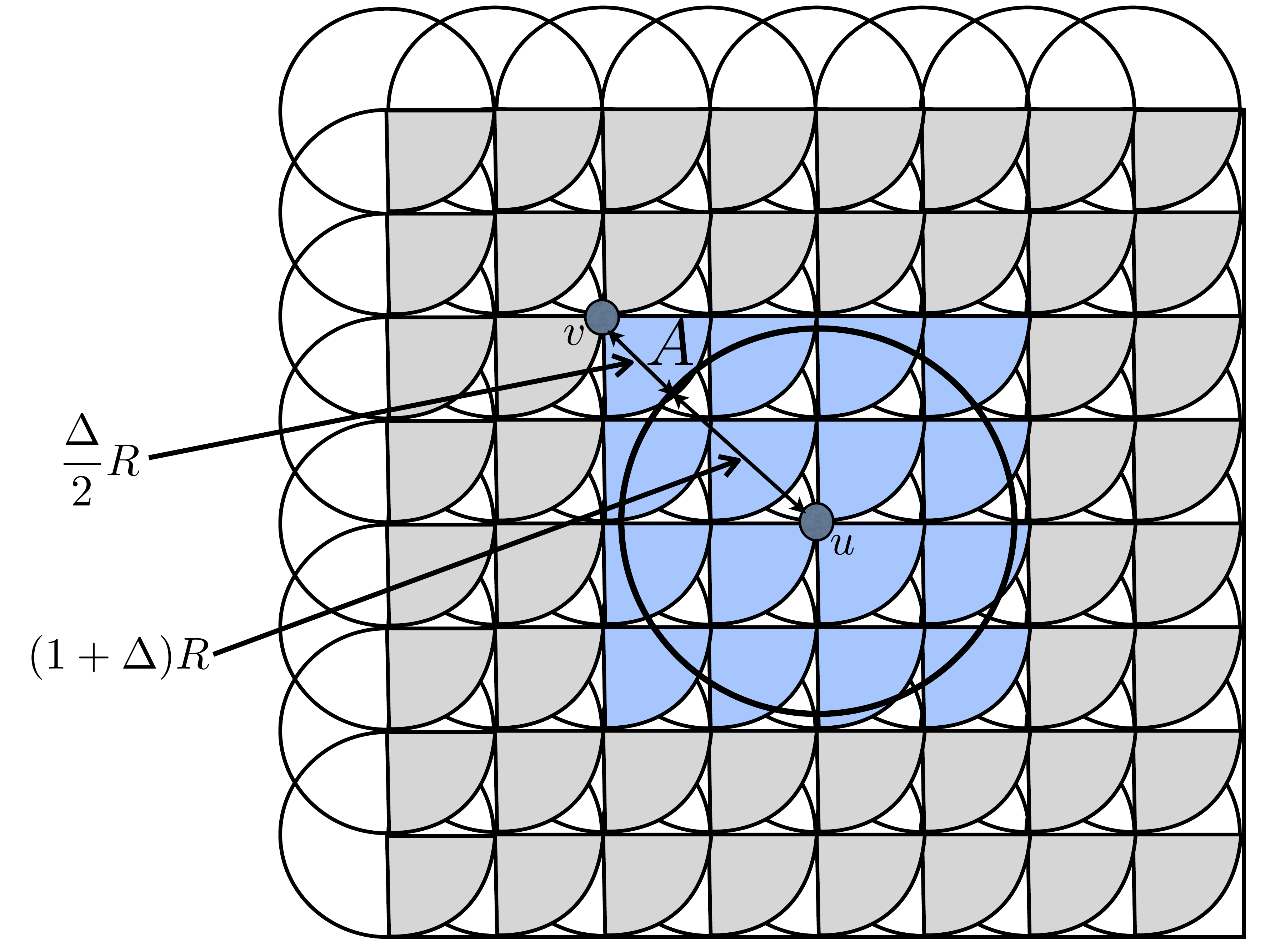}}
\caption{In this figure, $u$ and $v$ are receivers. The radius of each grey sector is $\frac{\Delta}{2}R$. Each grey sector is $\frac{1}{4}$ of each disk with radius $\frac{\Delta}{2}R$. The maximum area that are consumed by receiver $u$ is the disk centered at $u$, with radius $(1+\Delta)R$. The blue sectors are the maximum activated sectors that are caused by the active receiver $u$. $A$ indicates the sector containing receiver $v$.}
\label{fig: grid_2}
\end{figure}

Now we prove this lemma accurately. From the arguments in Appendix \ref{sec: the proof of theorem 1 and 2}, we know that all the disks with radius of $\frac{\Delta}{2}R$ have to be disjoint. Moreover, there is at least a fraction $\frac{1}{4}$ of the area of such disks inside the network. Therefore, to obtain an upper bound of the maximum concurrent transmissions, we maximumly pack such sectors\footnote{In the following, we denote the sector that is $\frac{1}{4}$ of the disk with radius of $\frac{\Delta}{2}R$ as ``sector".} inside the network as shown in Fig~\ref{fig: grid_2} (Of course, Fig~\ref{fig: grid_2} shows an over optimistic way of packing, since we cannot guarantee all the disks with radius $\frac{\Delta}{2}R$ are disjoint. However, at least all the sectors are disjoint.). Notice that
\begin{align}
\label{eq: grid 2}
\{ \mbox{Any disk} \cap U(R,\Delta, \Lsf) \} \subseteq \{\mbox{Any sector} \cap U(R,\Delta, \Lsf)\}.
\end{align}

Now we consider each such sector as an analogue of the squarelet considered before. This shows that if the receiver $u$ is activated, then the upper bound of the maximum number of sectors that can intersect with $U(R, \Delta, \Lsf)$ are the blue sectors. Now pick a arbitrary node $v$, if there is no any active receiver inside a disk centered at $v$ of radius $\left(1+\frac{3\Delta}{2}\right)R$, then the sector $A$ cannot intersect with $U(R, \Delta, \Lsf)$, which means that if there is at least one active receiver inside a disk centered at $v$ of radius $\left(1+\frac{3\Delta}{2}\right)R$, then the sector $A$ may intersect with $U(R, \Delta, \Lsf)$. Since $v$ is arbitrary, then
\begin{align}
\label{eq: grid 3}
&\{\mbox{Any sector} \cap U(R,\Delta, \Lsf)\} \notag\\
&\subseteq \{\exists \; \mbox{an active receiver in a disk of radius $\left(1+\frac{3\Delta}{2}\right)R$} \}.
\end{align}
By using (\ref{eq: grid 2}) and (\ref{eq: grid 3}), (\ref{eq: grid 1}) is proved. 

\section{Proof of Lemma~\ref{lemma: upper bound}}
\label{Proof of lemma: upper bound}

Using (\ref{eq: pL 1}), we are interested in the quantity
\begin{eqnarray}
\label{eq: new lemma 2 1}
\left(1-(p^{\rm lb}(g))^{\left(1+\frac{3\Delta}{2}\right)^2 g} \right) \frac{n}{g} 
&=& \left(1- \left(1 - \frac{\frac{1}{1-{\gamma}}(Mg)^{1-{\gamma}} - \frac{1}{1-{\gamma}} + 1}{\frac{1}{1-{\gamma}}m^{1-{\gamma}}-\frac{1}{1-{\gamma}}}\right)^{\left(1+\frac{3\Delta}{2}\right)^2 g} \right) \frac{n}{g}.
\end{eqnarray}
We consider three regimes for $g$, namely, 
$g = o(m^\alpha)$, $g = \omega(m^\alpha)$ and $g=\Theta(m^\alpha) = \rho m^\alpha$.

When $g = o(m^\alpha)$, by using (\ref{eq: new lemma 2 1}), we have
\begin{eqnarray}
\left(1-(p^{\rm lb}(g))^{\left(1+\frac{3\Delta}{2}\right)^2 g} \right) \frac{n}{g} &\buildrel (a) \over \leq& \left(1- \left(1 - \left(1+\frac{3\Delta}{2}\right)^2 g\frac{\frac{1}{1-{\gamma}}(Mg)^{1-{\gamma}} - \frac{1}{1-{\gamma}} + 1}{\frac{1}{1-{\gamma}}m^{1-{\gamma}}-\frac{1}{1-{\gamma}}}\right) \right) \frac{n}{g} \notag\\
& = & \left(1+\frac{3\Delta}{2}\right)^2 M^{1-{\gamma}} n\left(\frac{g}{m}\right)^{1-{\gamma}} + o\left(n\left(\frac{g}{m}\right)^{1-{\gamma}}\right)\notag\\
& = & o\left(\frac{n}{m^\alpha}\right),
\end{eqnarray}
where (a) is because that $(1-x)^t \geq 1- tx$ for $0 \leq x \leq 1$ and $t \geq 1$. 

When $g = \omega(m^\alpha)$, by using (\ref{eq: new lemma 2 1}), we obtain
\begin{eqnarray}
\left(1-(p^{\rm lb}(g))^{\left(1+\frac{3\Delta}{2}\right)^2 g} \right) \frac{n}{g}
&\buildrel (a) \over \leq & \frac{n}{g}
 = o\left(\frac{n}{m^\alpha}\right),
\end{eqnarray}
where (a) is because that $\left(1-(p^{\rm lb}(g))^{\left(1+\frac{3\Delta}{2}\right)^2 g} \right) \leq 1$. 

When $g = \rho m^\alpha$, by using (\ref{eq: new lemma 2 1}), we get
\begin{eqnarray}
\left(1-(p^{\rm lb}(g))^{\left(1+\frac{3\Delta}{2}\right)^2 g} \right) \frac{n}{g}
& = & \frac{n}{\rho m^\alpha} \left(1-\left(1-\rho^{1-{\gamma}} M^{1-{\gamma}} m^{(\alpha-1)(1-{\gamma})}\right)^{\left(1+\frac{3\Delta}{2}\right)^2 \rho m^\alpha}\right) \notag\\
&\buildrel (a) \over = & \frac{n}{\rho m^\alpha} \left(1-\left(1-\rho^{1-{\gamma}} M^{1-{\gamma}} m^{-\alpha}\right)^{\left(1+\frac{3\Delta}{2}\right)^2\rho m^\alpha} \right) \notag\\
& = & \frac{n}{\rho m^\alpha} \left(1-\left(\left(1-\rho^{1-{\gamma}}M^{1-{\gamma}}m^{-\alpha}\right)^{\rho^{-(1-{\gamma})}M^{-(1-{\gamma})}m^ \alpha}\right)^{\left(1+\frac{3\Delta}{2}\right)^2\rho^{2-{\gamma}}M^{1-{\gamma}}}\right) \notag
\end{eqnarray}
\begin{eqnarray}
\label{eq: new lemma 2 2}
&\buildrel (b) \over = & \frac{1}{\rho}\left(1-\exp\left(-\left(1+\frac{3\Delta}{2}\right)^2\rho^{2-{\gamma}}M^{1-{\gamma}}\right)\right)\frac{n}{m^{\alpha}} \notag\\
& = &  \left(1+\frac{3\Delta}{2}\right)^\frac{2}{2-{\gamma}} \frac{1}{\left(1+\frac{3\Delta}{2}\right)^\frac{2}{2-{\gamma}}\rho} \notag\\
& & \cdot \left(1-\exp\left(-\left(\left(1+\frac{3\Delta}{2}\right)^\frac{2}{2-{\gamma}}\rho\right)^{2-{\gamma}}M^{1-{\gamma}}\right)\right)\frac{n}{m^{\alpha}} \notag\\
&\buildrel (c) \over = & \left(1+\frac{3\Delta}{2}\right)^\frac{2}{2-{\gamma}} \frac{1}{\tilde{\rho}}\left(1-\exp\left(-\tilde{\rho}^{2-{\gamma}}M^{1-{\gamma}}\right)\right)\frac{n}{m^{\alpha}},
\end{eqnarray}
where (a) follows by using $(\alpha-1)(1-{\gamma}) = -\alpha$; 
(b) follows because $\lim_{x \rightarrow \infty} (1-x^{-1})^x = e^{-1}$;  
(c) is obtained by defining $\tilde{\rho} = \left(1+\frac{3\Delta}{2}\right)^\frac{2}{2-{\gamma}}\rho$.

We conclude that 
$\left(1-(p^{\rm lb}(g))^{\left(1+\frac{3\Delta}{2}\right)^2 g} \right) \frac{n}{g} = O\left(\frac{n}{m^\alpha}\right)$ and, 
when $g = \rho m^\alpha$, then $\left(1-(p^{\rm lb}(g))^{\left(1+\frac{3\Delta}{2}\right)^2 g} \right) \frac{n}{g} = \Theta\left(\frac{n}{m^\alpha}\right)$.
Now we compute the optimal constant $\tilde{\rho}$, which is shown in the following lemma.
\begin{lemma}
\label{lemma c}
The optimal value of $\tilde{\rho}^*$ to maximize 
$\left(1-(p^{\rm lb}(g))^{\left(1+\frac{3\Delta}{2}\right)^2 g} \right) \frac{n}{g}$ is the solution of 
\[ \tilde{\rho}^{2-{\gamma}}M^{1-{\gamma}} = \log\left(1+ (2-{\gamma})\tilde{\rho}^{2-{\gamma}}M^{1-{\gamma}}\right). \] 
Moreover, the solution satisfies ${\tilde{\rho}{^*}}^{2-{\gamma}}M^{1-{\gamma}} > \alpha$, 
and Eq. $x = \log\left(1+ (2-{\gamma})x\right)$ is a fixed point equation and
has a non-negative solution for $x > \alpha$.
\end{lemma}

\begin{IEEEproof}
From (\ref{eq: new lemma 2 2}), we know that to maximize 
$\left(1-(p^{\rm lb}(g))^{\left(1+\frac{3\Delta}{2}\right)^2 g} \right) \frac{n}{g}$ we need to maximize 
$\frac{1}{\tilde{\rho}}\left(1-\exp(-\tilde{\rho}^{2-{\gamma}}M^{1-{\gamma}})\right)$. 
Differentiating this expression with respect to $\tilde{\rho}$, and equating to zero, we find 
\be
\label{eq: c1}
\tilde{\rho}^{2-{\gamma}}M^{1-{\gamma}} = \log\left(1+ (2-{\gamma})\tilde{\rho}^{2-{\gamma}}M^{1-{\gamma}}\right).
\ee
This proves the first part of Lemma \ref{lemma c}.

Then, by letting $x = \tilde{\rho}^{2-{\gamma}}M^{1-{\gamma}}$, we get
\be
\label{eq: c2}
x = \log\left(1+ (2-{\gamma})x\right).
\ee
Let $f(x) = \log\left(1+ (2-{\gamma})x\right) - x$. We observe that if $f(x) = 0$, then there are two roots, one is $0$ which must b excluded 
since $x=\rho^{2-{\gamma}}M^{1-{\gamma}} > 0$ and the other root is greater than $0$. Differentiating with respect to $x$, we find
\begin{eqnarray}
\frac{d}{dx}f(x) & = & \frac{2-{\gamma}}{1+(2-{\gamma})x} - 1 \notag\\
& = & \frac{(2-{\gamma})(1-x)-1}{1+(2-{\gamma})x}.
\end{eqnarray}
We observe that $\frac{d}{dx}f(x) < 0$ for $x > \alpha$,   $\frac{d}{dx}f(x)>0$ for $x < \alpha$, and $\frac{d}{dx}f(x)=0$ for $x=\alpha$. 
Thus,  $f(x)$ achieves its maximum value when $x=\alpha$. 

Now we can see that 
\begin{eqnarray}
f\left(\alpha\right) & = & \log\left(1+ (2-{\gamma})\alpha\right)-\alpha \notag\\
& = & \log(2-{\gamma}) - \alpha > 0,
\end{eqnarray}
when $0 \leq {\gamma}<1$.
Thus, the positive root of $f(x)=0$ is greater than $\alpha$. This proves the second part of Lemma \ref{lemma c}.

Let $\phi(x) = \log\left(1+ (2-{\gamma})x\right)$, then if $\phi(x)$ is a contraction from $\mathbb{R}$ to $\mathbb{R}$, then we can show that $\phi(x) = x$ is a fixed point equation and can be solved by iterations numerically \cite{rudin1976principles}. Therefore, we need to show when $x>\alpha$, $\phi(x)$ is a 
contraction from $\mathbb{R}$ to $\mathbb{R}$.

Let $x,y \in \mathbb{R}$. With out loss of generality we assume $x>y$. When $x>\alpha$, we get
\begin{eqnarray}
\left|\phi(x) - \phi(y)\right| & = & \left|\log\left(1+ (2-{\gamma})x\right) - \log\left(1+ (2-{\gamma})y\right)\right| \notag\\
& = & \left|\log\left(\frac{1+(2-{\gamma})x}{1+(2-{\gamma})y}\right)\right| \notag\\
& = & \left|\log\left(1+\frac{2-{\gamma}}{1+(2-{\gamma})y}(x-y)\right)\right| \notag\\
&\buildrel (a) \over < & \frac{2-{\gamma}}{1+(2-{\gamma})y}|x-y| \notag\\
&\buildrel (b) \over < & k|x-y|,
\end{eqnarray}
where $k=\frac{2-{\gamma}}{1+(2-{\gamma})y}<1$. (a) is because that $\log(1+x)<x$, when $x \neq 0$. (b) is because when $y > \alpha$, then $k=\frac{2-{\gamma}}{1+(2-{\gamma})y}<1$.
Thus $\phi(x)$ is a contraction from $\mathbb{R}$ to $\mathbb{R}$ when $x>\alpha$. 
Therefore, we conclude that $\phi(x)=x$ is a fixed point equation for $x>\alpha$.
\end{IEEEproof}

\section{Proof of Lemma \ref{lemma: second term}}
\label{sec: Proof of Lemma second term}

We have
\begin{eqnarray}
\sum_{i=1}^{m^*} \left(P_r(i)^2(1-(1-P_c^*(i))^{M(g_c(m)-1)})\right) & \buildrel (a) \over\leq  & \sum_{i=1}^{m^*}P_r(i)^2\left(1-\left(\frac{P_r(m^*+1)}{P_r(i)}\right)^\frac{M(g_c(m)-1)}{M(g_c(m)-1)-1}\right) \notag\\
& = & \sum_{i=1}^{m^*}P_r(i)^2 - \sum_{i=1}^{m^*}P_r(i)^2\left(\frac{P_r(m^*+1)}{P_r(i)}\right)^\frac{M(g_c(m)-1)}{M(g_c(m)-1)-1} \notag\\
& = & \sum_{i=1}^{m^*}P_r(i)^2 - P_r(m^*+1) \sum_{i=1}^{m^*}P_r(i)\left(\frac{P_r(m^*+1)}{P_r(i)}\right)^\frac{1}{M(g_c(m)-1)-1} \notag
\end{eqnarray}
\begin{eqnarray}
\label{eq: puu 3}
& = & \frac{H(2{\gamma},1,m^*)}{H({\gamma},1,m)^2} - \frac{({m^*+1})^{-{\gamma}}}{H({\gamma},1,m)^2}\left(\sum_{i=1}^{m^*}i^{-{\gamma}}\left(\frac{i}{m^*+1}\right)^\frac{{\gamma}}{M(g_c(m)-1)-1}\right),
\end{eqnarray}
where $(a)$ is because $\nu \geq z_{m*+1}$.

Now, in order to compute an upper bound on (\ref{eq: puu 3}), we consider the first and the second term separately. 
In order to upper bound the first term, we have to consider the cases of
${\gamma} \neq \frac{1}{2}$ and ${\gamma} = \frac{1}{2}$. 
For ${\gamma} \neq \frac{1}{2}$, by using Lemma~\ref{lemma: H}, the first term in (\ref{eq: puu 3}) can be upper bounded as:
\begin{eqnarray}
\label{eq: puu 41}
\frac{H(2{\gamma},1,m^*)}{H({\gamma},1,m)^2} & \leq & \frac{\frac{1}{1-2{\gamma}}{m^*}^{1-2{\gamma}} - \frac{1}{1-2{\gamma}} + 1}{\left(\frac{1}{1-{\gamma}}(m+1)^{1-{\gamma}} - \frac{1}{1-{\gamma}}\right)^2}  \notag\\
& = & \frac{(1-{\gamma})^2}{1-2{\gamma}} \frac{{m^*}^{1-2{\gamma}} - 2{\gamma}}{((m+1)^{1-{\gamma}} - 1)^2} \notag\\
& \leq & \frac{(1-{\gamma})^2}{1-2{\gamma}} \frac{{m^*}^{1-2{\gamma}} - 2{\gamma}}{(m^{1-{\gamma}} - 1)^2} \notag\\
& = & \frac{(1-{\gamma})^2}{(1-2{\gamma})} \frac{\left(\frac{Mg_c(m)}{{\gamma}}\right)^{1-2{\gamma}} - 2{\gamma}}{(m^{1-{\gamma}} - 1)^2}.
\end{eqnarray}
For ${\gamma} = \frac{1}{2}$, by using Lemma~\ref{lemma: H}, the first term in (\ref{eq: puu 3}) can be upper bounded as:
\begin{eqnarray}
\label{eq: puu 44}
\frac{H(2{\gamma},1,m^*)}{H({\gamma},1,m)^2} & = & \frac{H(1,1,m^*)}{H(\frac{1}{2},1,m)^2} \notag\\
& \leq & \frac{\log(m^*) + 1}{\left(\frac{1}{1-\frac{1}{2}}(m+1)^{1-\frac{1}{2}} - \frac{1}{1-\frac{1}{2}}\right)^2} \notag\\
& = & \frac{\log(Mg_c(m))+\log 2 + 1}{\left(2(m+1)^\frac{1}{2}-2\right)^2}. 
\end{eqnarray}
By letting $g_c(m) = \frac{c_1{\gamma}m^\alpha}{M}$, we have
\begin{eqnarray}
\frac{H(2{\gamma},1,m^*)}{H({\gamma},1,m)^2}  & \leq & 
\frac{\frac{1}{3}\log\left(m\right) + \log(c_1) + 1}{\left(2m^{1-{\gamma}} - 2\right)^2}. 
\end{eqnarray}
The second term in (\ref{eq: puu 3}), for any ${\gamma}  < 1$, can be lower bounded as:
\begin{eqnarray}
& & \frac{({m^*+1})^{-{\gamma}}}{H({\gamma},1,m)^2}\left(\sum_{i=1}^{m^*}i^{-{\gamma}}\left(\frac{i}{m^*+1}\right)^\frac{{\gamma}}{M(g_c(m)-1)-1}\right) \notag \\
& \geq & \frac{({m^*+1})^{-{\gamma}}}{\left(\frac{1}{1-{\gamma}}m^{1-{\gamma}} - \frac{1}{1-{\gamma}} + 1\right)^2} \frac{1}{({m^*+1})^\frac{{\gamma}}{M(g_c(m)-1)-1}} \sum_{i=1}^{m^*}i^{-\frac{M(g_c(m)-1)-2}{M(g_c(m)-1)-1}{\gamma}} \notag\\
& \geq & \frac{({m^*+1})^{-{\gamma}}}{\left(\frac{1}{1-{\gamma}}m^{1-{\gamma}} - \frac{{\gamma}}{1-{\gamma}}\right)^2} \frac{1}{({m^*+1})^\frac{{\gamma}}{M(g_c(m)-1)-1}} \notag\\
& & \cdot \int_1^{m^*+1}x^{-\frac{M(g_c(m)-1)-2}{M(g_c(m)-1)-1}{\gamma}}dx \notag
\end{eqnarray}
\begin{eqnarray}
\label{eq: puu 4}
& = & \frac{({m^*+1})^{-{\gamma}}}{\left(\frac{1}{1-{\gamma}}m^{1-{\gamma}} - \frac{{\gamma}}{1-{\gamma}} \right)^2} \frac{1}{({m^*+1})^\frac{{\gamma}}{M(g_c(m)-1)-1}} \notag\\
& & \cdot \frac{1}{1-\frac{M(g_c(m)-1)-2}{M(g_c(m)-1)-1}{\gamma}} \left((m^*+1)^{1-\frac{M(g_c(m)-1)-2}{M(g_c(m)-1)-1}{\gamma}} - 1\right) \notag\\ 
& = & (1-{\gamma}) \frac{(m^*+1)^{1-2{\gamma}}}{(m^{1-{\gamma}}-{\gamma})^2}\left(1 - \frac{{\gamma}-\frac{M(g_c(m)-1)-2}{M(g_c(m)-1)-1}{\gamma}}{1-\frac{M(g_c(m)-1)-2}{M(g_c(m)-1)-1}{\gamma}}\right) \notag\\
& & - \frac{1}{1-\frac{M(g_c(m)-1)-2}{M(g_c(m)-1)-1}{\gamma}} \frac{({m^*+1})^{-\frac{M(g_c(m)-1)-2}{M(g_c(m)-1)-1}{\gamma}}}{\left(\frac{1}{1-{\gamma}}m^{1-{\gamma}} - \frac{{\gamma}}{1-{\gamma}} \right)^2}\notag\\
& \geq & (1-{\gamma}) \frac{(m^*+1)^{1-2{\gamma}}}{(m^{1-{\gamma}}-{\gamma})^2} - o\left(\frac{(m^*+1)^{1-2{\gamma}}}{(m^{1-{\gamma}}-{\gamma})^2}\right) \notag\\
& = & (1-{\gamma})\frac{\left(\frac{M}{{\gamma}}g_c(m)+1\right)^{1-2{\gamma}}}{(m^{1-{\gamma}}-{\gamma})^2} - o\left(\frac{\left(\frac{M}{{\gamma}}g_c(m)+1\right)^{1-2{\gamma}}}{(m^{1-{\gamma}}-{\gamma})^2} \right)
\end{eqnarray}
In order to obtain the scaling behavior of (\ref{eq: puu 3})
we consider the cases of ${\gamma} < \frac{1}{2}$, ${\gamma} > \frac{1}{2}$ and ${\gamma} = \frac{1}{2}$. 

For  ${\gamma} < \frac{1}{2}$, let $g_c(m) = \frac{c_1{\gamma}m^\alpha}{M}$, by using Lemma~\ref{lemma: H}, (\ref{eq: puu 41}) and (\ref{eq: puu 4}), we have
\begin{eqnarray} \label{eq: puu 5}
& & \sum_{i=1}^{m^*} \left(P_r(i)^2(1-(1-P_c^*(i))^{M(g_c(m)-1)})\right) \notag\\
& & \leq \frac{(1-{\gamma})^2}{1-2{\gamma}} \frac{{c_1^{1-2{\gamma}}m^{\frac{(1-{\gamma})(1-2{\gamma})}{2-{\gamma}}}}}{(m^{1-{\gamma}} - 1)^2} - (1-{\gamma})\frac{({c_1m^{\frac{1-{\gamma}}{2-{\gamma}}}})^{1-2{\gamma}}}{m^{2(1-{\gamma})}} - \frac{2{\gamma}}{(m^{1-{\gamma}} - 1)^2} + o\left(\frac{({c_1m^{\frac{1-{\gamma}}{2-{\gamma}}}}+1)^{1-2{\gamma}}}{(m^{1-{\gamma}}-{\gamma})^2}\right) \notag
\end{eqnarray}
\begin{eqnarray}
& & = \frac{(1-{\gamma})^2}{1-2{\gamma}}c_1^{1-2{\gamma}}m^{-\frac{3(1-{\gamma})}{2-{\gamma}}}\left(1+\frac{1}{m^{1-{\gamma}}-1}\right)^2 - (1-{\gamma})c_1^{1-2{\gamma}}m^{-\frac{3(1-{\gamma})}{2-{\gamma}}} \notag\\
& & - \frac{2{\gamma}}{(m^{1-{\gamma}} - 1)^2} + o\left(\frac{({c_1m^{\frac{1-{\gamma}}{2-{\gamma}}}}+1)^{1-2{\gamma}}}{(m^{1-{\gamma}}-{\gamma})^2}\right)\notag \\
& & = \frac{{\gamma}(1-{\gamma})}{1-2{\gamma}}c_1^{1-2{\gamma}}m^{-\frac{3(1-{\gamma})}{2-{\gamma}}} + o\left(m^{-\frac{3(1-{\gamma})}{2-{\gamma}}}\right).
\end{eqnarray}
For ${\gamma} > \frac{1}{2}$, let $g_c(m) = \frac{c_1{\gamma}m^\alpha}{M}$, by using Lemma~\ref{lemma: H}, (\ref{eq: puu 41}) and (\ref{eq: puu 4}), we have
\begin{eqnarray}
\label{eq: puu 6}
& & \sum_{i=1}^{m^*} \left(P_r(i)^2(1-(1-P_c^*(i))^{M(g_c(m)-1)})\right) \notag\\
& & \leq \frac{(1-{\gamma})^2}{1-2{\gamma}} \frac{{c_1^{1-2{\gamma}}m^{\frac{(1-{\gamma})(1-2{\gamma})}{2-{\gamma}}}} - 2{\gamma}}{(m^{1-{\gamma}} - 1)^2} - (1-{\gamma})\frac{({c_1m^{\frac{1-{\gamma}}{2-{\gamma}}}})^{1-2{\gamma}}}{m^{2(1-{\gamma})}} + o\left(\frac{(m^*+1)^{1-2{\gamma}}}{(m^{1-{\gamma}}-{\gamma})^2}\right) \notag\\
& & \leq \frac{2{\gamma}(1-{\gamma})^2}{2{\gamma}-1}m^{-2(1-{\gamma})} - (1-{\gamma})c_1^{1-2{\gamma}}m^{-\frac{3(1-{\gamma})}{2-{\gamma}}} + o\left(\frac{(m^*+1)^{1-2{\gamma}}}{(m^{1-{\gamma}}-{\gamma})^2}\right) \notag\\
& & = \frac{2{\gamma}(1-{\gamma})^2}{2{\gamma}-1}m^{-2(1-{\gamma})} - O\left(m^{-\frac{3(1-{\gamma})}{2-{\gamma}}}\right).
\end{eqnarray}
This settles the scaling behavior of the term $\sum_{i=1}^{m^*} \left(P_r(i)^2(1-(1-P_c^*(i))^{M(g_c(m)-1)})\right)$ for ${\gamma} \neq \frac{1}{2}$.

For the case ${\gamma} = \frac{1}{2}$, let $g_c(m) = \frac{c_1{\gamma}m^\alpha}{M}$, we use (\ref{eq: puu 44}) and (\ref{eq: puu 4}) to obtain
\begin{eqnarray}
\label{eq: puu 7}
& & \sum_{i=1}^{m^*} \left(P_r(i)^2(1-(1-P_c^*(i))^{M(g_c(m)-1)})\right)  \notag\\
& & = \frac{1}{12} \frac{\log\left(m\right) + \log(c_1) + 1}{\left(m^{\frac{1}{2}} - 1\right)^2} - \frac{1}{2m} + o\left(\frac{1}{m}\right)\notag\\
& & = \frac{1}{12} \frac{\log m}{m} + O\left(\frac{1}{m}\right).
\end{eqnarray}

From (\ref{eq: puc 1-1}), and by using (\ref{eq: puu 6}) and (\ref{eq: puu 6}), we obtain the desired result.

\section{Proof of Lemma \ref{lemma: puuc upper bound}}
\label{sec: Proof of Lemma lemma: puuc upper bound}

By using (\ref{eq: puu 2}) and (\ref{eq: puu 3}), 
we have two cases of ${\gamma}$ to consider, namely,  ${\gamma} \neq \frac{1}{2}$ and ${\gamma} = \frac{1}{2}$. 
When ${\gamma} \neq \frac{1}{2}$, by using (\ref{eq: puc 1}), (\ref{eq: puc 2}), (\ref{eq: puu 41}) and (\ref{eq: puu 4}), 
we have
\begin{eqnarray}
\label{eq: puu 21}
p_{uu'}^c 
& \leq & \left({\gamma}^{{\gamma}} \left(\frac{Mg_c(m)}{m}\right)^{1-{\gamma}} + o\left(\left(\frac{Mg_c(m)}{m}\right)^{1-{\gamma}}\right)\right)^2 + \frac{(1-{\gamma})^2}{1-2{\gamma}} \frac{\left(\frac{M}{{\gamma}}g_c(m)\right)^{1-2{\gamma}} - 2{\gamma}}{(m^{1-{\gamma}} - 1)^2} \notag\\
& & - (1-{\gamma})\frac{\left(\frac{M}{{\gamma}}g_c(m)+1\right)^{1-2{\gamma}}}{(m^{1-{\gamma}}-{\gamma})^2} + o\left(\frac{\left(\frac{M}{{\gamma}}g_c(m)+1\right)^{1-2{\gamma}}}{(m^{1-{\gamma}}-{\gamma})^2}\right) \notag
\end{eqnarray}
\begin{eqnarray}
& = & {\gamma}^{2{\gamma}} \left(\frac{Mg_c(m)}{m}\right)^{2(1-{\gamma})} + \frac{(1-{\gamma})^2}{(1-2{\gamma})} \frac{\left(\frac{Mg_c(m)}{{\gamma}}\right)^{1-2{\gamma}} - 2{\gamma}}{(m^{1-{\gamma}} - 1)^2} \notag\\
& & - \frac{\frac{1-{\gamma}}{{\gamma}^{1-2{\gamma}}}\left(Mg_c(m)+{\gamma}\right)^{1-2{\gamma}}}{(m^{1-{\gamma}} - 1)^2} + o\left(\left(\frac{Mg_c(m)}{m}\right)^{2(1-{\gamma})}\right) \notag\\
& \leq & {\gamma}^{2{\gamma}} \left(\frac{Mg_c(m)}{m}\right)^{2(1-{\gamma})} + o\left(\left(\frac{Mg_c(m)}{m}\right)^{2(1-{\gamma})}\right).
\end{eqnarray}
When ${\gamma} = \frac{1}{2}$, by using (\ref{eq: puc 1}), (\ref{eq: puc 2}), (\ref{eq: puu 44}) and (\ref{eq: puu 7}), 
we have
\begin{eqnarray}
p_{uu'}^c 
& \leq & \left( \left(\frac{1}{2}\right)^{\frac{1}{2}} \left(\frac{Mg_c(m)}{m}\right)^{1-\frac{1}{2}} + o\left(\frac{Mg_c(m)}{m}\right)^{\frac{1}{2}}\right)^2 + \frac{\log\left(\frac{M}{\frac{1}{2}}g_c(m)\right)+1}{\left(2(m+1)^\frac{1}{2}-2\right)^2} \notag\\
& & - \frac{1}{2}\frac{\left(\frac{M}{\frac{1}{2}}g_c(m)\right)^{1-2\frac{1}{2}}}{m^{2(1-\frac{1}{2})}} + o\left(\frac{\left(\frac{M}{\frac{1}{2}}g_c(m)\right)^{1-2\frac{1}{2}}}{m^{2(1-\frac{1}{2})}}\right)\notag
\end{eqnarray}
\begin{eqnarray}
\label{eq: puu 31}
& = & \frac{1}{2} \left(\frac{Mg_c(m)}{m}\right) + \frac{\log(Mg_c(m))+\log 2 + 1}{\left(2(m+1)^\frac{1}{2}-2\right)^2} - \frac{1}{2m} \notag\\
& & + o\left(\frac{Mg_c(m)}{m}\right) \notag\\
& \leq & \frac{1}{2} \left(\frac{Mg_c(m)}{m}\right) + o\left(\frac{Mg_c(m)}{m}\right).
\end{eqnarray}

Thus, (\ref{eq: puu 21}) and (\ref{eq: puu 31}) give the desired result.

\section{Continuity and perturbations}  \label{continuity-perturbation}


In this section, under the condition that 
\be
T_{\rm sum}^*(p) \leq f_{\rm ub}(\rho^*) \frac{n}{m^{\alpha}},
\ee
and
\be
p^{\rm lb} (g^*) = 1 - {(M \rho^*)}^{1-{\gamma}} m^{-\alpha} + o(m^{-\alpha}),
\ee 
we want to show that
\be
T_{\rm sum}^*(p) \leq f_{\rm ub}(\rho^*) \frac{n}{m^{\alpha}} + no(m^{-\alpha}), 
\ee
where 
$p = p^{\rm lb}(g^*) = 1 - {(M\rho^*)}^{1-{\gamma}}m^{-\alpha}$.

{


From calculus, We know that
\begin{align}
\label{eq: perturbation 1}
& T^{\rm ub}_{\rm sum}( 1 - {(M\rho^*)}^{1-{\gamma}} m^{-\alpha}) \notag\\
& = T^{\rm ub}_{\rm sum}( 1 - {(M\rho^*)}^{1-{\gamma}} m^{-\alpha} + o(m^{-\alpha})) \\
& + \left . \frac{\frac{d T^{\rm ub}_{\rm sum}}{dg}}{\frac{dp^{\rm lb}}{dg}} \right |_{g = \rho^* m^{-\alpha}} \times o(m^{-\alpha}) + o(o(m^{-\alpha})) \notag\\
& = f_{\rm ub}(\rho^*)\frac{n}{m^{\alpha}} + \left . \frac{\frac{d T^{\rm ub}_{\rm sum}}{dg}}{\frac{dp^{\rm lb}}{dg}} \right |_{g = \rho^* m^{-\alpha}} \times o(m^{-\alpha}) + o(o(m^{-\alpha})).
\end{align} 

Thus, the goal is to compute $\left . \frac{\frac{d T^{\rm ub}_{\rm sum}}{dg}}{\frac{dp^{\rm lb}}{dg}} \right |_{g = \rho^* m^{-\alpha}}$, which requires to compute $\frac{d T^{\rm ub}_{\rm sum}}{dg}$ and $\frac{dp^{\rm lb}}{dg}$.
To obtain $\frac{d T^{\rm ub}_{\rm sum}}{dg}$, we first need to compute the derivative in the form of $F(x) = f(x)^{g(x)}$, which is given by
\begin{eqnarray}
\frac{dF(x)}{dx} & = & F(x) \left(g'(x)\log f(x) + \frac{g(x)}{f(x)}f'(x)\right) \notag\\
& = & f(x)^{g(x)}\left(g'(x)\log f(x) + \frac{g(x)}{f(x)}f'(x)\right).
\end{eqnarray}

Then, by denoting $g_R(m)$ as $g$, we obtain
\begin{align}
\frac{d T^{\rm ub}_{\rm sum}}{dg} &= \frac{\partial}{\partial g} \frac{16}{\Delta^2} \cdot \left(\left(1 - (p^{\rm lb}(g))^{\left(1+\frac{3\Delta}{2}\right)^2 g} \right) \frac{n}{ g}\right) \notag\\
& =  \frac{16}{\Delta^2}\left(- \frac{n}{g} \frac{\partial}{\partial g}\left((p^{\rm lb}(g))^{\left(1+\frac{3\Delta}{2}\right)^2 g} \right)  + \left(1 - (p^{\rm lb}(g))^{\left(1+\frac{3\Delta}{2}\right)^2 g} \right)  \frac{\partial}{\partial g} \left(\frac{n}{g}\right)\right) \notag\\
& = \frac{16}{\Delta^2} \left(- \frac{n}{g}(p^{\rm lb}(g))^{\left(1+\frac{3\Delta}{2}\right)^2 g}\left(1+\frac{3\Delta}{2}\right)^2\left(\log(p^{\rm lb}(g)) + \frac{g}{p^{\rm lb}(g)}\left(\frac{\partial  p^{\rm lb}(g)}{\partial g}\right)\right)  \right. \notag\\
&\quad \left. + \left(1 - (p^{\rm lb}(g))^{\left(1+\frac{3\Delta}{2}\right)^2 g} \right) \frac{\frac{d n}{d g}g - n}{g^2} \right) \notag\\
& = \frac{16}{\Delta^2} \left(- \frac{n}{g}(p^{\rm lb}(g))^{\left(1+\frac{3\Delta}{2}\right)^2 g}\left(1+\frac{3\Delta}{2}\right)^2\left(\log(p^{\rm lb}(g)) + \frac{g}{p^{\rm lb}(g)}\left(\frac{\partial p^{\rm lb}(g)}{\partial g}\right)\right)  \right. \notag\\
&\quad \left. + \left(1 - (p^{\rm lb}(g))^{\left(1+\frac{3\Delta}{2}\right)^2 g} \right) \frac{ - n}{g^2} \right).
\end{align}

Then, we get
\begin{align}
 \frac{\partial p^{\rm lb}(g)}{\partial g} & =  \frac{\partial}{\partial g} \left(1 - \frac{\frac{1}{1-{\gamma}}(Mg)^{1-{\gamma}} - \frac{1}{1-{\gamma}} + 1}{\frac{1}{1-{\gamma}}m^{1-{\gamma}}-\frac{1}{1-{\gamma}}}\right) \notag\\
 & = - \frac{(Mg)^{-{\gamma}}M\left(\frac{1}{1-{\gamma}}m^{1-{\gamma}}-\frac{1}{1-{\gamma}}\right) - \left(\frac{1}{1-{\gamma}}(Mg)^{1-{\gamma}} - \frac{1}{1-{\gamma}} + 1\right)m^{-{\gamma}} \cdot \frac{d m}{d g}}{\left(\frac{1}{1-{\gamma}}m^{1-{\gamma}}-\frac{1}{1-{\gamma}}\right)^2} \notag\\
 &= - \frac{(Mg)^{-{\gamma}}M}{\frac{1}{1-{\gamma}}m^{1-{\gamma}}-\frac{1}{1-{\gamma}}}.
\end{align} 

Therefore, we obtain
\begin{align}
\frac{\frac{d T^{\rm ub}_{\rm sum}}{dg}}{\frac{dp^{\rm lb}}{dg}} &= \frac{16}{\Delta^2}\frac{\left(- \frac{n}{g}(p^{\rm lb}(g))^{\left(1+\frac{3\Delta}{2}\right)^2 g}\left(1+\frac{3\Delta}{2}\right)^2\left(\log(p^{\rm lb}(g)) + \frac{g}{p^{\rm lb}(g)}\left(\frac{\partial  p^{\rm lb}(g)}{\partial g}\right)\right)  \right)}{- \frac{(Mg)^{-{\gamma}}M}{\frac{1}{1-{\gamma}}m^{1-{\gamma}}-\frac{1}{1-{\gamma}}}} \notag\\
&\quad + \frac{16}{\Delta^2}\frac{\left(p^{\rm lb}(g))^{\left(1+\frac{3\Delta}{2}\right)^2 g} \right) \frac{ - n}{g^2}}{- \frac{(Mg)^{-{\gamma}}M}{\frac{1}{1-{\gamma}}m^{1-{\gamma}}-\frac{1}{1-{\gamma}}}} \notag\\
& = \frac{16}{\Delta^2} \frac{- \frac{n}{g}(p^{\rm lb}(g))^{\left(1+\frac{3\Delta}{2}\right)^2 g}\left(1+\frac{3\Delta}{2}\right)^2\log(p^{\rm lb}(g))}{- \frac{(Mg)^{-{\gamma}}M}{\frac{1}{1-{\gamma}}m^{1-{\gamma}}-\frac{1}{1-{\gamma}}}} \notag\\
&\quad + \frac{16}{\Delta^2}\frac{- \frac{n}{g}(p^{\rm lb}(g))^{\left(1+\frac{3\Delta}{2}\right)^2 g}\left(1+\frac{3\Delta}{2}\right)^2\frac{g}{p^{\rm lb}(g)}\left(\frac{\partial  p^{\rm lb}(g)}{\partial g}\right)}{- \frac{(Mg)^{-{\gamma}}M}{\frac{1}{1-{\gamma}}m^{1-{\gamma}}-\frac{1}{1-{\gamma}}}} \notag\\
&\quad + \frac{16}{\Delta^2}\frac{\left(1 - (p^{\rm lb}(g))^{\left(1+\frac{3\Delta}{2}\right)^2 g} \right) \frac{ - n}{g^2}}{- \frac{(Mg)^{-{\gamma}}M}{\frac{1}{1-{\gamma}}m^{1-{\gamma}}-\frac{1}{1-{\gamma}}}}.
\end{align} 

By letting $m \rightarrow \infty$, we obtain
\begin{align}
& \left . \frac{\frac{d T^{\rm ub}_{\rm sum}}{dg}}{\frac{dp^{\rm lb}}{dg}} \right |_{g = \rho^* m^{-\alpha}} \notag\\
& = \frac{16}{\Delta^2}\left(1+\frac{3\Delta}{2}\right)^2\frac{1}{\rho^*} \frac{M^{{\gamma}-1}}{1 - {\gamma}}{\rho^*}^{{\gamma}}\left(e^{-\zeta(\rho^*)} + o(1)\right)\frac{n}{m^\alpha} \log(p^{\rm lb}(g)) m^{\alpha {\gamma}}m^{1-{\gamma}} \notag\\
&\quad - \frac{16}{\Delta^2}\left(1+\frac{3\Delta}{2}\right)^2 \frac{1}{{\rho^*}}\left(e^{-\zeta({\rho^*})} + o(1)\right)\frac{n}{m^\alpha} \frac{{\rho^*} m^\alpha}{p^{\rm lb}(g)} \notag\\
&\quad + \frac{16}{\Delta^2} \left(1-e^{-\zeta({\rho^*})} + o(1) \right) \frac{1}{{\rho^*}^2} \frac{n}{m^{2\alpha}}\frac{M^{{\gamma}-1}}{1 - {\gamma}}{\rho^*}^{{\gamma}}m^{\alpha {\gamma}}m^{1-{\gamma}} \notag\\
& = \frac{16}{\Delta^2}\left(1+\frac{3\Delta}{2}\right)^2 \frac{M^{{\gamma}-1}}{1 - {\gamma}}{\rho^*}^{{\gamma}-1} \left(e^{-\zeta({\rho^*})} + o(1)\right)\frac{n}{m^\alpha} (-(1-p^{\rm lb}(g)) + O\left({(1-p^{\rm lb}(g))}^2\right) m^{\alpha {\gamma}}m^{1-{\gamma}} \notag\\
&\quad - \frac{16}{\Delta^2}\left(1+\frac{3\Delta}{2}\right)^2 \left(e^{-\zeta({\rho^*})} + o(1)\right)\frac{n}{p^{\rm lb}(g)} \notag\\
&\quad + \frac{16}{\Delta^2} \left(1-e^{-\zeta({\rho^*})} + o(1) \right)\frac{M^{{\gamma}-1}}{1 - {\gamma}}{\rho^*}^{{\gamma}-2}\frac{n}{m^{2\alpha}}m^{\alpha {\gamma}}m^{1-{\gamma}} \notag\\
& = -\frac{16}{\Delta^2}\left(1+\frac{3\Delta}{2}\right)^2 \frac{M^{{\gamma}-1}}{1 - {\gamma}}{\rho^*}^{{\gamma}-1} \left(e^{-\zeta({\rho^*})} + o(1)\right) \notag\\
& \quad \cdot \left(M^{1-{\gamma}}{\rho^*}^{1-{\gamma}}\frac{n}{m^\alpha} m^{\alpha(1-{\gamma}) - (1-{\gamma})} m^{\alpha {\gamma}}m^{1-{\gamma}} + O\left(\frac{n}{m^\alpha}m^{\alpha {\gamma}}m^{1-{\gamma}} \frac{1}{m^{2\alpha}}\right) \right) \notag\\
&\quad - \frac{16}{\Delta^2}\left(1+\frac{3\Delta}{2}\right)^2 \left(e^{-\zeta({\rho^*})} + o(1)\right)\frac{n}{p^{\rm lb}(g)} \notag\\
&\quad + \frac{16}{\Delta^2} \left(1-e^{-\zeta({\rho^*})} + o(1)\right)\frac{M^{{\gamma}-1}}{1 - {\gamma}}{\rho^*}^{{\gamma}-2}\frac{n}{m^{2\alpha}}m^{\alpha {\gamma}}m^{1-{\gamma}} \notag\\
& = -\frac{16}{\Delta^2}\left(1+\frac{3\Delta}{2}\right)^{2} \frac{1}{1 - {\gamma}} \left(e^{-\zeta({\rho^*})} + o(1)\right)\left(1+O\left(\frac{1}{m^\alpha}\right)\right)n \notag\\
&\quad - \frac{16}{\Delta^2}\left(1+\frac{3\Delta}{2}\right)^{2} \left(e^{-\zeta({\rho^*})} + o(1)\right)n \notag\\
&\quad + \frac{16}{\Delta^2}\left(1-e^{-\zeta({\rho^*})} + o(1) \right)\frac{M^{{\gamma}-1}}{1 - {\gamma}}{\rho^*}^{{\gamma}-2}n \notag\\
& = \frac{16}{\Delta^2}\left(\frac{M^{{\gamma}-1}}{1 - {\gamma}}{\rho^*}^{{\gamma}-2}\left(1 - e^{-\zeta({\rho^*})} + o(1)\right) \right. \notag\\
&\quad \left. - \left(1+\frac{3\Delta}{2}\right)^{2}\left(e^{-\zeta({\rho^*})} + o(1)\right) - \left(1+\frac{3\Delta}{2}\right)^{2}\frac{1}{1 - {\gamma}}\left(e^{-\zeta({\rho^*})} + o(1)\right) + O\left(\frac{1}{m^\alpha}\right) \right)n \notag\\
& = \frac{16 }{\Delta^2}\left(\frac{M^{{\gamma}-1}}{1 - {\gamma}}{\rho^*}^{{\gamma}-2}\left(1 - e^{-\zeta({\rho^*})}\right) \right. \notag\\
&\quad \left. - \left(1+\frac{3\Delta}{2}\right)^{2}e^{-\zeta({\rho^*})} - \left(1+\frac{3\Delta}{2}\right)^{2}\frac{1}{1 - {\gamma}}e^{-\zeta({\rho^*})} + o\left(1\right) \right)n \notag\\
& = O(n).
\end{align}

Thus, we obtain
\begin{align}
\label{eq: perturbation 2}
& T^{\rm ub}_{\rm sum}( 1 - {(M{\rho^*})}^{1-{\gamma}} m^{-\alpha}) \notag\\
& = T^{\rm ub}_{\rm sum}( 1 - {(M{\rho^*})}^{1-{\gamma}} m^{-\alpha} + o(m^{-\alpha})) + \left . \frac{\frac{d T^{\rm ub}_{\rm sum}}{dg}}{\frac{dp^{\rm lb}}{dg}} \right |_{g = {\rho^*} m^{-\alpha}} \times o(m^{-\alpha}) + o(o(m^{-\alpha})) \notag\\
& = f_{\rm ub}({\rho^*})\frac{n}{m^{\alpha}} + \left . \frac{\frac{d T^{\rm ub}_{\rm sum}}{dg}}{\frac{dp^{\rm lb}}{dg}} \right |_{g = {\rho^*} m^{\alpha}} \times o(m^{-\alpha}) + o(o(m^{-\alpha})) \notag\\
& = f_{\rm ub}({\rho^*})\frac{n}{m^{\alpha}} + O(n) \cdot o(m^{-\alpha}) +  o(o(m^{-\alpha}))  \notag\\
& \leq f_{\rm ub}({\rho^*})\frac{n}{m^{\alpha}} + no(m^{-\alpha}).
\end{align} 
}

\bibliographystyle{IEEEbib}
\bibliography{references}

\end{document}